\documentclass[%
 aip,
amsmath,amssymb,
preprint,%
]{revtex4-1}

\usepackage{graphicx}
\usepackage{dcolumn}
\usepackage{bm}
\usepackage[mathlines]{lineno}

\usepackage[utf8]{inputenc}
\usepackage[T1]{fontenc}
\usepackage{mathptmx}
\usepackage{etoolbox}

\makeatletter
\def\@email#1#2{%
 \endgroup
 \patchcmd{\titleblock@produce}
  {\frontmatter@RRAPformat}
  {\frontmatter@RRAPformat{\produce@RRAP{*#1\href{mailto:#2}{#2}}}\frontmatter@RRAPformat}
  {}{}
}%
\makeatother
\begin{document}

\preprint{AIP/PoP}

\title{Self-pulsing of Dielectric Barrier Discharges at Low Driving Frequencies}

\author{Shanti K. Thagunna}
\affiliation{Department of Space Science, The University of Alabama in Huntsville, Huntsville, AL 35805, USA}
\email{skt0009@uah.edu}

\author{Vladimir I. Kolobov}
\affiliation{Department of Space Science, The University of Alabama in Huntsville, Huntsville, AL 35805, USA}
\affiliation{CFD Research Corporation, Huntsville, AL 35806, USA }

\author{Gary P. Zank}
\affiliation{Department of Space Science, The University of Alabama in Huntsville, Huntsville, AL 35805, USA}

\date{\today}

\begin{abstract}

This paper investigates the self-pulsing of Dielectric Barrier Discharges (DBDs) at low driving frequencies. In particular, (a) the dependence of current on the product \textit{pd} of gas pressure $p$ and the gas gap length $d$, (b) the effects of lossy dielectrics (in resistive discharges) and large dielectric permittivity (in ferroelectrics) on current dynamics, (c) the transition from Townsend to a dynamic Capacitively Coupled Plasma (CCP) discharge with changing \textit{pd} values, and (d) the transition from Townsend to a high-frequency CCP regime with increasing the driving frequency. A one-dimensional fluid model of Argon plasma is coupled to an equivalent RC circuit for lossy dielectrics. Our results show multiple current pulses per AC period in Townsend and CCP discharge modes which are explained by uncoupled electron-ion transport in the absence of quasineutrality and surface charge deposition at dielectric interfaces. The number of current pulses decreases with an increasing applied frequency when the Townsend discharge transforms into the  CCP discharge. The resistive barrier discharge with lossy dielectrics exhibits Townsend and glow modes for the same \textit{pd} value (7.6 Torr cm) for higher and lower resistances, respectively. Finally, we show that ferroelectric materials can amplify discharge current in DBDs. Similarities between current pulsing in DBD, Trichel pulses in corona discharges, and subnormal oscillations in DC discharges are discussed.
\end{abstract}

\maketitle

\section{Introduction}
The Dielectric Barrier Discharge (DBD) is an alternating current (AC) discharge between electrodes covered by dielectrics \cite{Brandenburg_2017}. They have certain properties of Capacitively Coupled Plasma (CCP) sources at low gas pressures and glow discharges at atmospheric gas pressures. DBDs find numerous applications for ozone generation\cite{zhang2016ozone}, decontaminating of biological samples \cite{laroussi1999images}, thin-film deposition \cite{gherardi2000new,borcia2003dielectric}, excimer emission \cite{gellert1991generation,eliasson1988uv}, cold plasma processing \cite{yun2010inactivation,liao2018application}, and many more \cite{brandenburg2023barrier, laroussi2021resistive}. 

Depending upon the operating conditions, gas type, and pressure, a DBD can be spatially homogeneous or filamented. Also, they can operate in a glow or Townsend mode \cite{massines2005glow}. The glow mode, also known as an atmospheric pressure glow discharge (APGD), is distinguished by a higher discharge current reaching hundreds of milliamperes, non-uniform distribution of the electric field within the discharge gap, and plasma quasi-neutrality in the discharge bulk \cite{golubovskii2006effect}. The glow discharge has been studied in air, nitrogen, helium, and other noble gases under atmospheric pressure and is usually characterized by one current pulse per half cycle \cite{WANG2006384, Golubovskii_2003, mangolini2002radial}. APGDs were used for nanomaterial synthesis \cite{shirai2005synthesis}, thin film deposition \cite{girard2005atmospheric}, inactivation of bacteria \cite{vleugels2005atmospheric}, and microplasma thrusters \cite{takao2006miniature}. Conversely, the Townsend mode of discharge at atmospheric pressure, also known as an atmospheric pressure Townsend discharge (APTD), has a much lower discharge current of a few milliamperes, an almost uniform electric field along the discharge axis, and a lower electron density compared to the ion density \cite{golubovskii2006effect}. In this mode, the electron and ion densities within the gap could differ substantially,  and the Debye length is larger than the gas gap \cite{golubovskii2006effect}.
The Townsend discharge mode has applications for detecting volatile organic compounds \cite{muller2009ms}, ultraviolet light emission \cite{nygaard1965ultraviolet}, and thin film deposition \cite{massinens2005atmospheric} to name a few.

The trains of current pulses (or self-pulsation) in Townsend discharges have been observed in experiments and numerical simulations (see Ref. \cite{kolobov2013advances} for further references). The theory presented by Nikandrov and Tsendin, \cite{nikandrov2005low} explained the formation of current pulses in Townsend mode by deposition of surface charges on dielectric surfaces. The Townsend  mode occurs when the ion drift time $\tau_i$ through the discharge gap is much less than the inverse frequency of the applied voltage, i.e.,

\begin{equation}
    \tau_i = \frac{d}{E_{br} \mu_{i}} \ll \frac{2\pi}{\omega},
    \label{eqn:Ion_transition}
\end{equation}
where $E_{br}$ is the breakdown electric field, $d$ is gas gap length, and $\mu_{i}$ is the ion mobility. The opposite extreme, $\omega \tau_{i} > 1$, corresponds to typical conditions of CCP operation with a plasma-sheath discharge structure \cite{lieberman1994principles}.

The self-pulsing of current in filament-free discharges was initially found by Bartnikas et al. \citep{bartnikas1969discharge, Baetnikas} while studying corona discharges in the 1960s.
Akishev et al. \citep{akishev2001pulsed,akishev2002pulsed, akishev2003pulsed} studied the self-pulsing regime of DBD numerically and corona discharges (both numerically and experimentally) and suggested that negative differential resistance is responsible for self-pulsation of current.
Levko et al. \citep{levko2021self} simulated the self-pulsing of subnormal DC discharges using a 2D fluid model and attributed the observed oscillation to ion transit time instability.
Raizer et al. \cite{raizer2006self} studied the self-pulsing current in a Townsend DBD with semiconductor layers. They emphasized the dependence of the secondary emission coefficient on the electric field strength as a source of instability.
Zhang et al. \citep{zhang2018numerical,zhang2018influence} studied the evolution mechanism of self-pulsing and the influence of impurities.

This paper aims to clarify the physics of self-pulsing of DBDs at low driving frequencies and its possible relation to subnormal oscillations in DC discharges\cite{Robert_R_Arslanbekov_2003} and Trichel pulses in corona discharges \cite{zhang2017trichel, sattari2011trichel}. We illustrate that the self-pulsing of the filament-free DBD occurs because of the decoupling of electron and ion transport in the absence of quasineutrality in Townsend discharge. We show that self-pulsing can also occur in CCP operating in a dynamic regime due to the absence of quasineutrality in the sheath. For the purpose of this paper, a one-dimensional fluid model of DBD is adequate. We solve the appropriate set of fluid equations using COMSOL 6.0 and attach an equivalent RC circuit to naked electrodes to model lossy dielectrics. We identify different regimes of discharge operation including quasi-DC (Townsend) discharges and CCP discharges operating in dynamic and high-frequency modes. 

In this paper, we also discuss the dynamic regime of discharge operation. The dynamic regime is characterized by an almost constant (in time) value of plasma density in the gap but a substantial oscillation in electron temperature (mean energy) over an AC period. The dynamic regime was first introduced in Ref. \cite{nikandrov2006collisionless} for the collisionless sheath operation in CCP.  Later, it was identified for Inductively Coupled Plasma (ICP) at low driving frequencies \cite{kolobov2017inductively} and more recently, for a positive column in AC discharged \cite{humphrey2023electron} and CCP \cite{denpoh2020effects}. The dynamic regime occurs when the applied frequency is between the inverse of the energy relaxation time and the ambipolar diffusion time (see Section \ref{sec:time_scale} below for details).

The DBD characteristics depend also on the type of barrier materials used. The use of Resistive barrier discharge (RBD) and ferroelectric barrier discharge (FBD) have been discussed (see Ref.\citep{laroussi2021resistive,laroussi2003plasma,navascues2019large}). The RBD uses high-resistive layers (aka a lossy dielectric), to prevent arcing. The FBD has ferroelectric layers with very high dielectric permittivity for applications in thin film deposition \cite{khan2015negative}, plasma actuators \cite{johnson2014ferroelectric}, seed treatment, and plasma medicine \cite{fridman2008plasma}.

This paper is organized as follows. Section \ref{sec:numerical_model} describes the numerical model, which includes model geometry, fluid-Poisson model, boundary conditions, and reaction process. Section \ref{sec:time_scale} discusses the characteristic scales and introduces a classification of AC discharges. Section \ref{sec:results} contains our study's results, which are followed by the conclusions in Section \ref{sec:conclusions}.

\section{NUMERICAL MODEL}\label{sec:numerical_model}

\subsection{\label{sec:level2} Model Geometry}
\begin{figure}[h!]
    \centering 
    \includegraphics[scale=0.35]{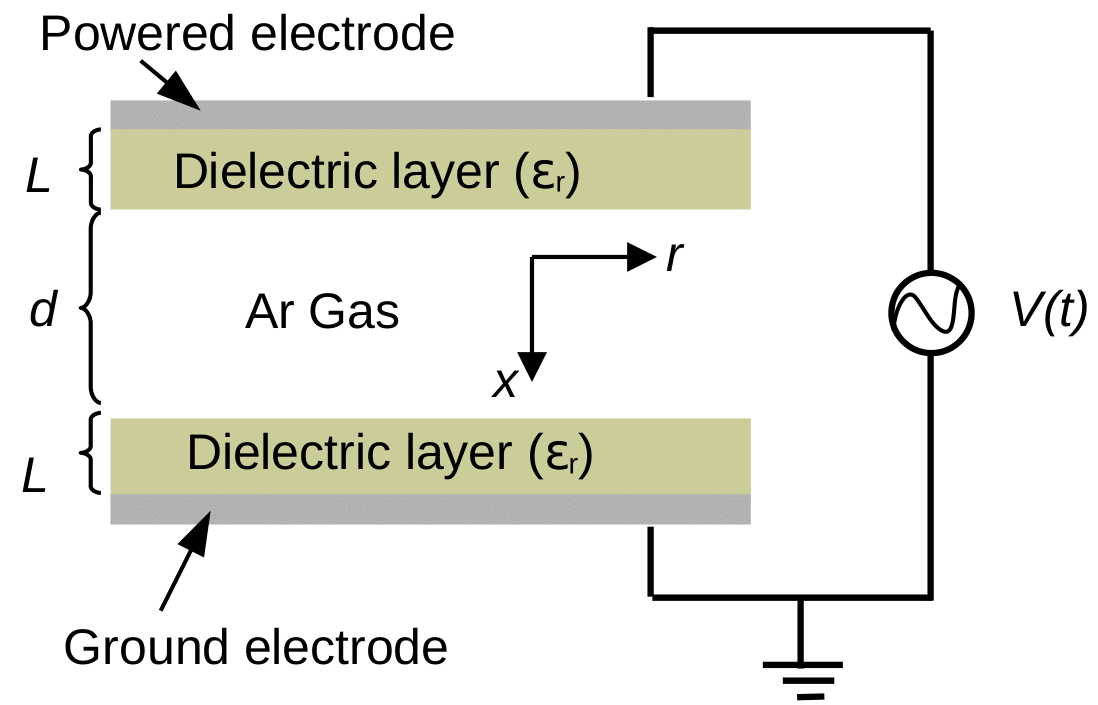}
    \caption{Schematic diagram of a DBD cell. Two dielectric layers of permittivity $\epsilon_{r}$ and thickness $L = 0.1$ mm (for both) are placed in between two metal electrodes (gray rectangle). Ar gas with a separation length $d$ is placed between the two dielectrics at atmospheric pressure. A sinusoidal voltage source is connected to the electrodes.}
    \label{fig:Scheme_DBD}
\end{figure}

A schematic diagram of the DBD used in our simulations is shown in Figure \ref{fig:Scheme_DBD}. Two dielectric layers of thickness $L = 0.1$ mm and dielectric constants $\epsilon_{r}$ are connected to a powered and the ground metal electrodes.  An Ar gas at 1 atmospheric pressure (760 Torr) is placed between the two dielectrics of thickness $L$. An external sinusoidal voltage $V(t)$ of amplitude $V_{0}$ is connected to two metal electrodes. There are a total of 268 grid points spanning between the electrodes, with 200 grid points within the gas gap. We solve the Fluid-Poisson model for 1D geometry.
\subsection{\label{sec:level3} Fluid-Poisson Model}

For weakly ionized plasma, the time evolution of every plasma particle species \textit{p} is obtained by solving the continuity equation \citep{hagelaar2000modeling}:
\begin{equation}
    \frac{\partial n_{p}}{\partial t} +  \frac{\partial \Gamma_{p}}{\partial x} = S_{p},
    \label{eq:cont_ele}
\end{equation}
where $n_{p}$ is the number density, ${\Gamma_{p}}$ is the density flux, $S_{p}$ is the source term, and index $\textit{p}$ denotes electron, ion, or excited atoms.

The density flux is given by
\begin{equation}
    \Gamma_{p} = sgn(q_{p})\mu_{p}n_{p} E  - D_{p} \frac{\partial n_{p}}{\partial x},
    \label{eq:particle_flux}
\end{equation}
where $\mu_{p}$ is the mobility, $q_{p}$ is the particle charge, ${E}$ is the electric field, and $D_{p}$ is the diffusion coefficient. For neutral particles, the first term on the right-hand side of the Eq. \ref{eq:particle_flux} is zero as $q_{p} = 0$.

The source term $S_{p}$ is written in terms of a rate coefficient given by:
 \begin{equation}
     S_p = \sum_{r} c_{p,r}R_{r},
 \end{equation}
where index $r$ refers to a reaction type; $c_{p,r}$ is the net number of particles of species p created in the reaction of type $r$, and $R_{r}$ is the reaction rate for reaction $r$ which is proportional to the densities of the reacting particles, i.e.,
$R_{r} = k^{f}n_{n}n_{p}$ for two-body reactions \citep{lieberman1994principles}, $n_{n}$ is neutral gas density. 

For electron-induced reactions, the rate coefficients have been computed from cross-section data by the following integral:
\begin{equation}
    k^{f} = \sqrt{\frac{2q}{m}}\int_{0}^{\infty} \epsilon \sigma_k(\epsilon) f_0(\epsilon) d\epsilon
\end{equation}
where $m$ is the electron mass, $\epsilon$ is electron kinetic energy, $\sigma_k$ is the collision cross-section, and $f_{0}$ is an electron energy distribution function (EEDF), which is assumed Maxwellian in the present paper.

The continuity equation for electron energy density ($n_{\epsilon} = n_{e} \overline{\epsilon}$), is calculated from the energy balance equation \citep{hagelaar2000modeling}
\begin{equation}
    \frac{\partial n_{\epsilon}}{\partial t} +  \frac{\partial \Gamma_{\epsilon}}{\partial x} =  S_{\epsilon},
    \label{eq:energy_equation}
\end{equation}
where $n_{\epsilon}$ is the electron energy density, $S_{\epsilon}$ is the energy loss or gain due to inelastic collisions, and $\Gamma_{\epsilon}$ is electron energy flux calculated as\citep{hagelaar2005solving}
\begin{equation}
    \Gamma_{\epsilon} =  -\mu_{\epsilon}E n_{\epsilon} - D_{\epsilon}\frac{\partial {n_\epsilon}}{\partial x}.
\end{equation}
where $\mu_{\epsilon}$ and $D_{\epsilon}$ are the electron energy mobility and the energy diffusion coefficients. The electron energy loss is obtained by summing the collisional energy loss overall reactions:
\begin{equation}
    S_\epsilon = -E\Gamma_{e} -n_{e}\sum_{r} k_r n_r \Delta \epsilon_r, 
\end{equation}
where the first term on RHS represents heating by the electric field and the second term represents energy loss in
collisions. Here, $n_{r}$ is the density of the target particles and the $\Delta \epsilon_r$ is the threshold energy.

The electrostatic potential $V$ is calculated from the Poisson equation: 
\begin{equation}
    -\epsilon_{m}\frac{\partial^2 V}{\partial x^2} = \sum\limits_{p}q_{p}n_{p},
\end{equation}
where $\epsilon_{m}$ is the permittivity of the medium.

\subsection{Boundary Conditions}
The boundary and surface conditions used in our simulations are described below. 

The net surface charge $\sigma_{s}$ at dielectric surfaces is obtained from the particle fluxes as
\begin{equation}
\frac{\partial\sigma_{s}}{\partial t} = j_{i} +  j_{e},
\end{equation}
where $j_{e}$ and $j_{i}$ are total electron and ion current densities. 

The boundary condition for the electric potential on a dielectric surface is
\begin{equation}
    \epsilon_{1} E_{1} -  \epsilon_{2} E_{2} = -\sigma_{s},
    \label{eqn:boundary_surface}
\end{equation}
where the subscripts 1 and 2 represent the gas gap and solid dielectric material, respectively. 

The electron flux normal to the electrodes or walls, assuming no reflection, is given by
\begin{equation}
    \Gamma_{e} = \frac{1}{2}n_{e}v_{th,e} - \gamma\sum_{p} \Gamma_{p},
\end{equation}
where the electron thermal velocity is defined as $ v_{th,e} = \sqrt{(\frac{8K_{B}T_{e}}{\pi m})}$, and $\gamma$ is the secondary electron emission coefficient. 

The normal component of the electron energy flux is given by
\begin{equation}
     \Gamma_{\epsilon} = \frac{2}{3}n_{\epsilon}v_{th,e} - \gamma\sum_{p}\epsilon_{p} \Gamma_{p},
\end{equation}
where $\epsilon_{p}$ is the mean energy of the secondary electron emitted by $p^{th}$ species. 

For ions, and excited neutral species, which are lost to the wall due to surface reactions\citep{hagelaar2000modeling}, the flux is:
\begin{equation}
  \Gamma_{p} =  \frac{1}{2}n_{p}v_{th,p}.
\end{equation}

The boundary conditions for the electric potential at the two electrodes are
\begin{equation}
    \phi_{\rm powered}  =  V  =  V_{0} sin(\omega t),
\end{equation}
and
\begin{equation}
    \phi_{\rm ground}  = 0,
\end{equation}
where $\phi_{\rm powered}$ is the potential at the powered electrode and $\phi_{\rm ground}$ is the potential at the ground electrode.

The initial conditions used for the simulations are a gas temperature of 400 K, an electron density of $10^6 \, \rm m^{-3}$, a mean electron energy of 5 eV, and a mean energy of secondary electrons is 2.5 eV.  We define the electron temperature as 
\begin{equation}
    T_{e} = \frac{2}{3} \overline\epsilon ,
\end{equation}
where $\overline \epsilon $ denotes the electron mean energy \cite{kolobov2006simulation} and 1 eV = 11,600 K. Note that the results of the simulation do not change with changes in the initial conditions such as changing the initial electron density to $10^6 \, \rm{m}^{-3}$ from $10^5 \, \rm{m}^{-3}$.

\subsection{Transport Coefficients and Reaction Processes}
The set of volume and surface reactions considered in our simulations are shown in Table \ref{tab:table1} and Table \ref{tab:table2}.
\begin{table}[h]
    \caption{\label{tab:table1} Table of Collisions and Reactions Modeled (electron impact cross section are obtained from Ref. \citep{phelps_database})  }
    \begin{ruledtabular}
    \begin{tabular}{ccccc}
    REACTION & FORMULA & TYPE & $\nabla\epsilon (eV )$ & Rate constant $({m^3}/{s.mol})$\\
    \hline
    1 & e + Ar --> e + Ar & Elastic & 0 & Cross section\\
    2 & e + Ar --> e + $\rm Ar^*$ & Excitation & 11.5 & Cross section\\  
    3 & e + $\rm Ar^*$ --> e + Ar & Superelastic & -11.5 & Cross section\\
    4 & e + Ar --> 2e + $\rm Ar^{+}$ & Direct Ionization & 15.8 & Cross section\\
    5 & e + $\rm Ar^{*}$ --> 2e + $\rm Ar^{+}$ & Stepwise Ionization & 4.3 & Cross section\\
    6 & $\rm Ar^*$ + $\rm Ar^*$ --> e + Ar + $\rm Ar^{+}$ & Penning ionization & -  &3.4$\times 10^{8}$\\
    7 & $\rm Ar^*$ + Ar --> Ar + Ar & Metastable quenching & -  &1807\\
    \end{tabular}
    \end{ruledtabular}
\end{table}
\begin{table}[h]
    \caption{\label{tab:table2} Table of Surface Reactions Modeled}
    \begin{ruledtabular}
    \begin{tabular}{ccc}
    REACTION & FORMULA & STICKING COEFFICIENT\\
    \hline
    1 & $\rm Ar^*$ --> Ar & 1\\
    2 & $\rm Ar^{+}$ --> Ar & 1\\  
    \end{tabular}
    \end{ruledtabular}
\end{table}

 In our simulation, the electron diffusivity ($D_{e}$), energy mobility ($\mu_{\epsilon}$), and energy diffusivity (or thermal diffusivity) ($D_{\epsilon}$) are computed from the electron mobility using:
 $D_{e} = \mu_{e}T_{e}$, $\mu_{\epsilon} = \frac{5}{3}\mu_{e}$ , $D_{\epsilon} = \mu_{\epsilon} T_{e}$.

The electron mobility is calculated by using the expression, $\mu_{e} = \frac{e}{m\nu_{ea}}$, and the ion mobility is given by $\mu_{i} = 0.12 \frac{m^{2}}{V s} \frac{1 Torr}{p}$, taken from book \cite{raizer1991gas, humphrey2023electron}.
\section{Characteristics scales and Classification of AC Discharges}\label{sec:time_scale}

\begin{figure}[h!]
    \centering
    \includegraphics[scale = 0.2]{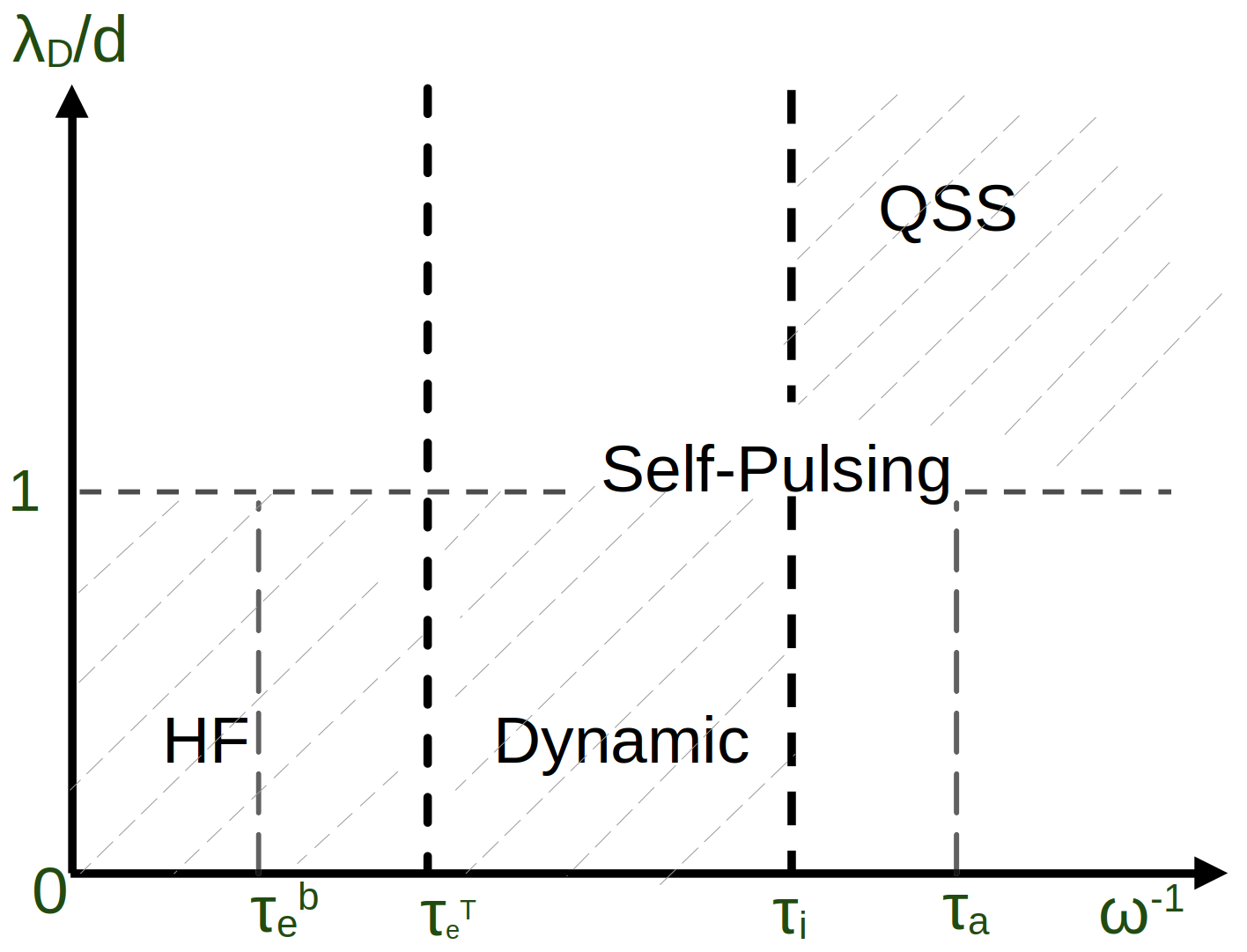}
    \caption{A schematic diagram representing different characteristic scales used for the classification of AC discharges where $\tau_{e}^{b}$ the electron drift time scale, $\tau_{e}^{D}$ is the electron diffusion time scale, $\tau_{i}$ ion time scale, and $\tau_{a}$ is the ambipolar diffusion time scale. The abbreviation QSS is for Quasi-stationary state and HF is for High frequency. The self-pulsing is a common feature that we observed in dynamic and QSS regimes in our paper. }
    \label{fig:scale_characterstics}
\end{figure}
The ion drift time across the gap is characterized by the time ($\tau_{i}$) defined in Equation (\ref{eqn:Ion_transition}). Ion diffusion can usually be neglected because the ion thermal velocity is small compared to its drift velocity when the electric field is not close to zero. The time scale corresponding to the electron drift \cite{lieberman1994principles}
\begin{equation}
    \tau_{e}^{b} = \frac{d}{E_{br}\mu_{e}},
    \label{eq:tau_drift}
\end{equation}
is much shorter because of the large electron mobility.

The time scale for electron diffusion across the gap is given by \cite{raizer1991gas} 
\begin{equation}
    \tau_{e}^{D} = \frac{d^{2}}{2 D_{e}},
    \label{eq:tau_diff}
\end{equation}
where $D_{e}$ is the electron diffusion coefficient. 

 The characteristics time scales coming from the electron energy equation include the electron thermal diffusion time scale ($\tau_{e}^{T}$) which is of the order ($\tau_{e}^{D}$) and the energy relaxation time ($\tau_{\epsilon}$) which describes the electron cooling rate in collisions with neutrals. The latter is calculated by using the expression \cite{avtaeva2009effect}
\begin{equation}
    \tau_{\epsilon}  \approx  \frac{M}{2 m \nu_{ea}},
\end{equation}
where $M$ is the mass of the atom.
AC discharges can operate in different regimes depending on the value of the driving frequency with respect to the time scales introduced above.

In quasi-neutral plasma, the electric field forces the electrons and ions to diffuse at equal rates. The charged particle dynamics are characterized by the slow ambipolar diffusion time scale $\tau_{a}$, defined as: 
\begin{equation}
    \tau_{a} = \frac{d^{2}}{D_{a}},
\end{equation}
where $D_{a}$ is the ambipolar diffusion coefficient. As a result, the plasma density does not change substantially during the AC period, but the electron temperature oscillates substantially over time. 
The electron temperature is nearly uniform over space because the electron thermal diffusion occurs fast (at the free electron diffusion time scale) in both the sheath and plasma. 

The dynamic regime of a collisionless sheath in CCP was first considered in detail by Nikandrov and Tsendin\cite{nikandrov2006collisionless} for the frequency interval $\omega \tau_{i} < 1 < \omega \tau_{B}$, where $\tau_{B} = \frac{d}{\sqrt{{T_e}/{M}}}$. In our case of the collisional sheath, the motion of the electrons and ions is controlled by drift. Hence, the dynamic regime for the collisional sheath occurs in the frequency range $\omega \tau_{e}^{b} < 1 < \omega \tau_{i}$. In the plasma, the dynamic regime corresponds to the frequency interval $\omega \tau_{\epsilon} < 1 < \omega \tau_{a} $\cite{kolobov2017inductively}. In CCP discharge with a plasma-sheath structure, both the sheath and the plasma can be in the dynamic regime at low driving frequencies.

By increasing $\omega$, the high-frequency discharge regime corresponds to ($\omega \tau_{\epsilon} > 1 $). In this regime, the
electrons and ions respond to the average field in plasma, but electrons respond to the instantaneous field in the sheath. Electron temperature and the ionization rate are controlled by the time-average field in the plasma.

We performed simulations for different regions of Figure \ref{fig:scale_characterstics} by changing gas gap length (\textit{d}),
pressure (\textit{p}), applied voltage (\textit{V}), and driving frequencies (\textit{f}). The results of these simulations are
presented in section sections \ref{sec:results}. 

\section{Results and Discussions} \label{sec:results}

We investigate the characteristics of an Argon DBD at atmospheric pressure and the transition from Townsend mode to CCP  by changing \textit{pd}. In addition, we also investigate the quasi-DC discharge, lossy DBD discharge, and ideal DBD discharge modes using the circuit equivalent of a resistive barrier discharge. Moreover, we study the transition from the Townsend mode of discharge to a dynamic CCP mode with a change in frequency in an ideal DBD. We use 1D simulations using the fluid-Poisson model in COMSOL and illustrate the physics of current multi-pulses and discharge characteristics. The details of each investigation are described below.

\subsection{Oscillations of Current Density in the Townsend Mode} \label{subsec:oscillation_townsend}
\begin{figure}[h!]
    \centering 
    \includegraphics[scale=0.99]{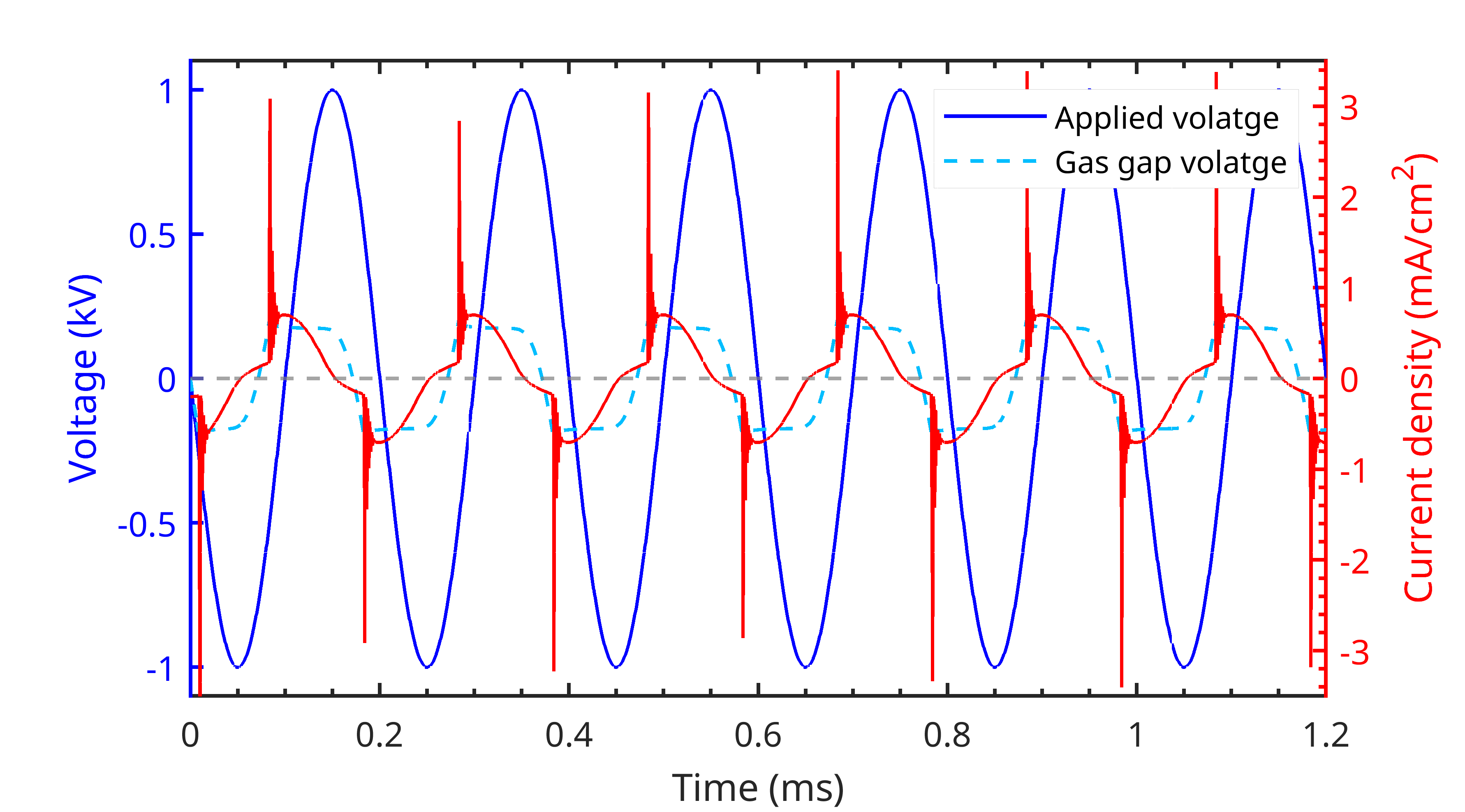}(a)\\
    \includegraphics[scale=0.48]{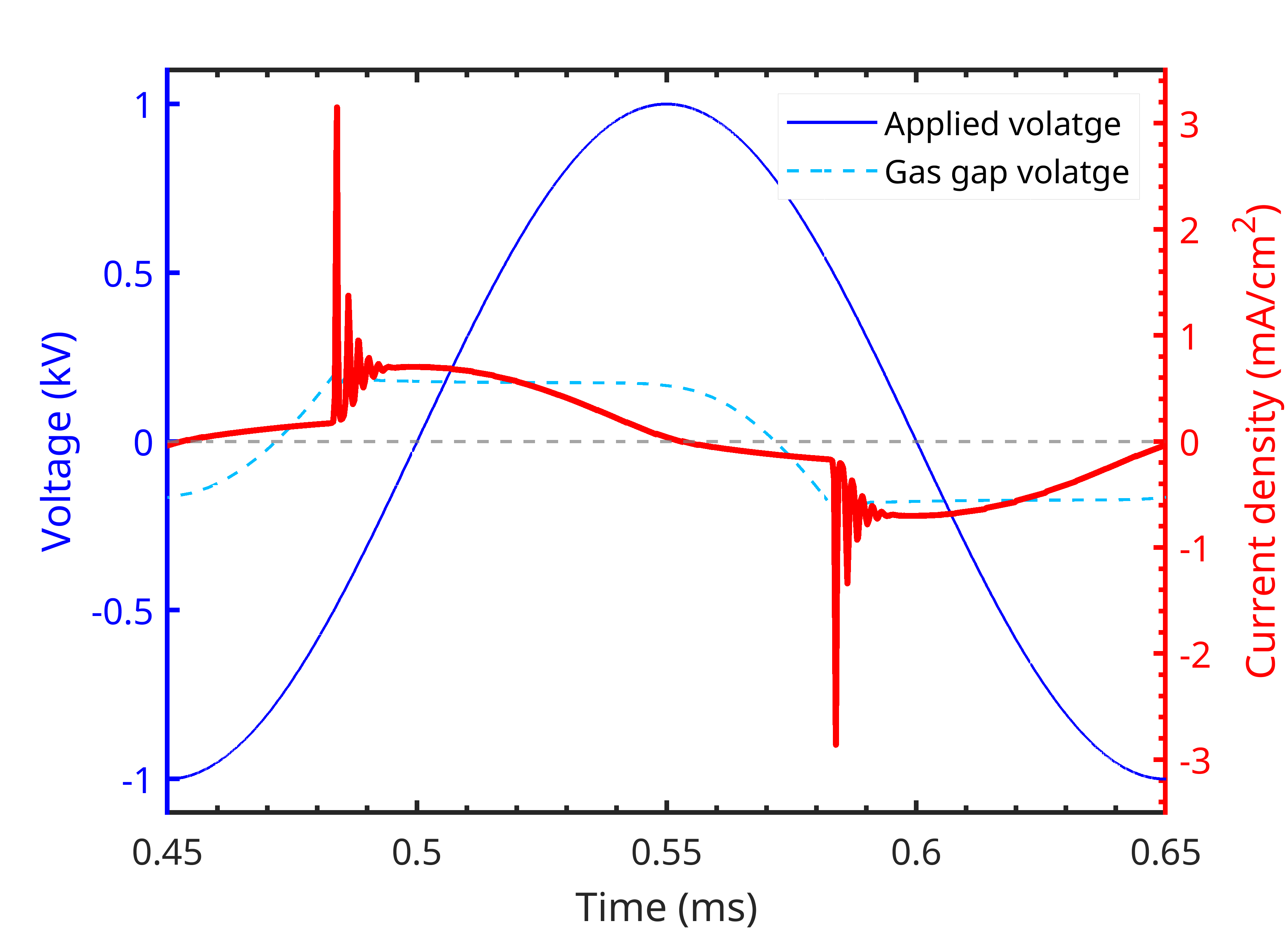}(b)
    \includegraphics[scale= 0.48]{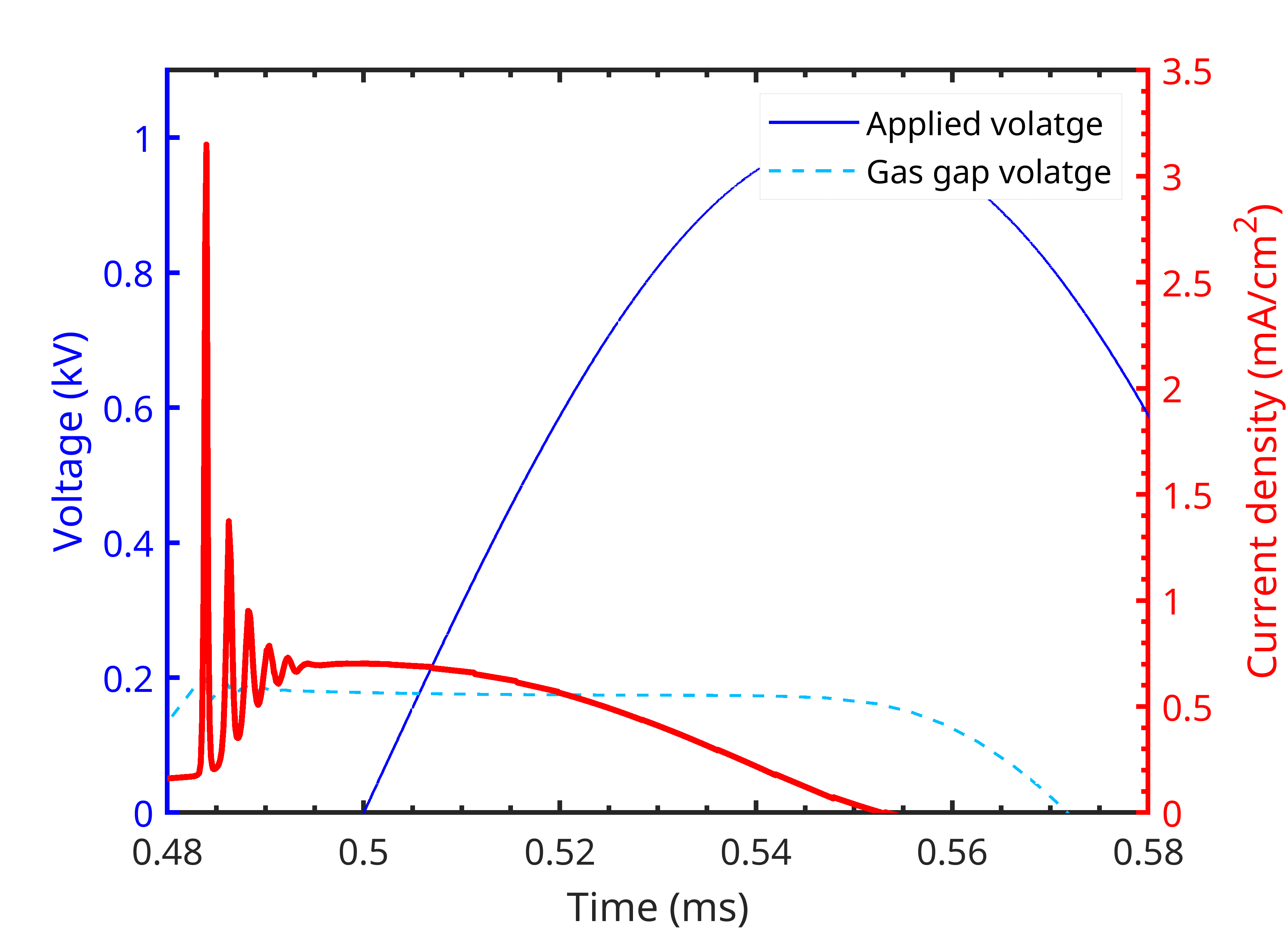}(c)
    \caption{(a) The time evolution of applied voltage (solid blue curve), gas gap voltage (dashed light blue curve), and terminal current density (red curve) over six cycles for Argon plasma at frequency 5 kHz, applied voltage amplitude 1000 V, and gas gap distance of 0.1 mm. (b) A single cycle over the time $0.45-0.65$ ms  showing current density multi-pulses in positive and negative half cycles with a breakdown voltage of approximately 500 V, and (c) time-resolved pulses of the first half cycle}
    \label{fig:multipeaks_basecase}
\end{figure}
In this section, we discuss the oscillating behavior, i.e., the self-pulsing nature of current density (measured near the powered electrode) in the 1D DBD simulation of Ar gas. The discharge parameters are a frequency of 5 kHz, applied voltage amplitude 1000 V, gas gap distance 0.1 mm, secondary emission coefficient 0.05, and relative permittivity 5, i.e., $\epsilon_{r} = 5$. Figure \ref{fig:multipeaks_basecase} (a) shows the time evolution of applied voltage and terminal current density over six cycles with a period ($\tau$) of $0.2$ ms. The zoomed-in view of the current-voltage characteristics curves for one cycle (from 0.45-0.65 ms) in Figure \ref{fig:multipeaks_basecase} (b) shows six pulses. The current density pulses are more distinctly visible in Figure \ref{fig:multipeaks_basecase} (c). Note that the amplitude of the current density pulses decreases from the first to the sixth peak. The higher amplitude of the current density during pulses is because of the electron current and the lower value current density between the pulses comes from the ion current. Also, there is a phase shift for current density and applied voltage because of the capacitive coupling.

Now we investigate the mode of discharge in our simulation (Townsend vs. glow mode). For this, we theoretically estimate the ion transit time using equation (\ref{eqn:Ion_transition}) and ion mobility ($\mu_{i}$) as a function of $\frac{E}{N}$ in Td (1 Td = 1 Townsend = $10^{-21}$ \rm{V} $\rm{m^{2}}$) following Basurto et. al \cite{PhysRevE.61.3053}.
The breakdown electric field required for equation (\ref{eqn:Ion_transition}) is given by $E_{br} = \frac{U_{br}}{d}$, where $d$ is gas gap distance and $U_{br}$ is the breakdown voltage, which is calculated from the simulation. The breakdown voltage is calculated as the difference between voltages measured at the dielectric-plasma interface node of the computational domain, which is 200 V for this case. The theoretically calculated value of the ion transit time is 24 $\mu$s.
The ion transit time satisfies the condition, $\omega \tau_{i} < 1$, and the discharge is the Townsend mode\cite{nikandrov2005low}.
This is a quasi-static regime (quasi-stationary state) when the conduction current flows as a train of current pulses associated with the deposition of surface charges on dielectrics.

An interesting phenomenon that we observe in the Townsend discharge mode is the low-frequency oscillation of the discharge current density (self-pulsing), as shown in Figure \ref{fig:multipeaks_basecase}. These oscillations appear because of the time lag between ion formation near the instantaneous cathode (ground electrode for positive half cycle) and the current due to ion-electron emission. These pulses are a result of electron avalanches initiated by the background electrons and secondary electrons emitted from the instantaneous cathode due to ion bombardment.
A gas breakdown occurs when the gap voltage exceeds the breakdown voltage. Rapid multiplication of seed electrons is linked to neutral gas species being ionized by electron impact and the growth of an electron avalanche. The newly generated electron-ion pairs move to the electrodes, driven by an electric field and diffusion. 
In the absence of plasma, the transport of electrons and ions occurs independently. Electrons drift and diffuse much faster than ions, and electron drift dominates over diffusion. As the breakdown voltage is much greater than the electron temperature, electrons drift to the instantaneous anode, and a net negative surface charge is deposited at the anode surface. The gap voltage decreases below the breakdown voltage, and the electron multiplication process stops. The above process corresponds to the first current pulse in the discharge curve of Figure \ref{fig:multipeaks_basecase}. This is how a current pulse is formed. At low driving frequencies ($\omega \tau_{i} <1$) ions have enough time to reach the instantaneous cathode during the half-period of the applied voltage increasing the electric field in the gap. The second current pulse is formed when the electric field again exceeds the breakdown value. Thus, the distance between the peaks is about the ion transit time in the Townsend discharge mode. Self-pulsing of DBD is similar to the self-pulsing of DC and corona discharge. It happens in the Townsend regime when the electron and ion transport is decoupled because the electric field is weakly perturbed by space charges \cite{zhang2023numerical}.  We discuss further details of the discharge characteristics of this mode in Subsection \ref{subsec:sp_temp_evl} below.

In Figure \ref{fig:Helium_model}, we present the validation of our model for helium discharge. As experimental data for current pulsing in filament-free DBD in the Townsend mode was unavailable, we conducted a simulation based on the experimental results of Shin et al.,\citep{shin2003dynamics} and compared our results with the theoretical results presented by Nikandrove and Tsendin \cite{nikandrov2005low}. Though our model is oversimplified in terms of chemical and ionization kinetics and neglects photo-electron emission, we observe that our simulation results have reasonable agreement with both experiment and theory. We obtained four peaks at 500 Hz and ten peaks at 100 Hz. Additionally, we determined that the separation between peaks for 500 Hz applied frequency is approximately 2.8 $\mu$s, which is the same order as ion transition time, i.e., $\tau_{i} = 3.3  $ $\mu$s. This validates our statement that the distance between the peaks is about the ion transition time in a Townsend discharge.  Note that the ion transit time in Argon is larger than the ion transit time in Helium because of the mass difference. Therefore, achieving a Townsend regime in Argon is more difficult than in Helium.
\begin{figure}[h!]
    \centering
    \includegraphics[scale=1.3]{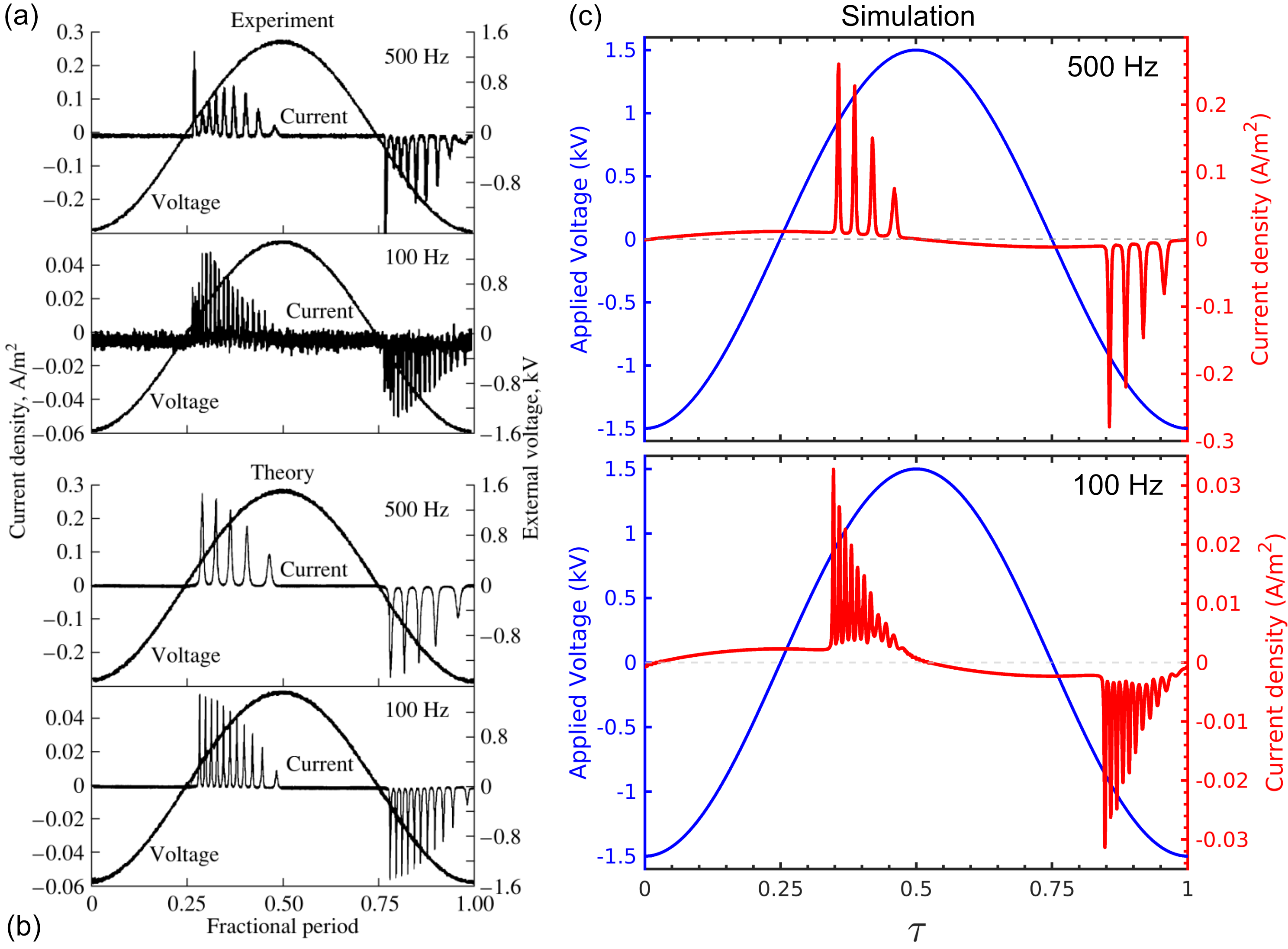}
    \caption{The model validation for Helium plasma in Townsend mode. (a) The experimental results by Shin et al., \citep{shin2003dynamics}, (b) theoretical results by Nikandrov and Tsendin \citep{nikandrov2005low}, and (c) simulation results at driving frequencies of 500 Hz and 100 Hz.}
    \label{fig:Helium_model}	
\end{figure}

\subsection{Spatiotemporal Evolution of a Discharge} \label{subsec:sp_temp_evl}
\subsubsection{Spatiotemporal Evolution of the Electric Field}
\begin{figure}[h!]
\centering
    \includegraphics[scale=.054]{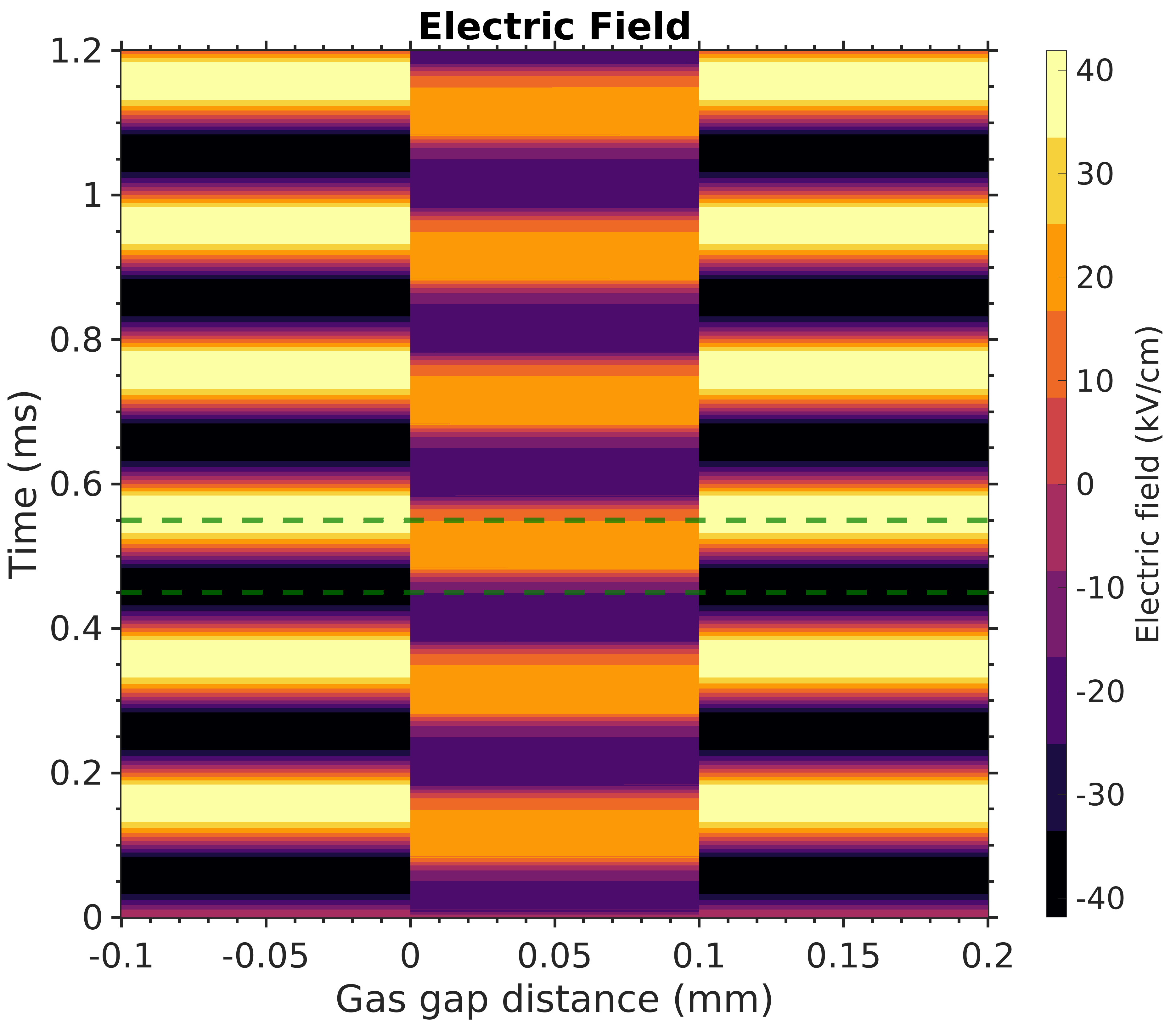}(a)
    \includegraphics[scale=.55]{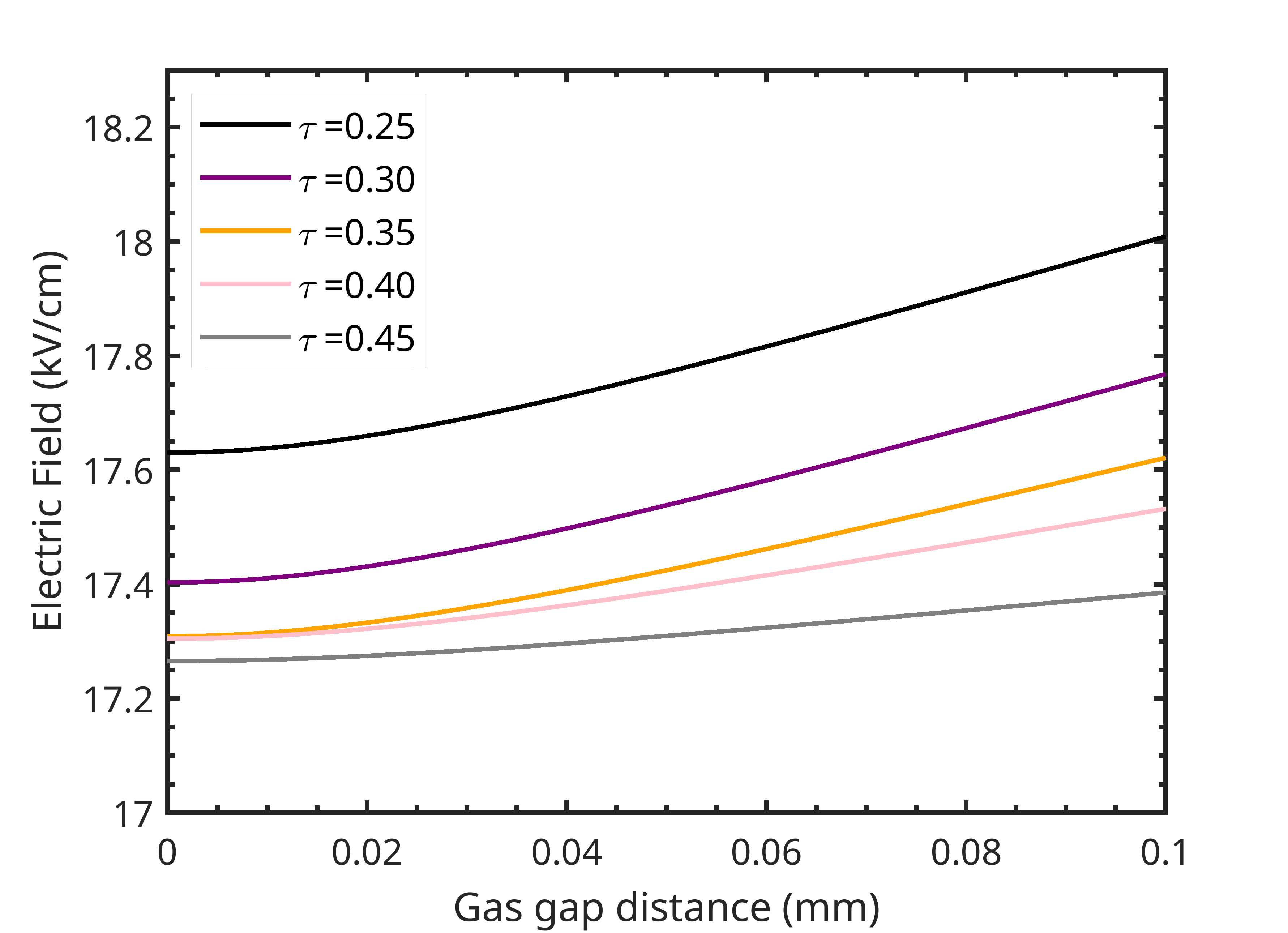}(b)
    \caption{(a) The spatiotemporal evolution of the electric field over the full simulation, including both dielectric and gas gaps, over the full six cycles. Here the dashed green line shows the time for the half cycle $0.45-0.55$ ms for which the electric field is plotted in panel (b). Panel (b) shows the spatial evolution of the electric field during five different phases of the half cycle of the AC period from $0.45-0.65$ ms, where $\tau = \frac{(\rm Time - 0.45)}{(0.65 - 0.45)}$ is the normalized time scale and represents different phases within the considered half cycle.}
    \label{fig:Electric_Field}	
\end{figure}

Figure \ref{fig:Electric_Field} (a) illustrates the spatiotemporal evolution of the electric field during the six cycles shown in Figure \ref{fig:multipeaks_basecase}. The electric field appears to be constant throughout the gas gap over all six cycles. There is a change in the magnitude of the electric field from a positive to a negative value over every half cycle. Panel (b) represents the spatial distribution of the electric field during five different phases of the half cycle of the AC period from time $0.45-0.55$ ms. The electric field remains constant throughout the gas gap for all five different phases. The uniform electric field over the gas gap distance indicates that the discharge is in Townsend mode.
\subsubsection{Spatiotemporal Evolution of the Electron and Ion Density}\label{subsec:sptemp_ele_ion}
\begin{figure}[h!]
\centering
     \includegraphics[scale=0.055]{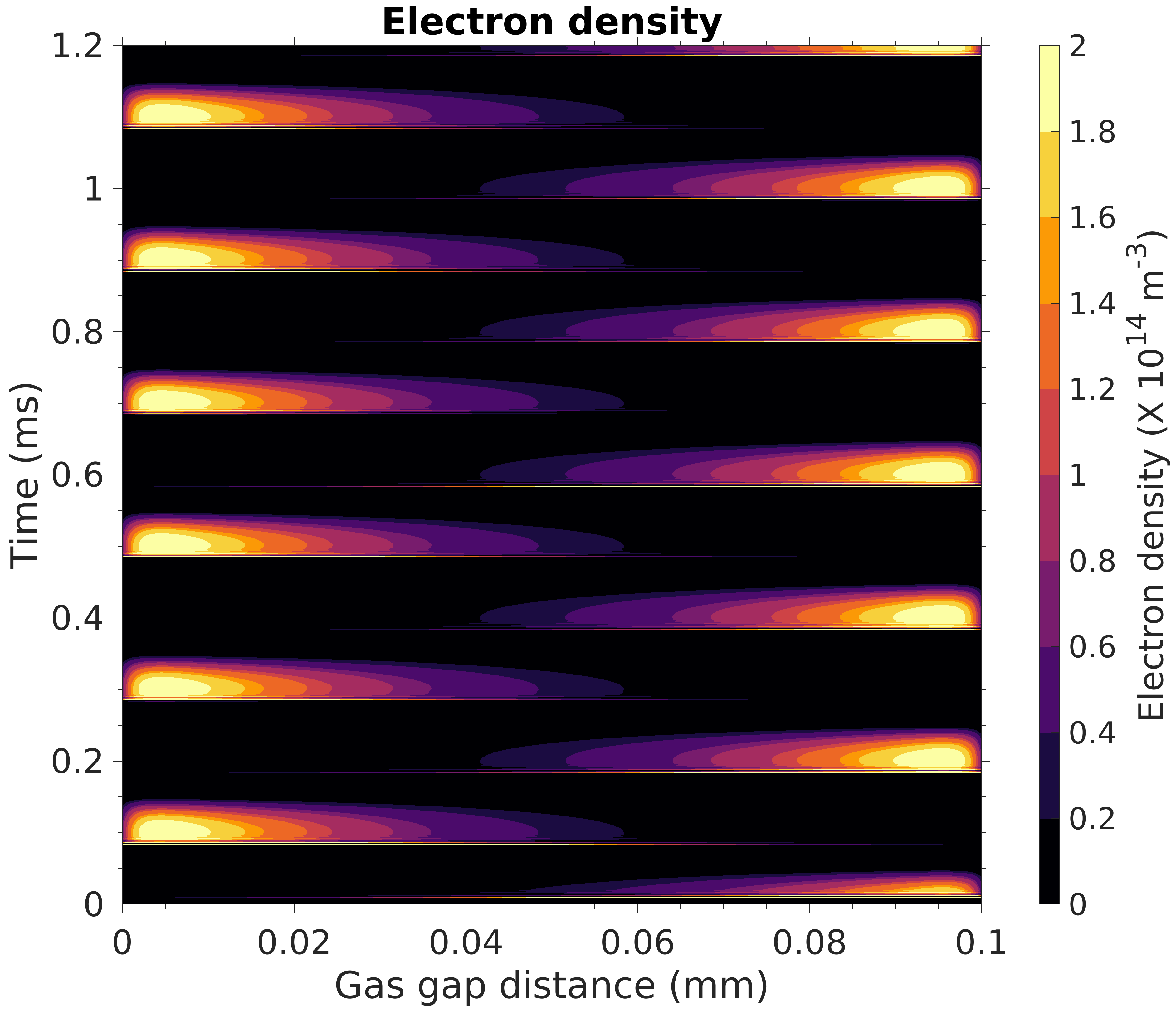}(a)
     \includegraphics[scale=.055]{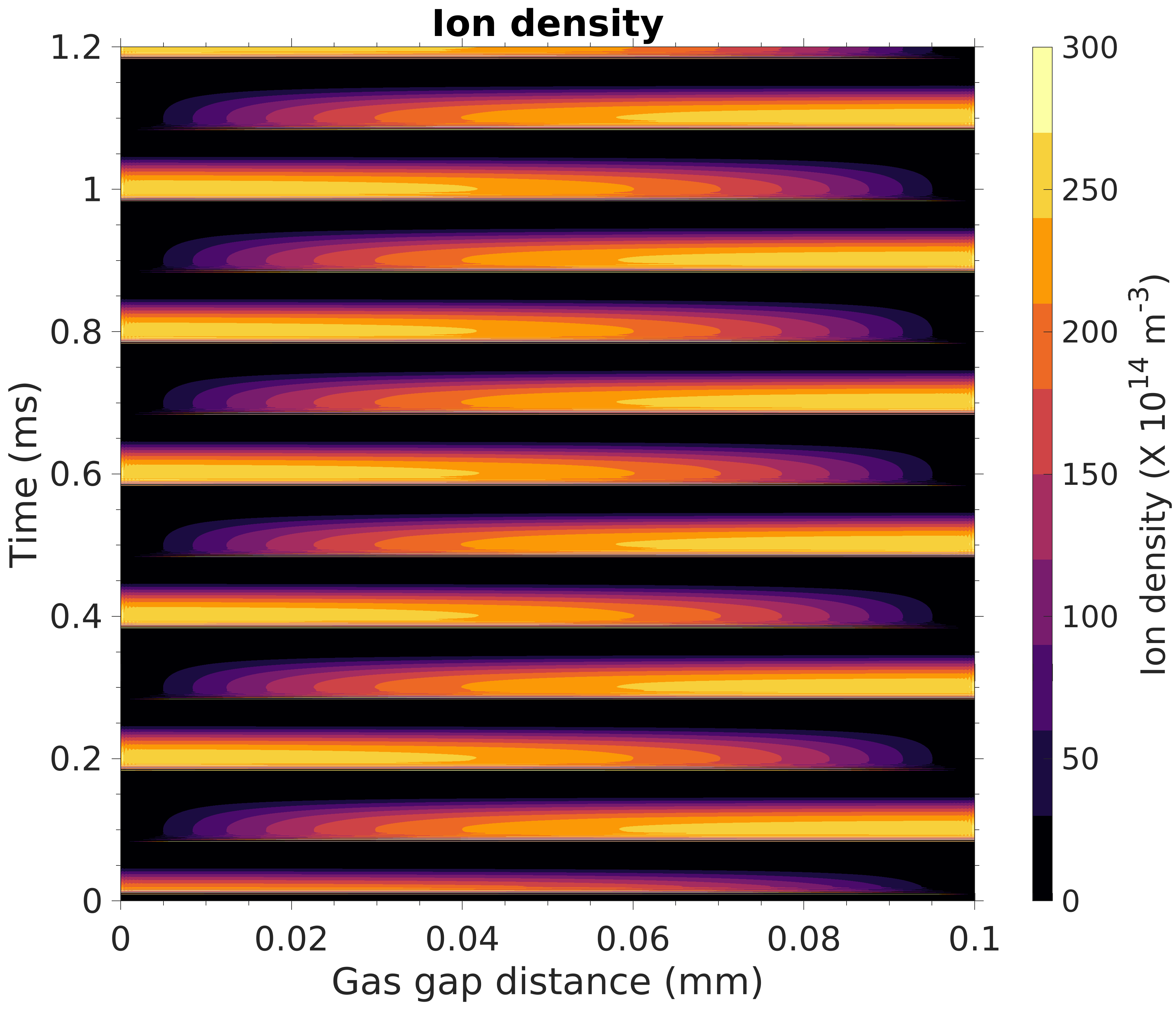}(b)
     \includegraphics[scale=.055]{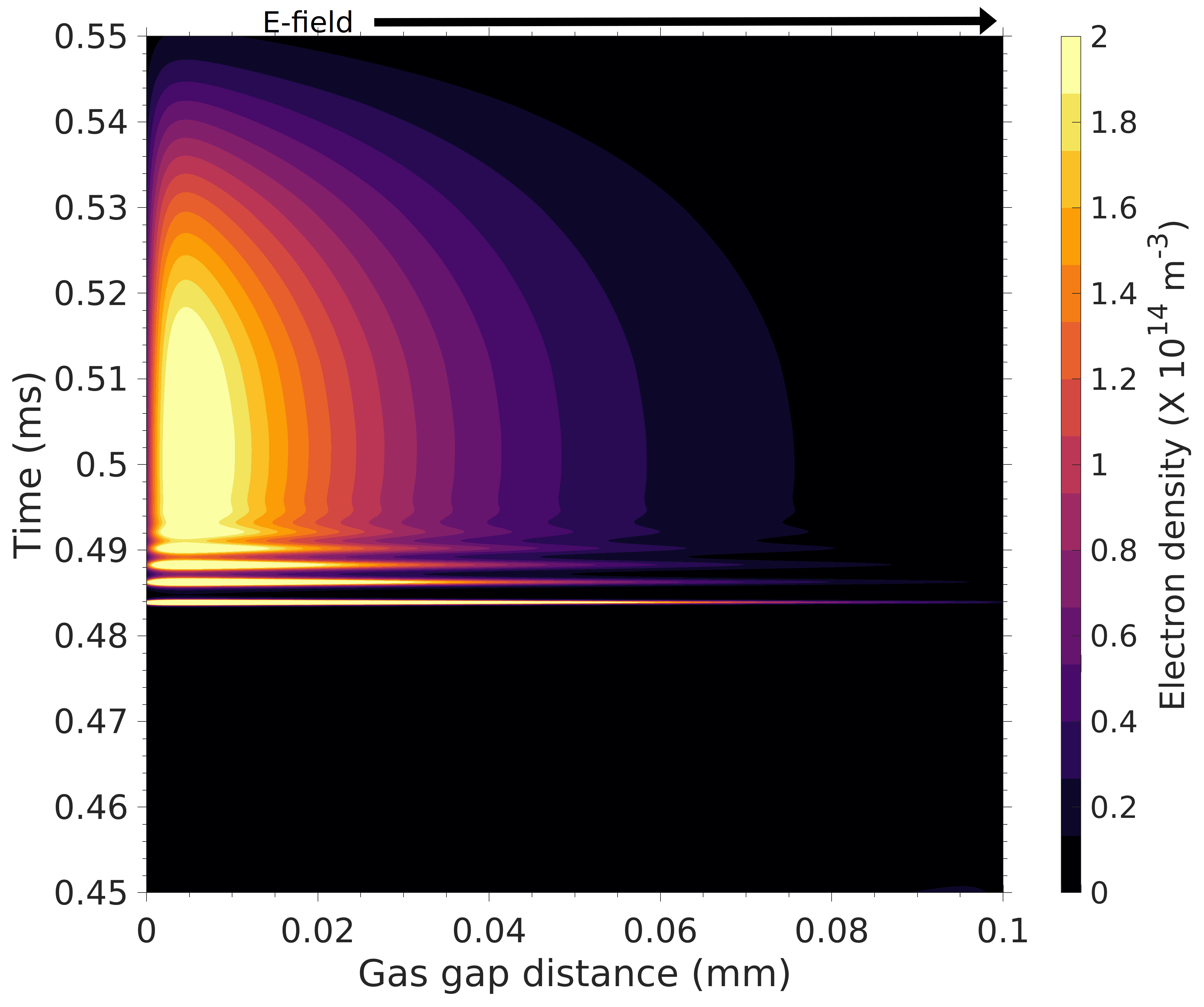}(c)
     \includegraphics[scale=.055]{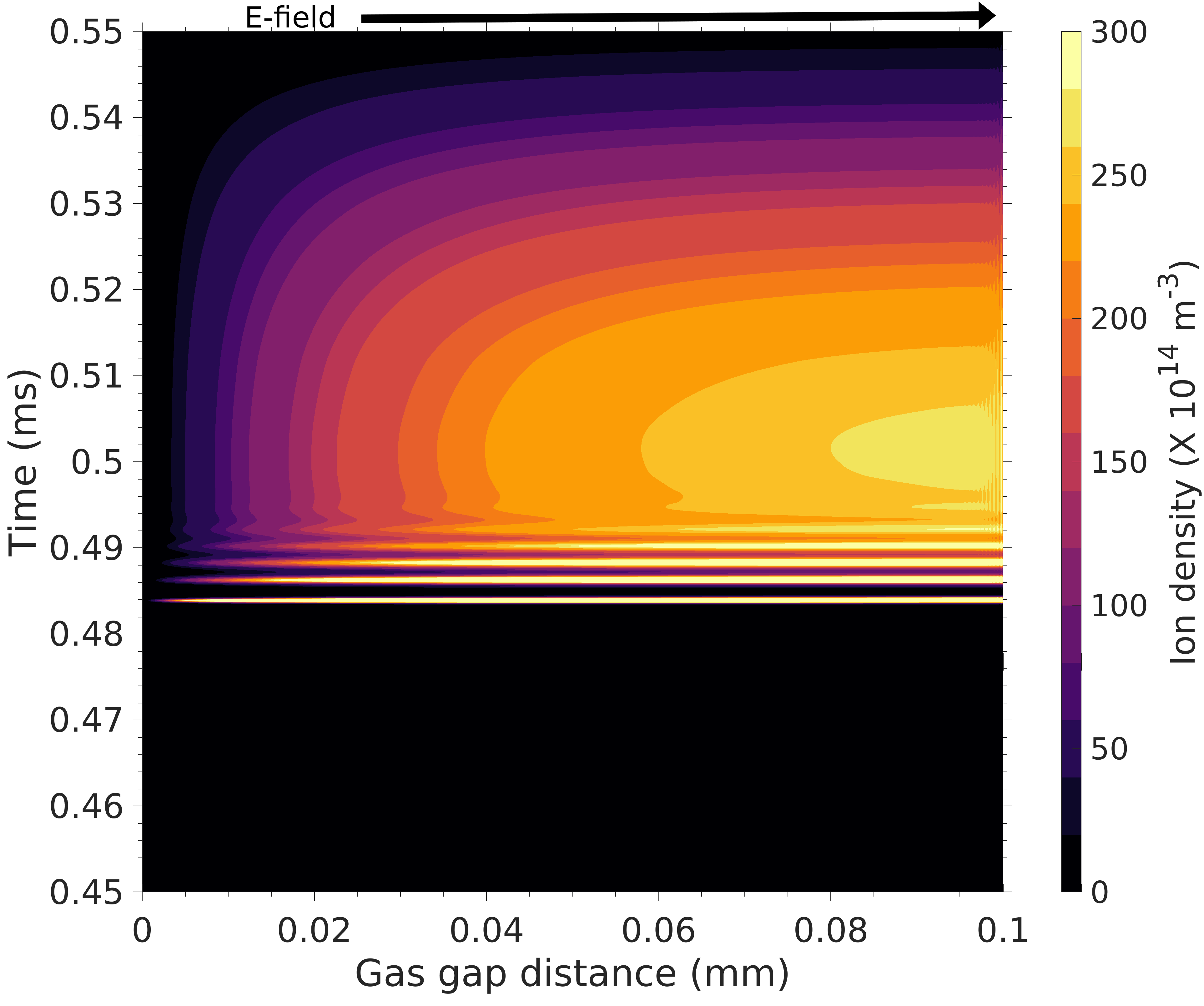}(d)
    \caption{The spatiotemporal evolution of the (a) electron and (b) ion density over the six cycles. The bottom panels show the spatiotemporal evolution of the (c) electron and (d) ion density over the half-cycle from $0.45-0.55$ ms.}
    \label{fig:electron_ion_density}	
\end{figure}
Figure \ref{fig:electron_ion_density} displays the spatiotemporal evolution of the electron density (left panels) and ion density (right panels). The top panels show the electron and ion densities over the six cycles and the bottom panels show the same quantities over the half-cycle from $0.45-0.55$ ms. Before the discharge, the electron density is uniform and has a very small value along the discharge axis (see panel (c)). During gas discharge, seed electrons become energized, which initiates an avalanche. The electron multiplication in the gap causes a gradual increase in the electron density towards the anode. After reaching the peak value, the electron density starts to decrease over time. The increase in electron density is seen during all six current density pulses.
Similarly, before the discharge, the ion density is uniform along the discharge axis and has a very small value as in the electron density. The ion density then swiftly increases during the discharge, near the instantaneous cathode, reaches a peak value, and then starts to decrease again. The increase in ion densities is seen during all six current density pulses. 

\begin{figure}[h!]
    \centering
    \includegraphics[scale=.9]{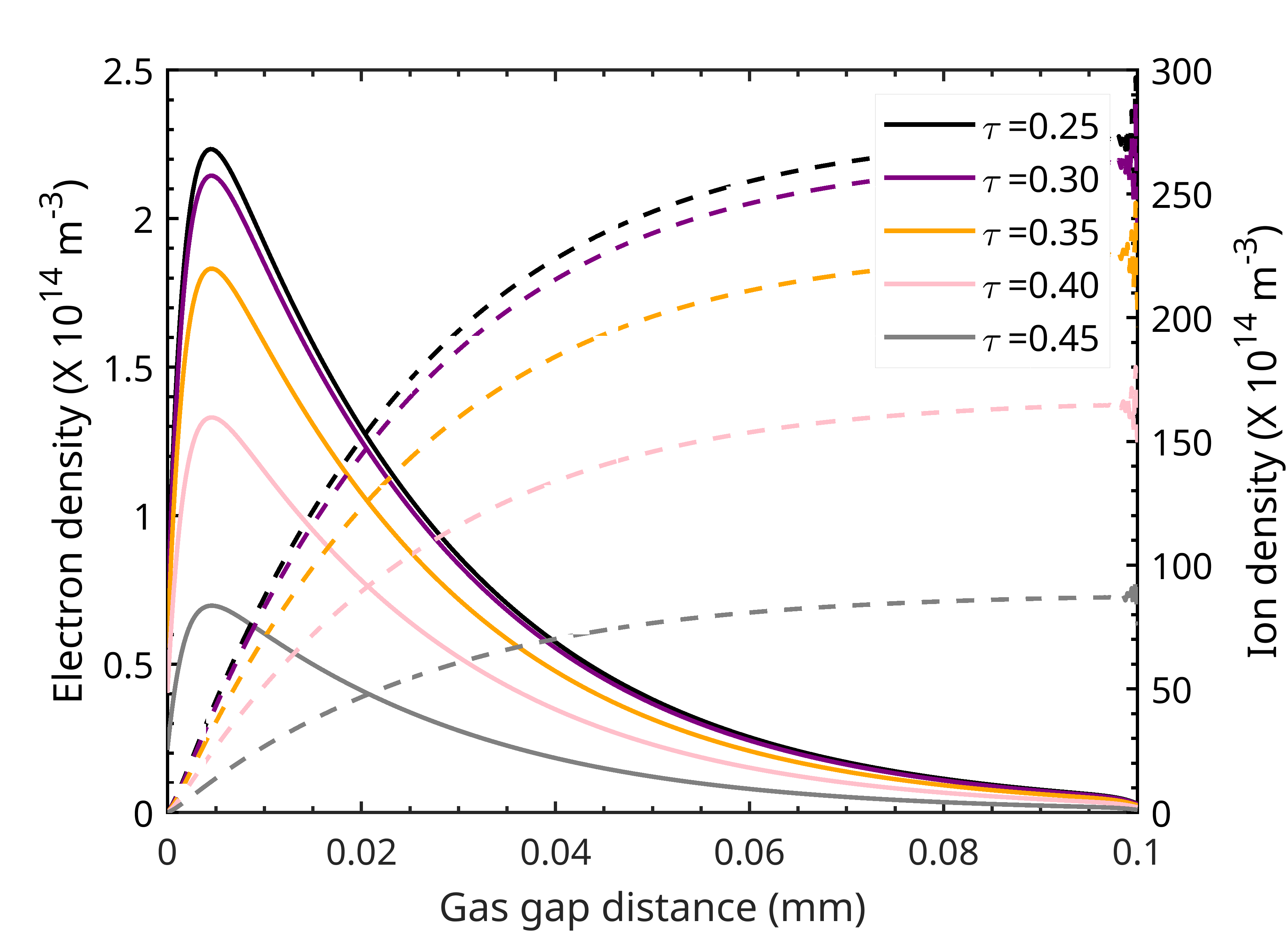}
    \caption{The spatial distribution of electron (solid curves) and ion densities (dashed curves) during the five different phases of the half cycle of the AC period from $0.45-0.65$ ms. The left ordinate corresponds to the electron density and the right ordinate to the ion density.}
    \label{fig:electron_ion_density_peaks}	
\end{figure}

Figure \ref{fig:electron_ion_density_peaks} displays the spatial evolution of the electron and ion densities during the five different phases of the half cycle of the AC period from $0.45-0.55$ ms. During each phase of half AC period, the electron density is maximum near the instantaneous anode (left side) and gradually decreases towards the instantaneous cathode (right side) whereas the ion density gradually increases towards the instantaneous cathode. In addition, the spatial ion density profile shows some oscillations along the discharge axis that are seen as spikes in the ion density profile near the instantaneous cathode. The ion density is around two orders of magnitude larger than the electron density during the pulse. This is a typical characteristic of the Townsend mode discharge where quasi-neutrality ($n_{i} = n_{e}$) is not observed. This behavior strongly supports the conclusion that the discharge is in the Townsend mode.   
\subsubsection{Temporal Variation of Surface Charge Density}
\begin{figure}[h!]
\centering
    \includegraphics[scale=.65]{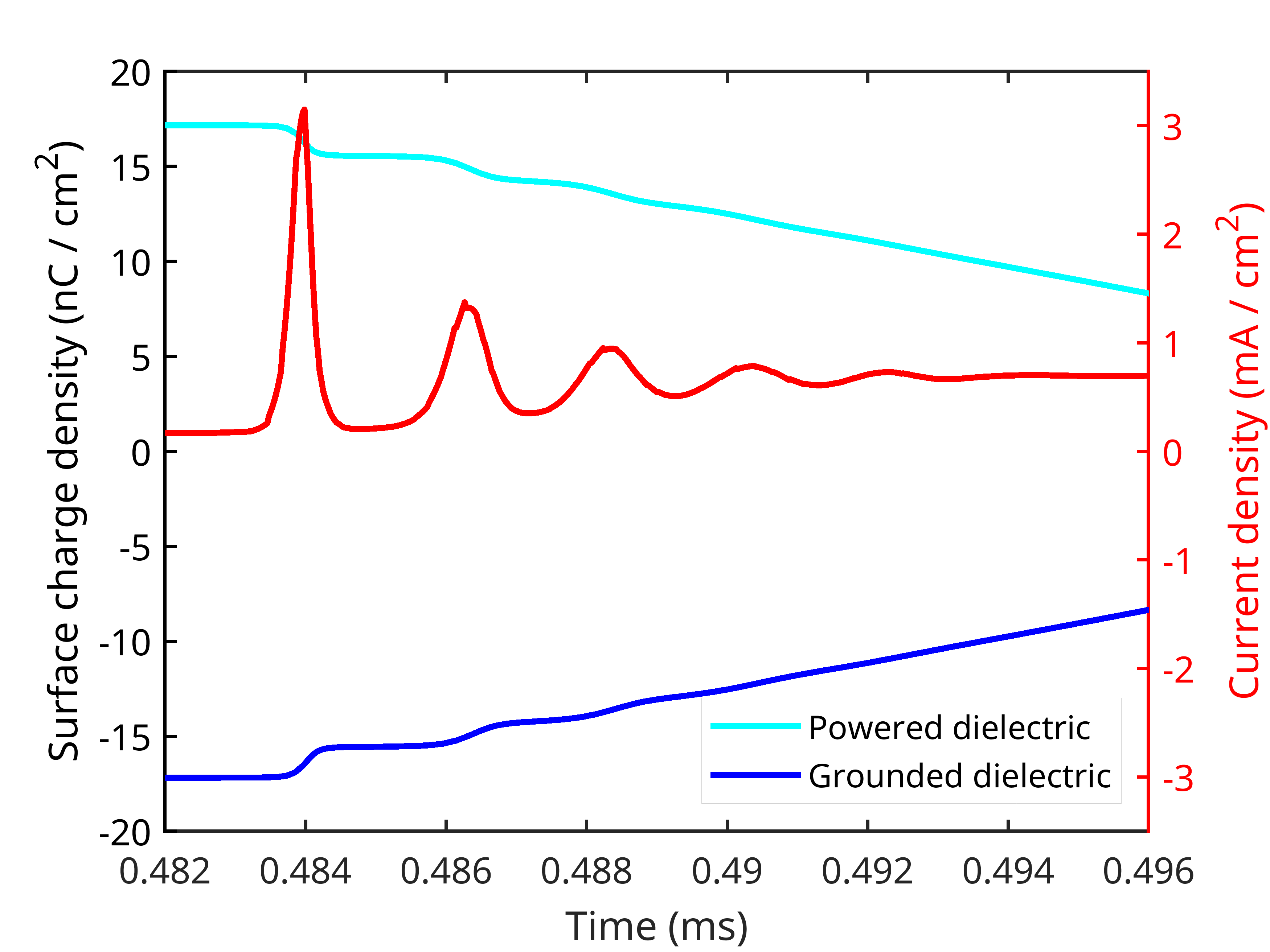}
    \caption{The temporal evolution of surface charge density at the left (cyan curve) and right (blue curve) dielectrics during the pulses from $0.482-0.496$ ms along with the terminal current density (red curve).}
    \label{fig:Surface_charge_density}	
\end{figure}
The surface charge accumulated at each plasma dielectric interface is calculated from Equation (\ref{eqn:boundary_surface}). Figure \ref{fig:Surface_charge_density} illustrates the surface charge accumulation on the left (powered dielectric) and right (grounded dielectric) interfaces. The surface charge plays an important role in the gas discharge. This is because the presence of surface charge modifies the value of the gas gap voltage $(V_{g})$ by altering the overall voltage drop across the dielectric $(V_{d})$. The relationship between the gas gap voltage and surface charge density is given by \cite{Zhang_2019}
\begin{equation}
    V_{g} = V_{a} - V_{d},
    \label{eq:volatge_diff}
\end{equation}
where $V_{a}$ is the applied voltage and $V_{d}$ is given by the relation
\begin{equation}
    |V_{d}| \propto \frac{(|\sigma_{1}| + |\sigma_{2}|)}{2 C_{d}}, \nonumber
\end{equation}
where $C_{d}$ stands for the capacitance of each dielectric. The $\sigma_{1}$ and $\sigma_{2}$ are the surface charge densities on the left and right dielectrics, respectively.

It is evident from Figure \ref{fig:Surface_charge_density} that before the beginning of the current pulses, the sum of the absolute values of the surface charge density on both dielectrics is a higher constant value, indicating that $V_{d}$ is larger and consequently $V_{g}$ is lower than the breakdown voltage. This situation prevents any discharge of the gas. As the surface charge density on both dielectrics begins to decrease over time, $V_{d}$ also decreases, leading to an increase in $V_{g}$ to a value larger than the breakdown voltage. Consequently, a breakdown in the gas gap is initiated. As discharge occurs, the surface charge density decreases first and then remains constant for a relatively longer time, causing $V_{g}$ to rise again, which in turn triggers the onset of the next discharge process as shown in Figure \ref{fig:Surface_charge_density}.

\subsection{Effect of Discharge Parameters on DBD}
Consider now the effect of different discharge parameters, such as \textit{pd} values, driving frequency, and properties of dielectric materials on discharge dynamics. We also classify the discharge based on different characteristic scales, especially in the case of \textit{pd} values and driving frequency.

\subsubsection{Dependence of Discharge Dynamics on the \textit{pd} Values} \label{subsubsec:pd_value}

\begin{figure} [h!]
    \centering 
    \includegraphics[scale=0.75]{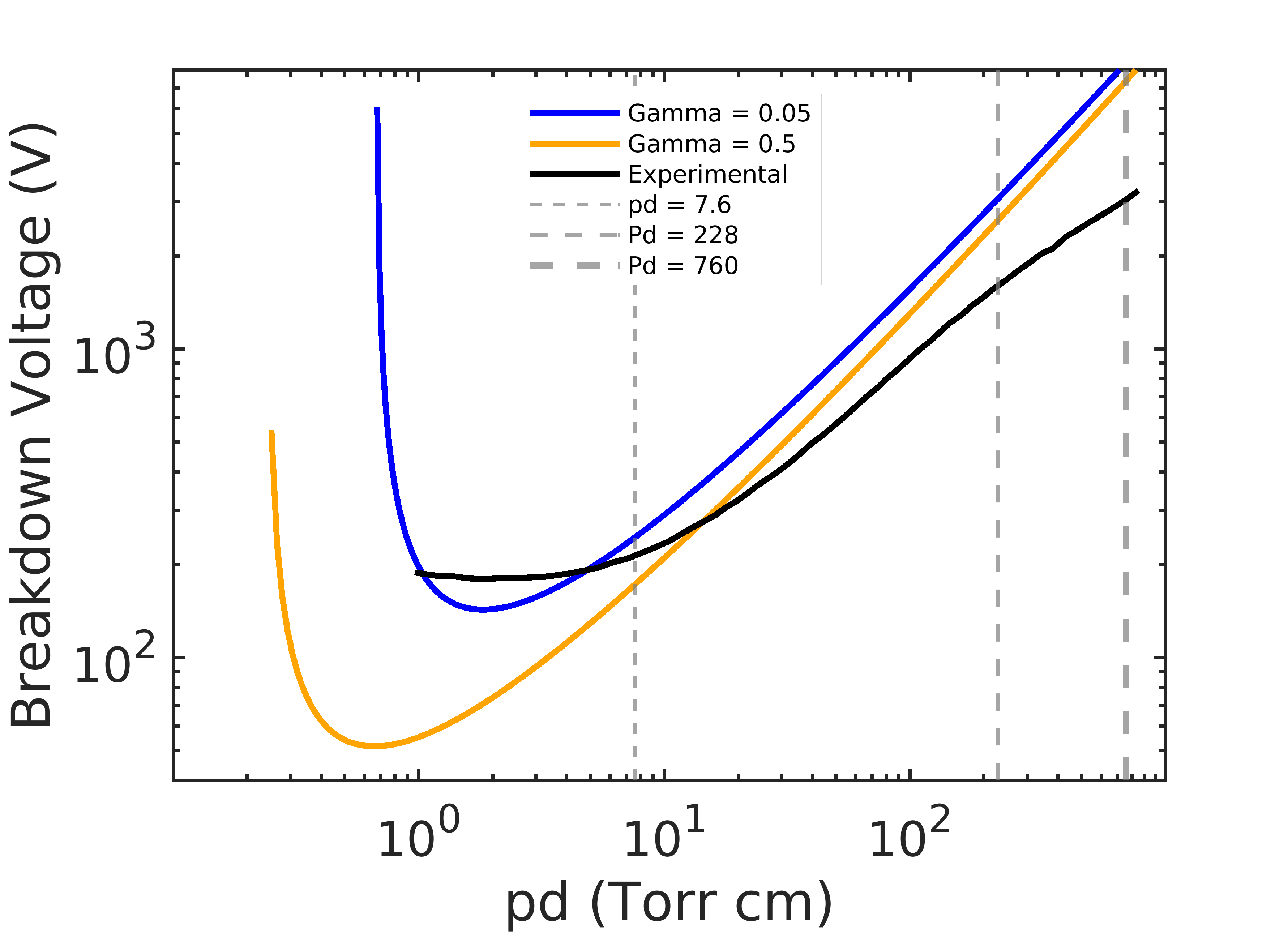}
    \caption{The Paschen curve for argon with $A = 4.53$, $B = 78.33$, and $\gamma = 0.05$ (solid blue curve). The Paschen curve with the same values of $A$, and $B$ but with different electron emission coefficients ($\gamma=0.5$) is also shown to illustrate how the Paschen curve can be modified based on $\gamma$. The experimental data (solid black curve) is extracted from the book \cite{raizer1991gas}. The dashed vertical gray lines with increasing thickness represent the \textit{pd} values for different gas gap distances at which simulations are performed.}
    \label{fig:Paschen_curve}
\end{figure}

The dependence of the gas breakdown voltage on the product of pressure ($\textit{p}$) and gap length ($\textit{d}$), i.e., $U_{br, pd} = f(pd)$, is known as a Paschen curve\cite{raizer1991gas}. The experimental Paschen curves for a static breakdown exhibit a minimum at a certain \textit{pd}, which corresponds to optimal breakdown conditions. The Paschen curve for a static breakdown can be approximated as \cite{lieberman1994principles},
\begin{equation}
    U_{br, pd} = \frac{Bpd}{ln(Apd) - ln[ln(1+\frac{1}{\gamma})]},
    \label{eqn:Paschen_law}
\end{equation}
where $\gamma$ is the secondary electron emission coefficient, and \textit{A} and \textit{B} are constants that describe the electron multiplication process and depend on the gas type. The experimental Paschen curve\cite{raizer1991gas} for Argon is shown in Figure \ref{fig:Paschen_curve}, illustrating that it has a minimum breakdown voltage of approximately 200 V at a $\textit{pd}$ of around 5 Torr cm (Paschen minimum). The theoretical Paschen curve for argon gas with $A = 4.53$ cm $\rm{Torr^{-1}}$, $B = 78.33$ $\rm{V Torr^{-1} cm^{-1}}$, and $\gamma = 0.05$ is also plotted on the same Figure (solid blue curve). The values of \textit{A} and \textit{B} used here are three times less than the value reported by \cite{lieberman1994principles}. These values are chosen to make the theoretical Paschen curve similar to the experimental curve. The theoretical Paschen curve also has a Paschen minimum at a $\textit{pd}$ of around 5 Torr cm where the breakdown voltage is around 143 V. Note that the shape of the Paschen curve depends on the secondary electron emission coefficient, e.g., the solid orange curve in Figure \ref{fig:Paschen_curve} with $\gamma = 0.5$. This curve has a significantly lower Paschen minimum breakdown voltage at a lower \textit{pd} value. \\
To study the effect of the \textit{pd} values on discharge dynamics, we performed simulations for three gap thicknesses (\textit{d}), keeping the pressure the same at 1 atm (760 Torr). The other simulation parameters are the same as for the base case. 
On the left branch of the Paschen curve, the electron velocity distribution is nonlocal in space and time, and electron "runaway" may occur \cite{levko2019modified}. Hence the fluid model does not work well near the Paschen minimum. Therefore, in this paper, we perform simulations for the \textit{pd} values on the right branch of the Paschen curve. At large values of \textit{pd}, ambipolar drift could become important, and a thermocurrent instability may occur \cite{aleksandrov2002hydrodynamic}.

\begin{figure}[!ht]
    \centering
    \includegraphics[scale =.5]{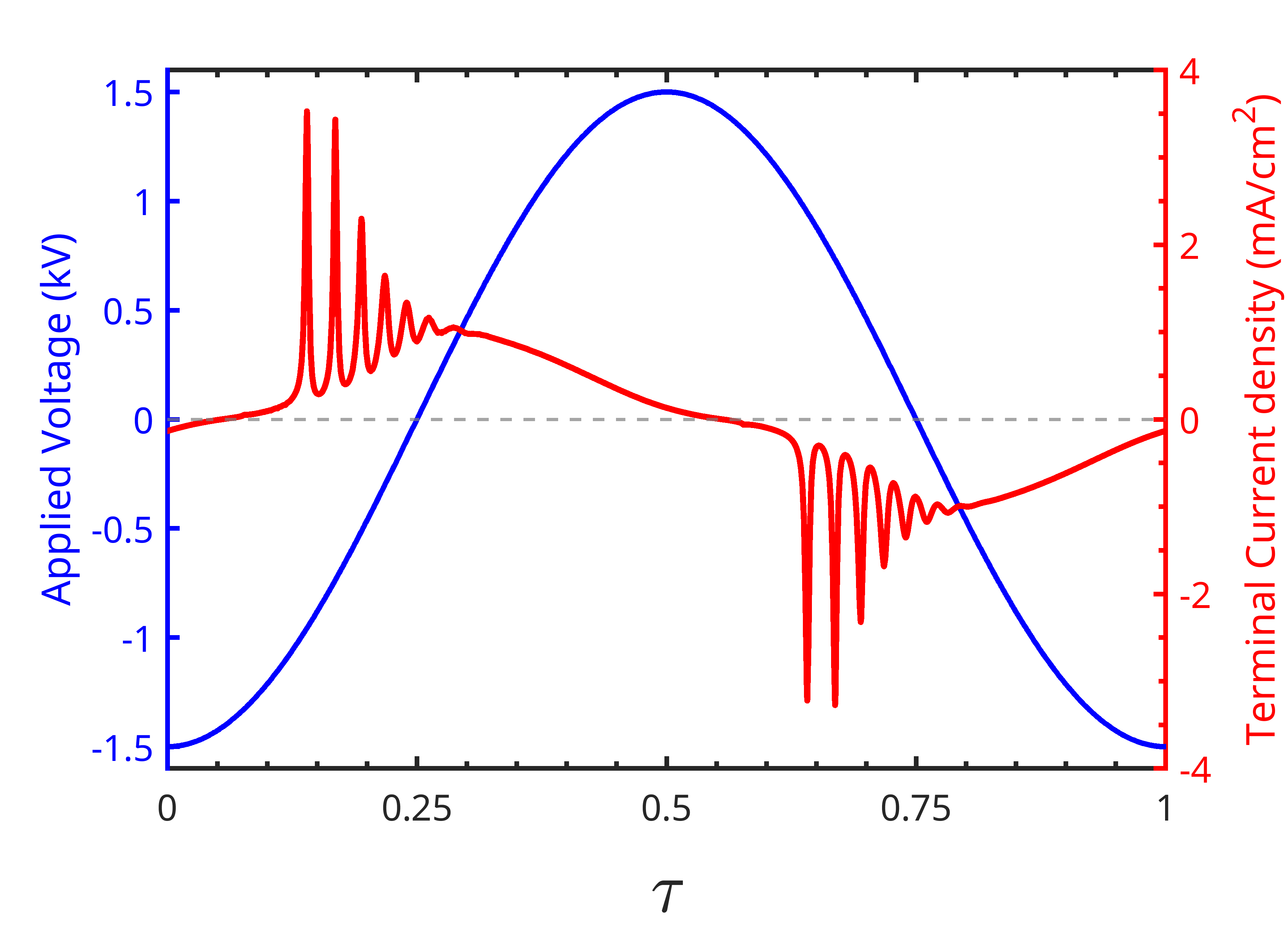}(a)
    \includegraphics[scale =.5]{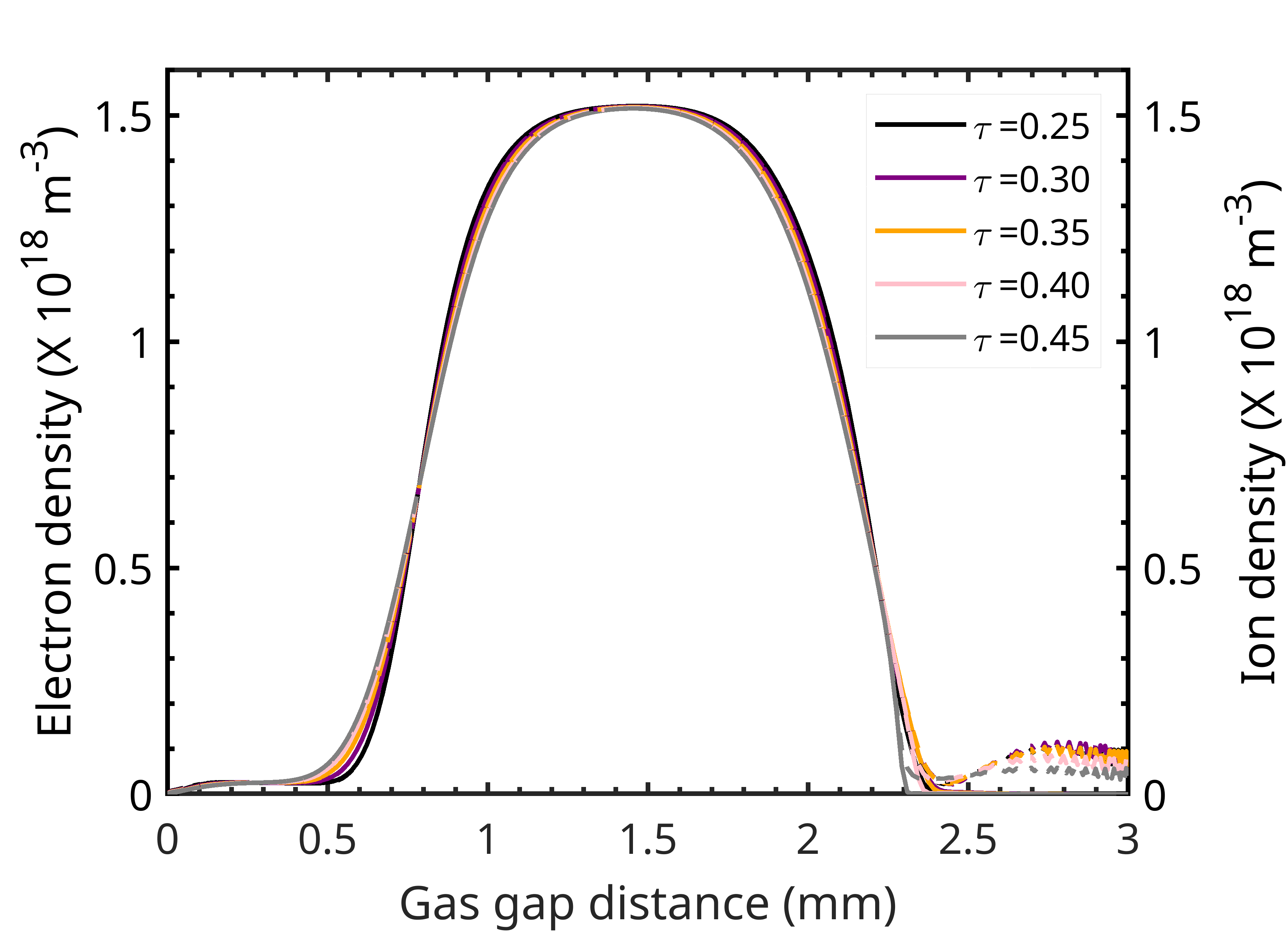}(b)
    \includegraphics[scale =.5]{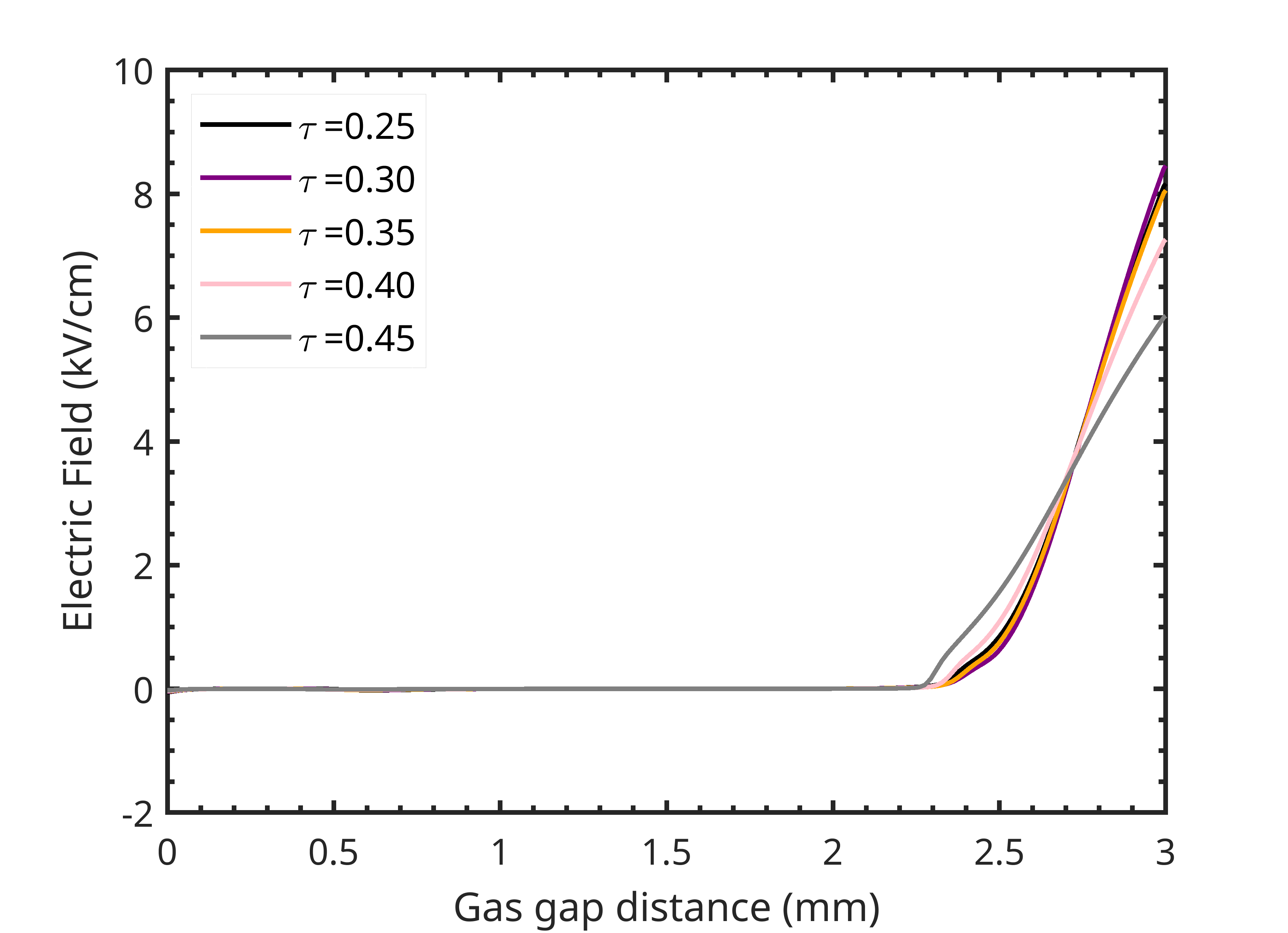}(c)
    \includegraphics[scale =.022]{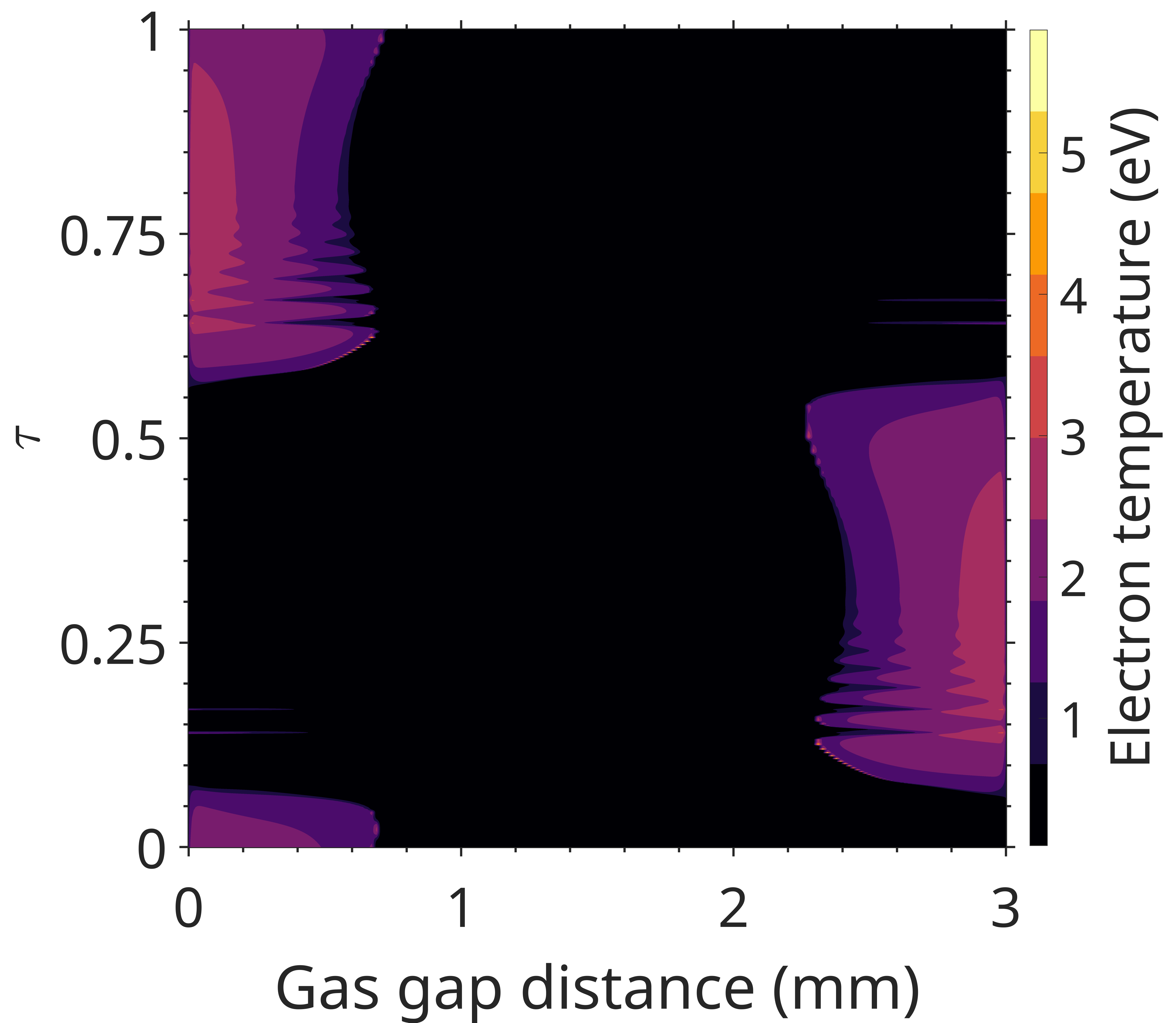}(d)
     \includegraphics[scale =.5]{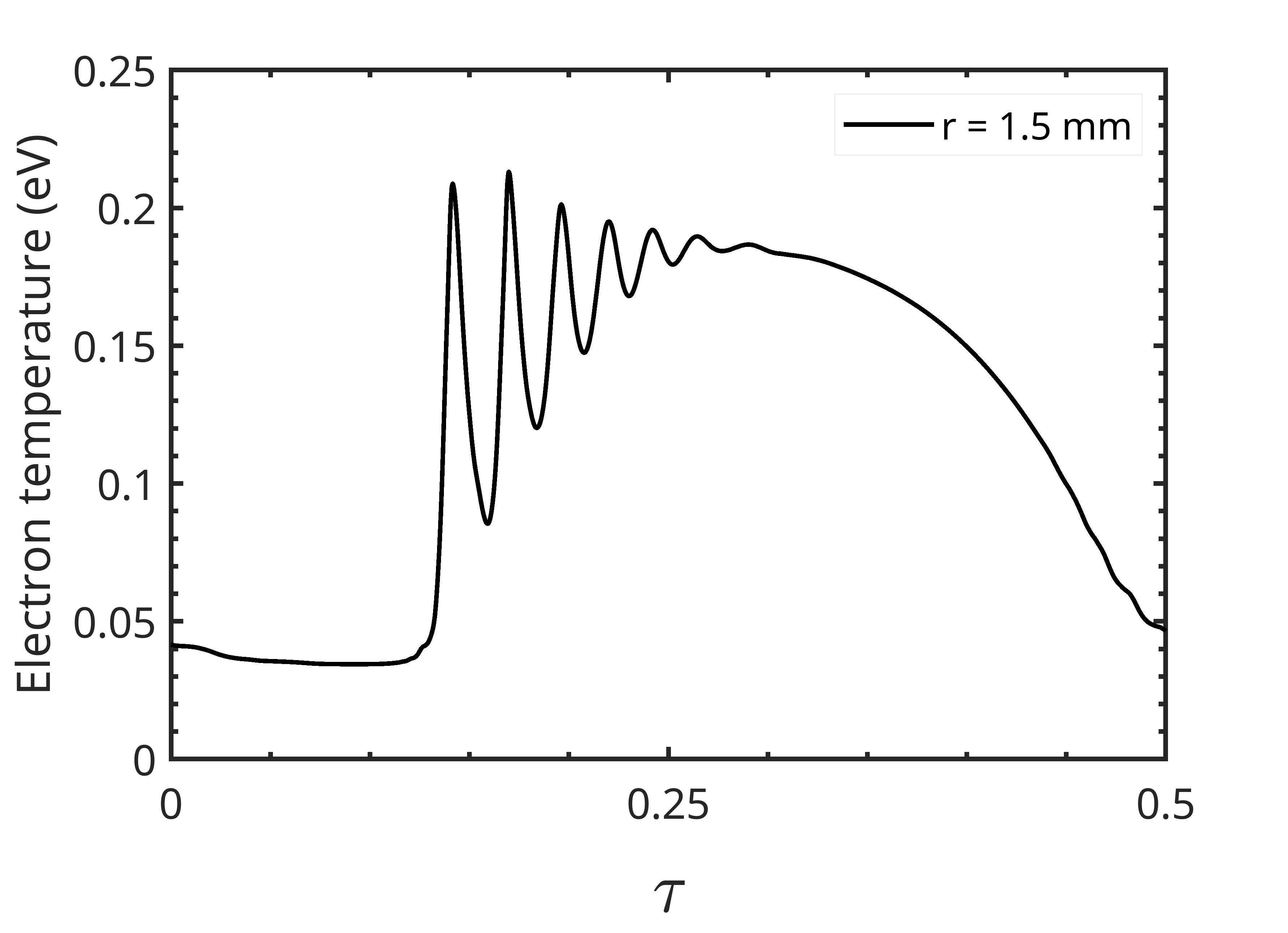}(e)
    \caption{Variation of the terminal current density showing multi-peaks (panel (a)), spatial variation of electron and ion density during five different phases of the half cycle of the considered third cycle (panel (b)), spatial variation of the electric field during five different phases of the half cycle of the AC period of the third cycle (panel (c)), the spatiotemporal evolution of electron temperature over one cycle (panel (d)), and spatial variation of the electron temperature over a half cycle (panel (e)) at 3 mm (\textit{pd = 228} Torr cm). The other discharge parameters are kept the same as the base-case simulation.}
    \label{fig:pd_3}
\end{figure}

Figure \ref{fig:pd_3} shows the simulation results for a gas gap distance of 3 mm (\textit{pd} value of 228 Torr cm). Panel (a) illustrates the multi-pulse current density throughout the considered third cycle. This case exhibits six current density pulses per half cycle. The spatial variation of the electron density (solid lines) and ion density (dashed lines) during five different phases of the half cycle of the considered cycle is shown in panel (b). As we can see, the electron density is equal to the ion density in most of the gas gap, except near the sheath region. The difference between the electron and ion density in the sheath region is more pronounced in the instantaneous cathode. In the sheath region, the ion density is greater than the electron density, and a sharp increase in the electric field near the instantaneous cathode is observed, as shown in panel (c). This implies the existence of the cathode fall region, where the electric field increases almost linearly towards the cathode \cite{shi2003cathode}. The quasineutrality of the plasma in the gas gap and the non-uniform nature of the electric field along the gas gap implies that the discharge is in the CCP glow mode. The spatiotemporal evolution of electron temperature over the considered cycle is shown in panel (d). The electron temperature is high in the sheath because of the Joule heating \cite{farouk2008atmospheric}, i.e., the electric field in the sheath accelerates electrons, leading to increased kinetic energy and, consequently, higher temperatures. The temporal variation of the electron temperature at the center of the gas gap ($r = 1.5$ mm) for the considered first half cycle is shown in panel (e). Here, the electron temperature exhibits a significant oscillation over the half-period of the driving AC source \cite{humphrey2023electron}.

Now we calculate the characteristic time scale to classify the discharge dynamics for the \textit{pd} value of 228 Torr cm. We obtain that $\omega \tau_{\epsilon} = 0.20$ and $\omega \tau_{a} = 1.04\times 10^{4}$, see table \ref{table:pd_values}. These values satisfy the condition, $\omega \tau_{\epsilon} < 1 < \omega \tau_{a}$, implying that the discharge is in the dynamic regime of CCP discharge.

\begin{figure}[!ht]
    \centering
    \includegraphics[scale =.5]{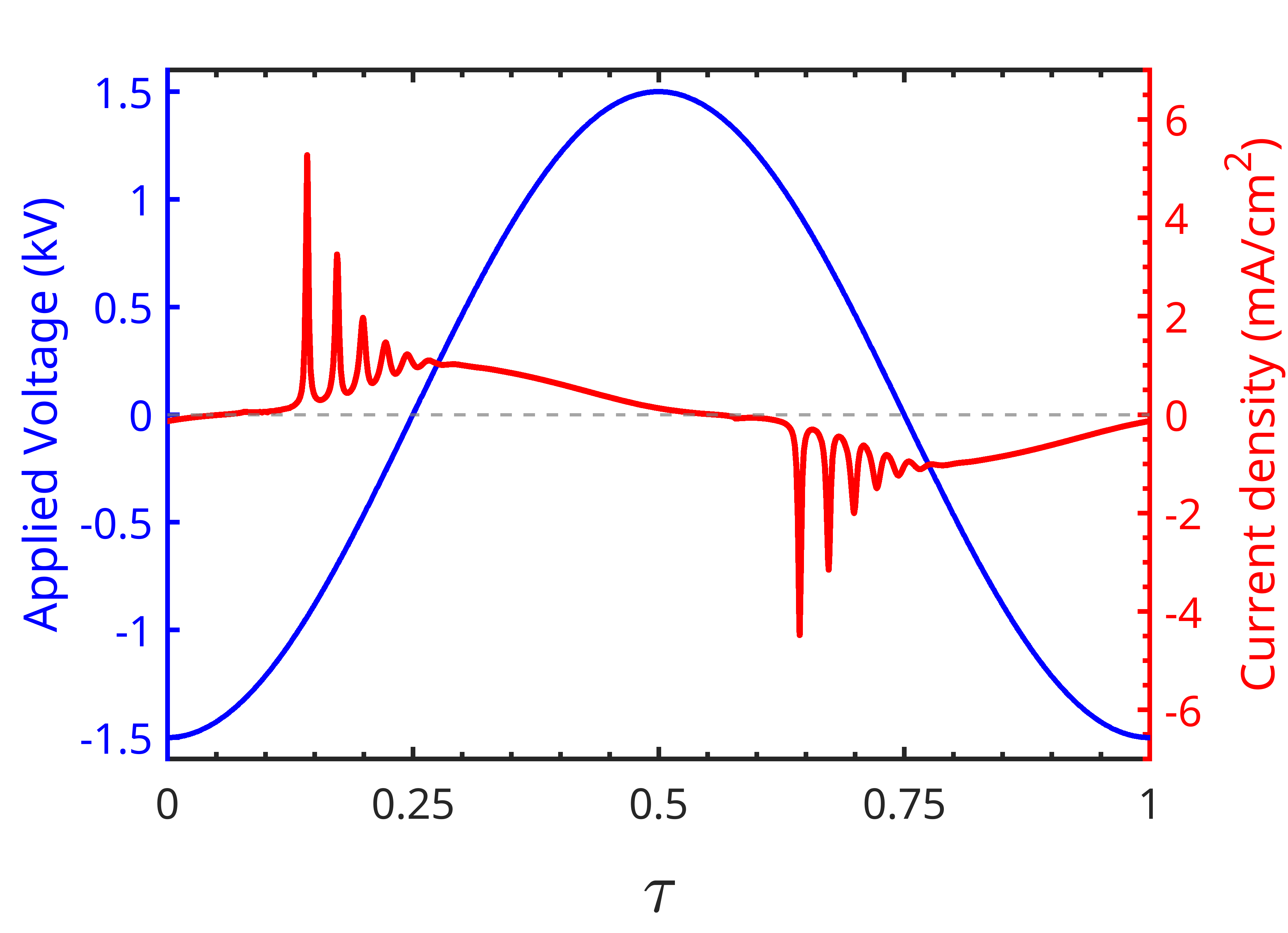}(a)
    \includegraphics[scale =.5]{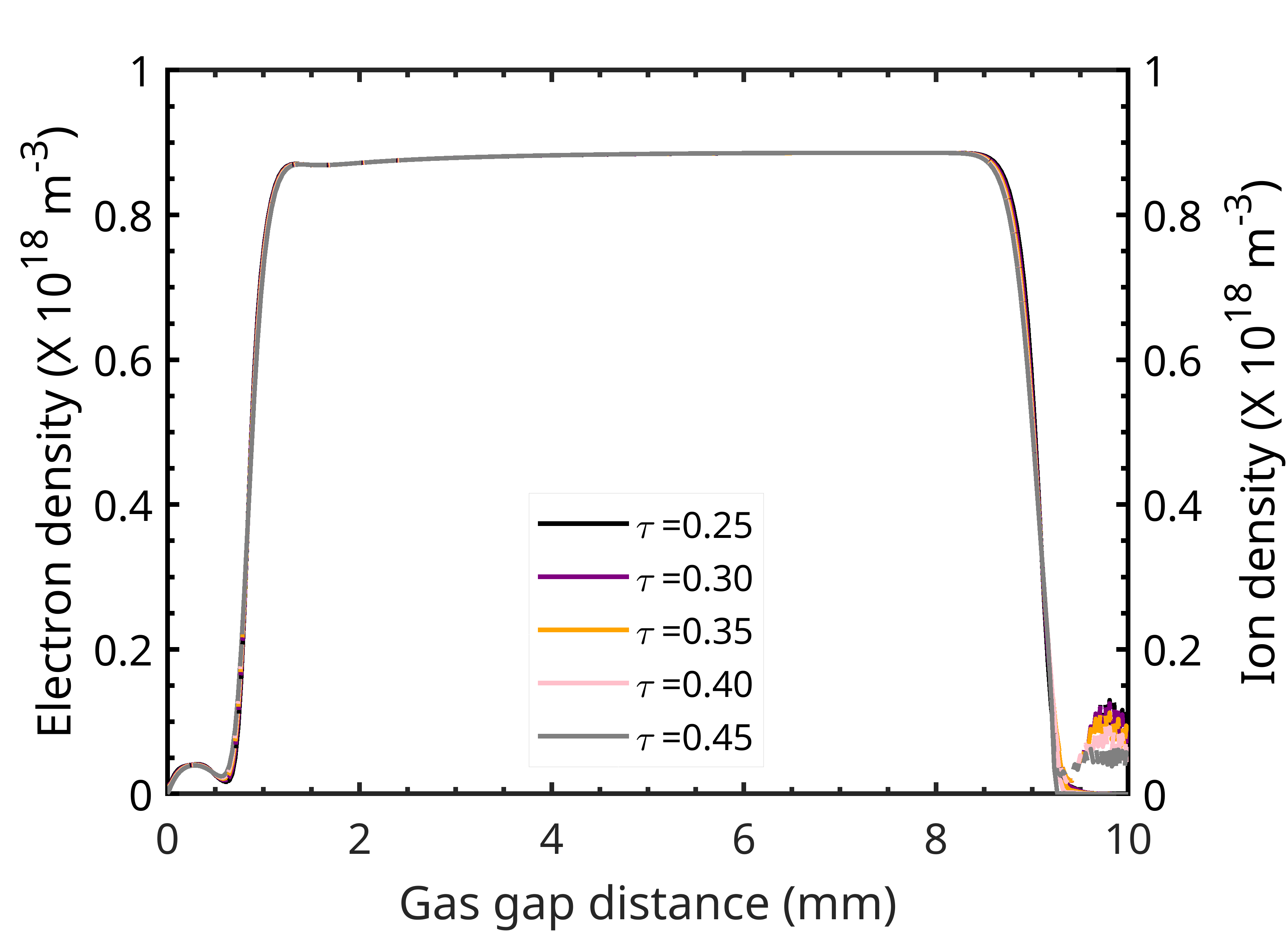}(b)
    \includegraphics[scale =.5]{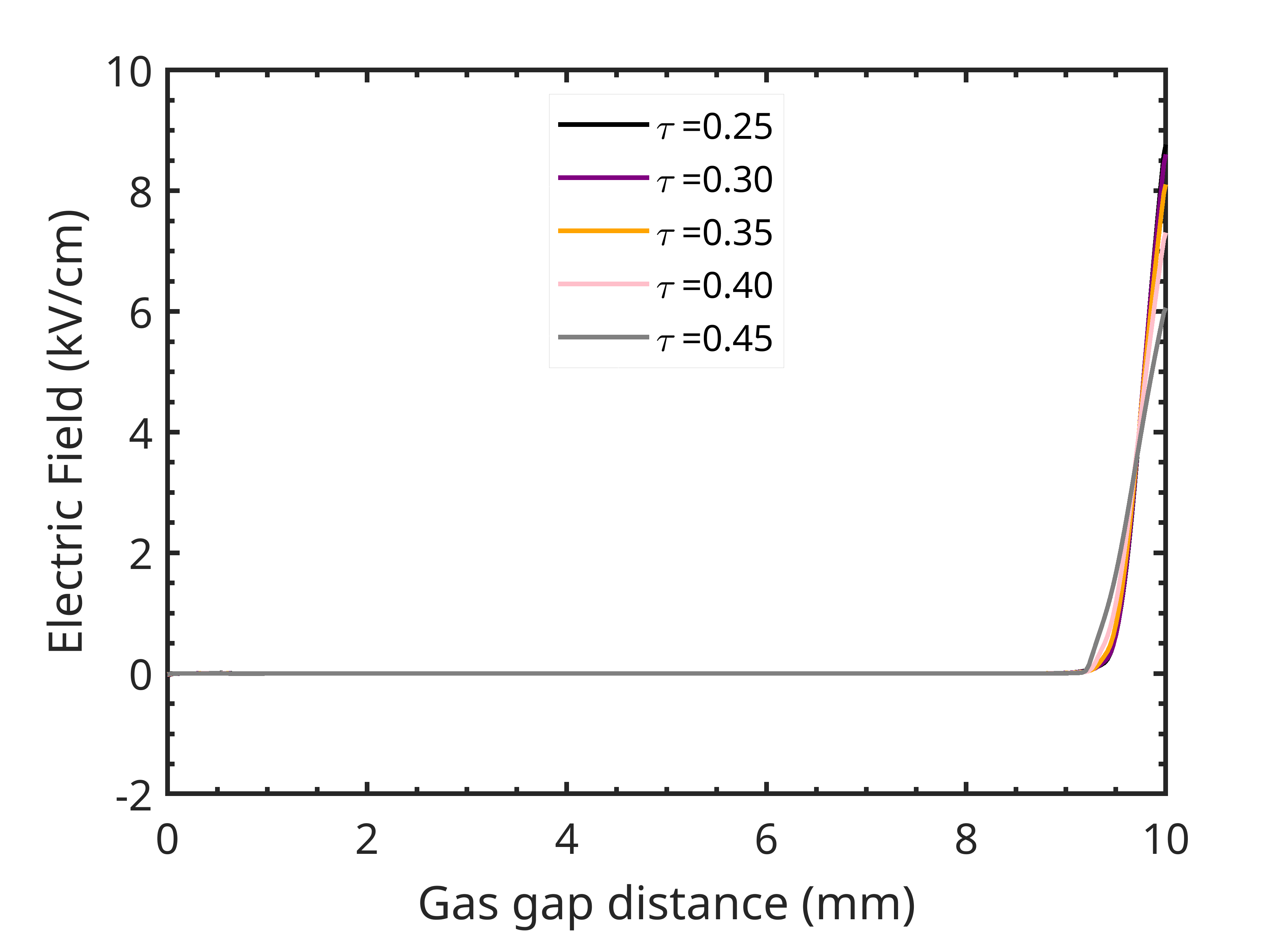}(c)
    \includegraphics[scale =.022]{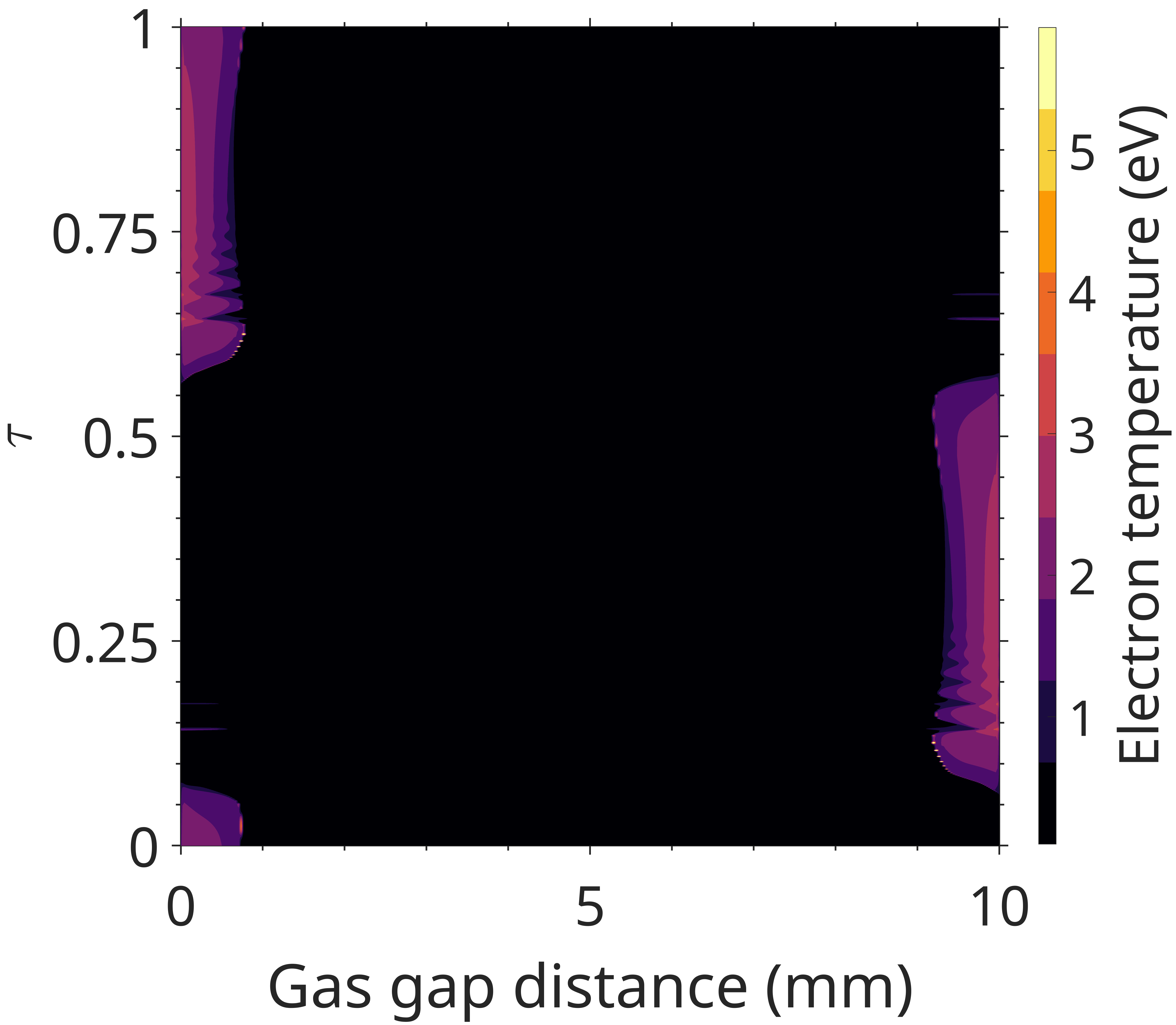}(d)
     \includegraphics[scale =.5]{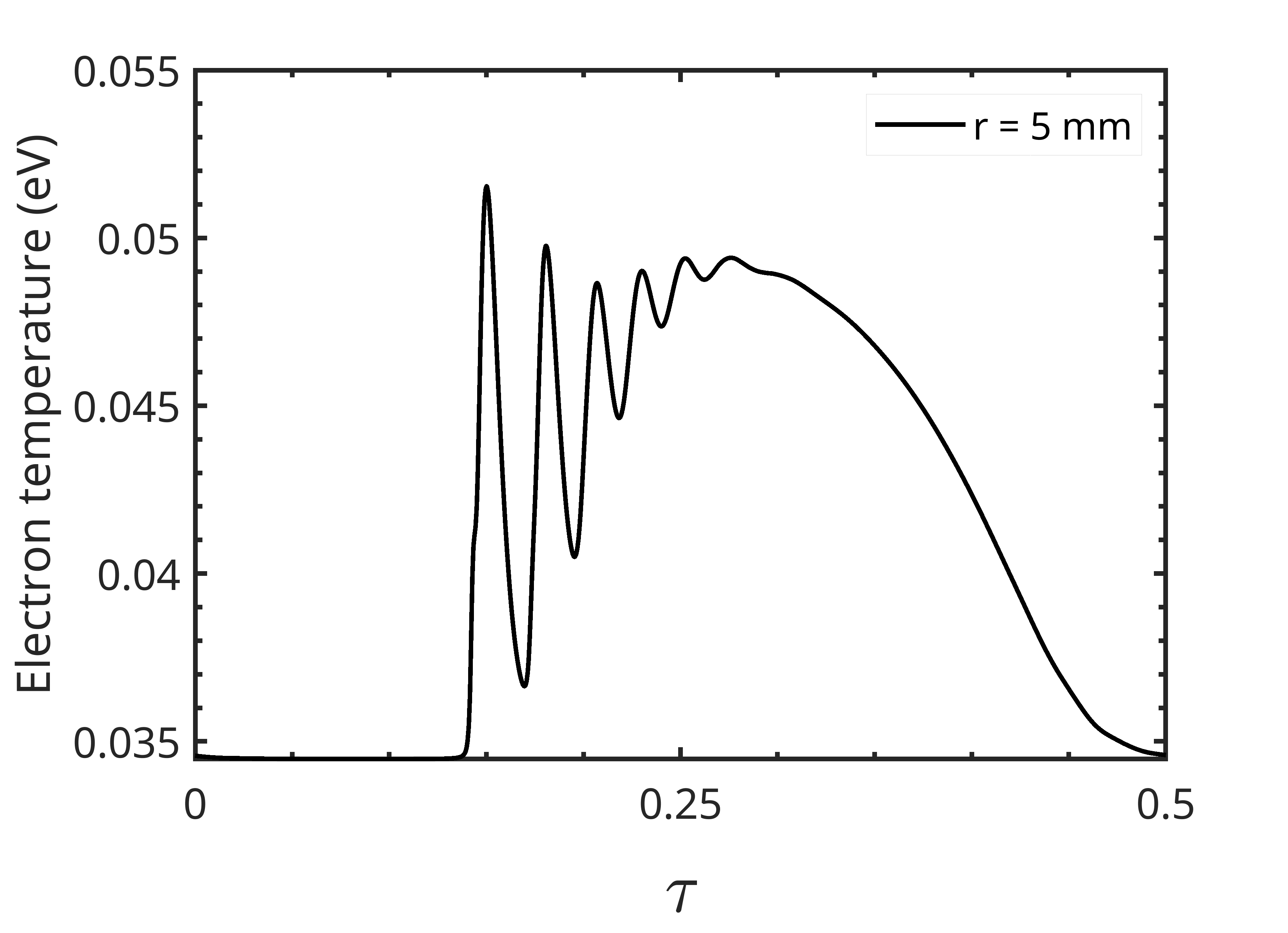}(e)
    \caption{Variation of the terminal current density showing multi-peaks (panel (a)),  spatial variation of electron and ion density during five different phases of the half cycle of the AC period from $0.45-0.65$ ms (panel (b)), spatial variation of the electric field during five different phases of the half cycle of the AC period from $0.45-0.65$ ms (panel (c)), the spatiotemporal evolution of electron temperature over one cycle (panel (d)) and spatial variation of the electron temperature (panel (e)) at 10 mm ($pd = 760$ Torr cm). The other discharge parameters are kept the same as the base-case simulation.}
    \label{fig:pd_10}
\end{figure}

Figure \ref{fig:pd_10} shows the simulation results for a gas gap distance of 10 mm (\textit{pd} value of 760 Torr cm). Panel (a) displays the current density for multiple pulses over the considered third cycle. In this case, six pulses are observed with relatively higher current density than the \textit{pd} value of 228 Torr cm.
The spatial variation of the electron density (solid lines) and ion density (dashed lines) during five different phases of the considered half cycle is shown in panel (b). As we can see, the electron density is equal to the ion density in most of the gas gap, except in the sheath region. As before, the difference between the electron and ion density in the sheath region is more pronounced in the instantaneous cathode. In the sheath, ion density is greater than electron density, and a sharp increase in the electric field near the instantaneous cathode is observed, as shown in panel (c). The sheath region, in this case, is even narrower than the previous case of \textit{pd} value 228 Torr cm. The quasineutrality of the plasma in the gas gap and the non-uniform nature of the electric field imply that the discharge is in CCP glow mode. The current density multi-pulses are observed in this case as well. The spatiotemporal evolution of electron temperature over the cycle is shown on panel (d). As before, the electron temperature is high near the dielectrics instead of the center of the gap. The temporal variation of the electron temperature at the center of the gas gap ($r = 5$ mm) for the first half cycle of panel (a) is shown on panel (e). Again, the electron temperature exhibits a significant oscillation over the half-period of the driving AC source. In addition, the overall electron temperature is slightly lower than in the previous case. As before, the characteristic time scales, $\omega \tau_{\epsilon} = 0.047$ and $\omega \tau_{a} = 4.4\times 10^{5}$ (see Table \ref{table:pd_values}) satisfy the condition `$\omega \tau_{\epsilon} < 1 < \omega \tau_{a}$', implying that the discharge is in the dynamic regime of CCP discharge mode.

\begin{table}[h!]
    \centering
    \caption{Table of plasma parameters for Gas Discharge at different $\textit{pd}$ values\footnote[1]{Note that all the plasma parameters: electron temperature, electron density, ion density, and Debye lengths are calculated at the gas gap center at the normalized time of $\tau = 0.25$ of the considered cycle}.}
    \begin{ruledtabular}
    \begin{tabular}{ccccccccccc} 
    \hline
    Gas gap & Product  & $T_{e}$ & $n_{e}$ & $n_{i}$ & $\lambda_{D}$ & $\tau_{e}^{b}$ & $\tau_{e}^{D}$ & $\tau_{\epsilon}$ & $\tau_{i}$ & $\tau_{a}$ \\ 
    (d) & $(\textit{pd})$ &  & ($\times 10^{16}$) & ($\times 10^{16}$) & & & & &  & \\ 
    
    (cm) & (Torr cm)  & (eV)  & ($\rm{m^{-3}}$) & ($\rm{m^{-3}}$) & (cm) & (ns)& (ns)& ($\mu$s)& (ns) & (s)\\
    \hline
    0.01 & 7.6   & 3.3 & 0.0039 & 2.4  & 0.22   & 0.52 & 31 & -   & 280 & -  \\
    \hline
    0.3 & 228 & 0.18  & 152  & 152 & 0.0003 & -    & -     & 6.4 & -   & 0.33\\
    \hline
    1 & 760 & 0.047 & 88   & 88  & 0.0002 & -    & -     & 1.5 & -   & 14\\
    \hline
    \end{tabular}
    \end{ruledtabular}
   \label{table:pd_values}
\end{table}
The physical reasoning for the dynamic regime of CCP glow mode for higher \textit{pd} values (228 and 760 Torr cm) can be explained as follows. Due to the slow ion time scale, ions do not have enough time to reach the cathode during the half-period of the applied voltage. They accumulate in the gap, and plasma forms during several AC periods. The ions and electrons are “glued” together by the electric field in the plasma. The electric field becomes inhomogeneous, and the typical plasma-sheath structure of a capacitively coupled plasma (CCP) is formed, with the potential drop in the plasma being about several electron volts. A slow, ambipolar time scale now characterizes the motion of electrons and ions in the plasma. However, electron heat transport does not depend on quasineutrality and occurs in the plasma and sheath at the same time scale as the free electron diffusion. Therefore, the plasma operates in a dynamic regime with small modulations of the plasma density and substantial modulations of the electron temperature and electron-induced ionization and chemical reactions. The current pulses are still observed in the dynamic regime of CCP operation as the processes in the anode sheath remain like those in Townsend discharges, i.e. electrons and ions are decoupled in the sheath, and current pulsing may still occur because of this decoupling.

\subsubsection{Dependence of Discharge Dynamics on Driving Frequency} \label{subsubsec:freq}
To study the effect of applied frequency on discharge dynamics, we perform the simulations of a 1D DBD discharge in Ar for frequencies of 10 kHz and 25 MHz, keeping the other parameters the same as for the base case. These results are classified based on the characteristic scales of the different regions of Figure \ref{fig:scale_characterstics}.\\
\begin{table}[]
    \caption{Table of plasma parameters for Gas Discharge at different frequencies\footnote[2]{Note that all the time scales along with electron temperature, electron density, ion density, and Debye lengths are calculated for the gas gap center at the normalized time of $\tau = 0.25$ of the considered cycle.}}
    \begin{ruledtabular}
    \begin{tabular}{cccccccccc} 
    \hline
    frequency  & $T_{e}$ & $n_{e}$ &  $n_{i}$  & $\lambda_{D}$& $\tau_{e}^{b}$ & $\tau_{e}^{D}$& $\tau_{\epsilon}$ & $\tau_{i}$ & $\tau_{a}$ \\ 
    (f) &  & ($\times 10^{17}$) &($\times 10^{17}$) & & & & & &\\ 
    
              & (eV)           & ($\rm m^{-3}$) & ($\rm m^{-3}$) & (cm)& (ns) &(ns)   & (ns) & (ns) &($\mu$s)\\
    \hline
    10 (kHz)  & 3.3           & 0.0008         & 0.48          & 0.15 & 0.71 & 31 & -   & 357 & - \\
    \hline
    25 (MHz)  & 1.5          & 493         & 493       & 0.0001& -& -& 91 &- & 45 \\
    \hline
    \end{tabular}
    \end{ruledtabular}
     \label{table:frequency_values}
\end{table}
\begin{figure}[h!]
\centering
   \includegraphics[scale=.5]{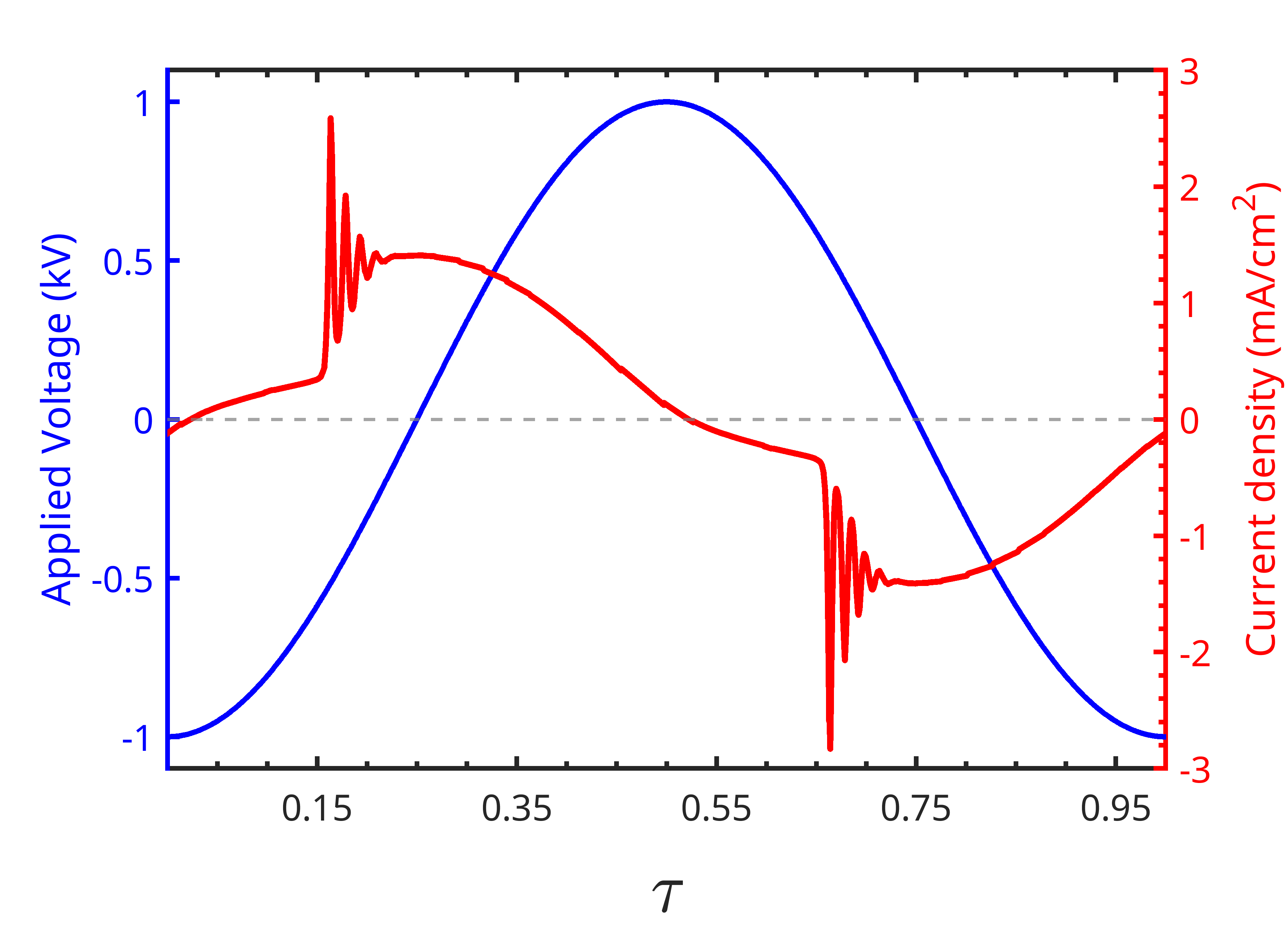}(a)
   \includegraphics[scale=.5]{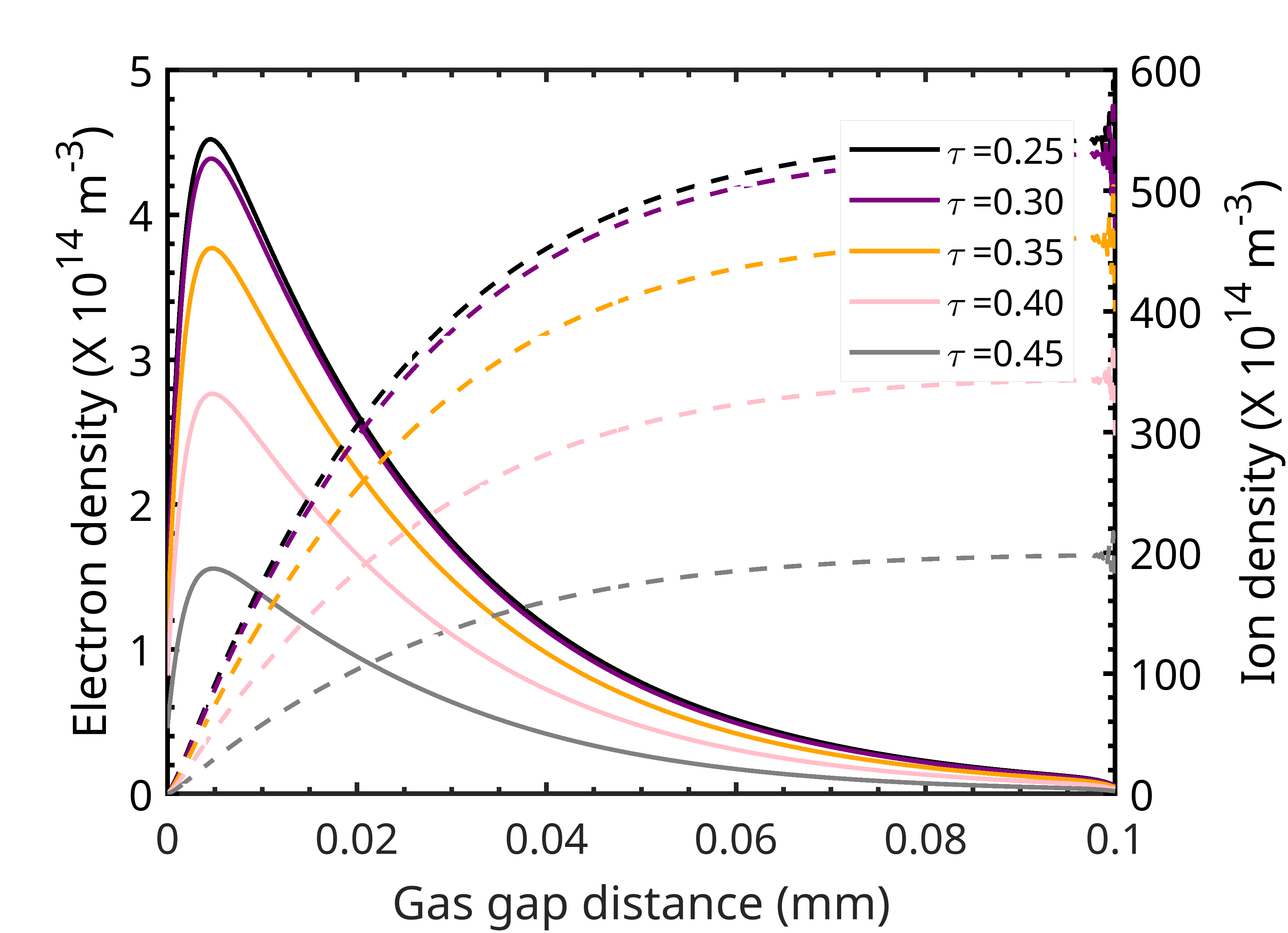}(b)
   \caption{Time evolution of current density multi-peaks (panel (a)), and spatial evolution of the electron (solid curve) and ion density (dashed curve) during five different phases of the half cycle of the AC period of the third cycle (panel (b)) at the applied frequency of 10 kHz. The other discharge parameters are kept the same as for the base case simulation.}
    \label{fig:terminal_curr_1000V_freq_10}	
\end{figure}
 
Figure \ref{fig:terminal_curr_1000V_freq_10} shows the results for a driving frequency of 10 kHz. Panel (a) shows the number of current pulses for the considered cycle and panel (b) shows the spatial variation of electron and ion density during five different phases of the half cycle of the AC period. The number of current density pulses in this case is four which is less than for the base case of 5 kHz (see Figure \ref{fig:multipeaks_basecase}). This is expected because the number of current density multi pulses is proportional to $\omega^{-1/2}$ for the Townsend mode of discharge \cite{nikandrov2005low}. Also, we observe that the electron density during the first pulse is much less than the ion density and the plasma scale length (0.01 cm) is less than the Debye length (0.0454 cm). This supports the conclusion that the discharge is in the Townsend mode.
\begin{figure}[h!]
    \centering   
    \includegraphics[scale=.5]{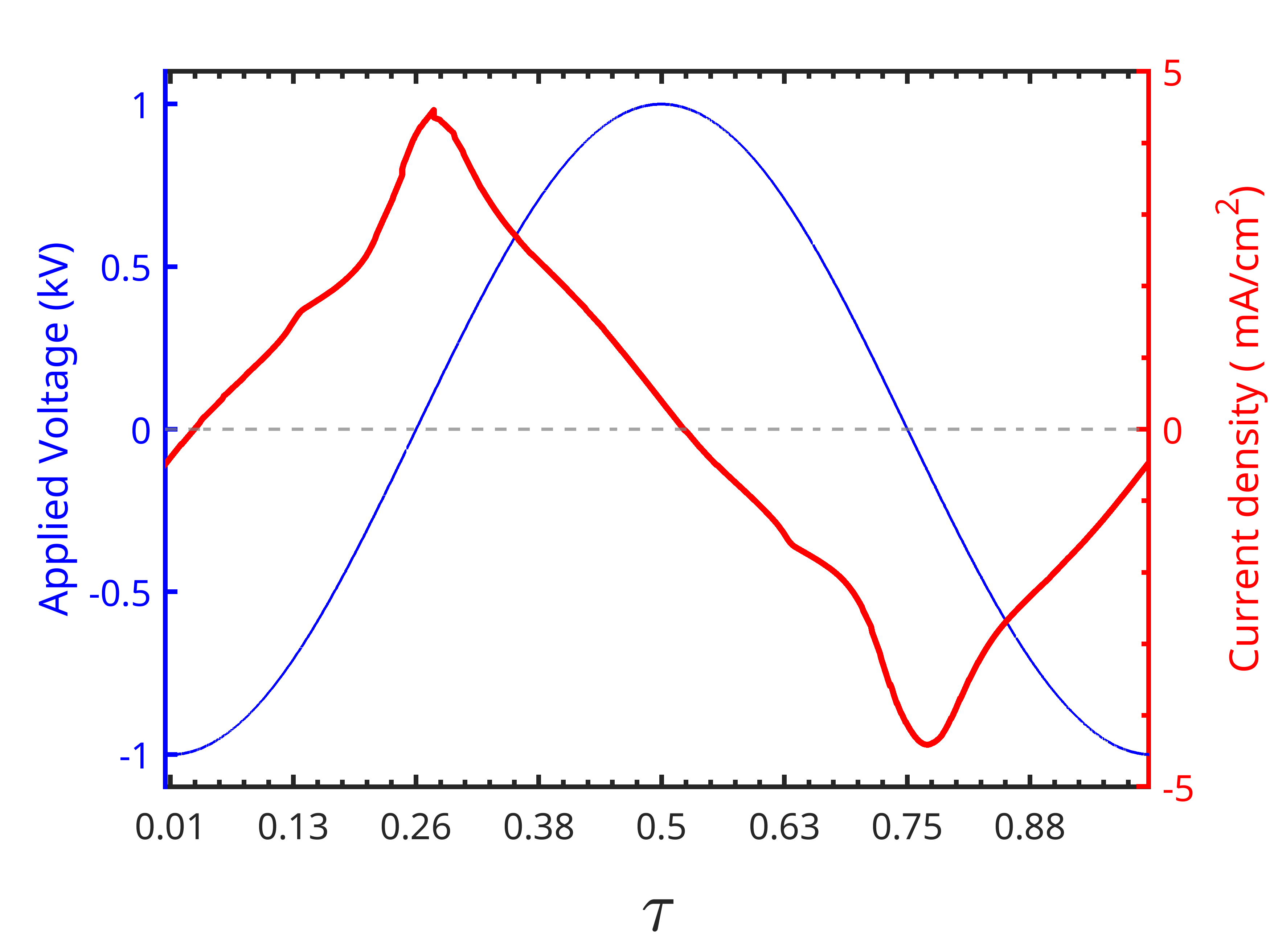}(a)
    \includegraphics[scale=.5]{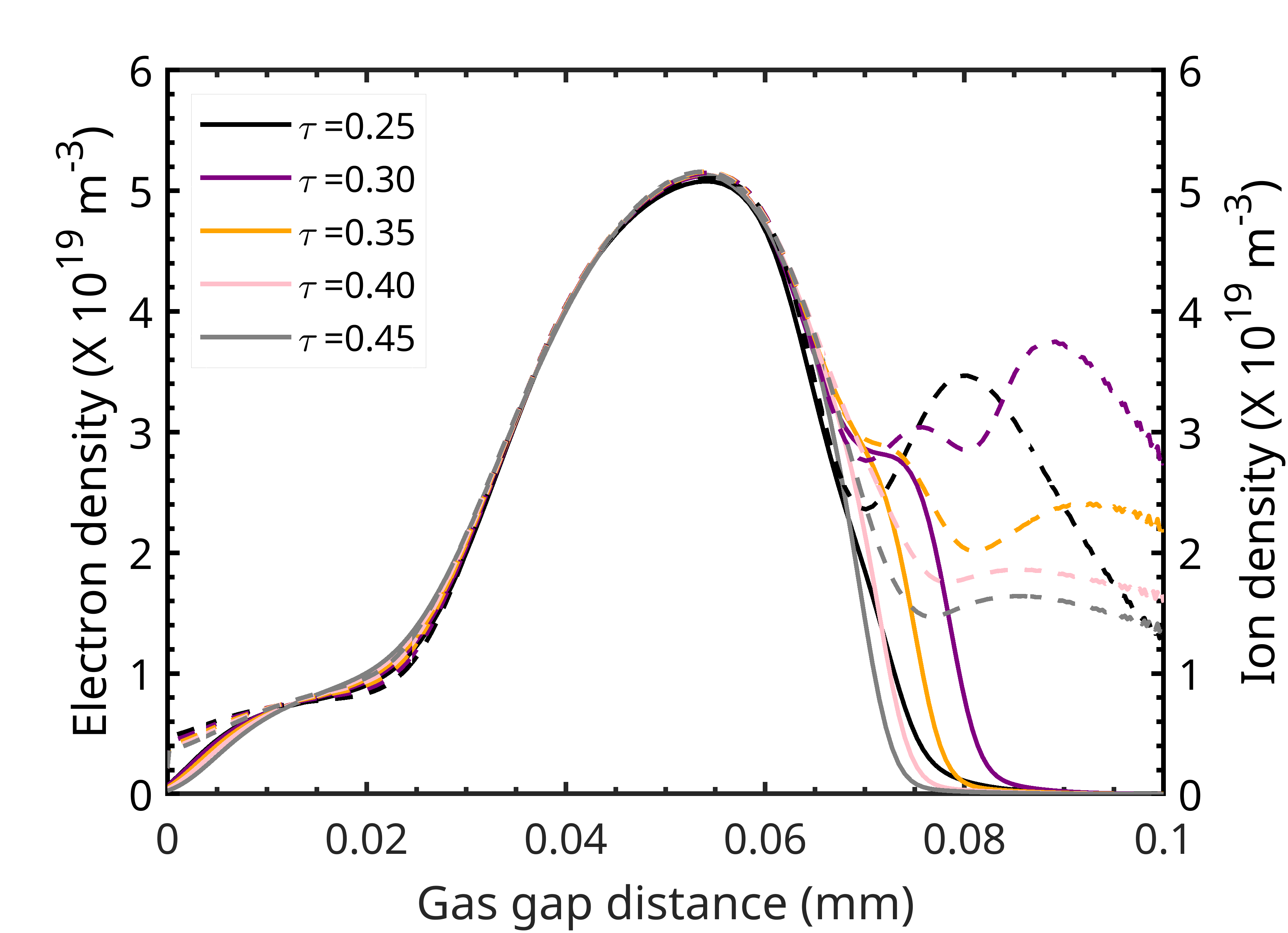}(b)
    \caption{Time evolution of current density multi-peaks (panel (a)), and spatial evolution of the electron (solid curve) and ion density (dashed curve) during five different phases of the half cycle of the AC period of the third cycle (panel (b)) at frequency 25 MHz. The other discharge parameters are kept the same as for the base case simulation.}
    \label{fig:terminal_curr_1000V_freq_25}	
\end{figure}

Figure \ref{fig:terminal_curr_1000V_freq_25} shows the results for a driving frequency of 25 MHz. Panel (a) shows the terminal current density pulse for the considered cycle and panel (b) shows the spatial variation of electron and ion density during five different phases of the half cycle of the AC period. We only observe a single current density pulse in this case, with a maximum current density of about 4.5 mA$\rm{cm^{-2}}$. This value is smaller than the previous  10 kHz case. In addition, the spatial variation of electron and ion density during the pulse shows that the electron and ion densities are equal during the pulse except at the sheath. The sheath region is most pronounced on the instantaneous cathode (right side). In addition, the plasma scale length (0.01 cm) is greater than the Debye length (0.0001 cm) and $\omega \tau_{a} > 1$ implies that the discharge characteristic corresponding to the 25 MHz case is in a high-frequency regime. The characteristic scale sizes for both 10 kHz and 25 MHz cases are shown in Table \ref{table:frequency_values}.

\subsubsection{Dependence of Pulses on Lossy Dielectrics in a Resistive Discharge}
\begin{figure}[h]
\centering
    \includegraphics[scale=.7]{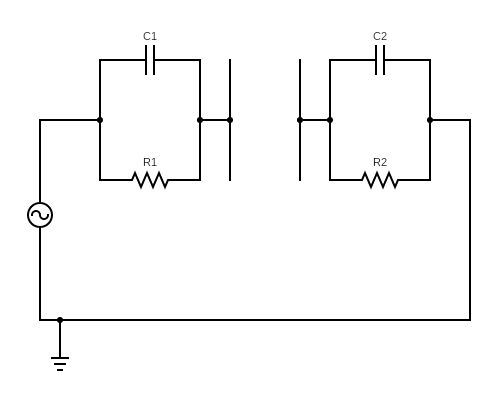}
    \caption{A schematic diagram of the circuit equivalent to the lossy dielectrics in the DBD simulations. Two RC parallel circuits are connected on each side of the gas gap to mimic the lossy dielectric \cite{wang2003study}.}
    \label{fig:Circuit_Lossy}	
\end{figure}
To study a lossy dielectric in resistive discharges, we used an equivalent circuit that emulates the resistive barrier discharge, as illustrated in Figure \ref{fig:Circuit_Lossy}. Here two RC parallel circuits are used on both sides of the middle capacitor.
The left and right RC circuit mimics the left and right dielectric, respectively and the middle part corresponds to the DBD with naked electrodes. The resistor connected to the capacitor in parallel models the lossy dielectric. We varied the dielectric resistance in the range 5  $\rm{\Omega}$ - 500 k$\rm{\Omega}$ in our simulation. The value of the two resistors is kept the same on both sides, i.e., $\rm{R_{1}} =  \rm{R_{2}}$. To obtain the value of the capacitor that mimics the dielectrics of the dielectric constant of 5 as in the base case simulation, we use the following expression \cite{griffiths2005introduction}
\begin{equation}
     C = \epsilon_{r} \epsilon_{0} \frac{A}{L},
\end{equation}
where C is the capacitance of each capacitor in Farad,  $\epsilon_{r} =5$ is the relative permittivity of the dielectric, and $\epsilon_{0}$  is the relative permittivity of the vacuum. The A and L are the area and width of each dielectric. We use the same dielectric on both sides, which gives $\rm{C_{1}} = \rm{C_{2}} = 3.447$ nF.
\begin{figure}[h]
    \centering
    \includegraphics[scale=.49]{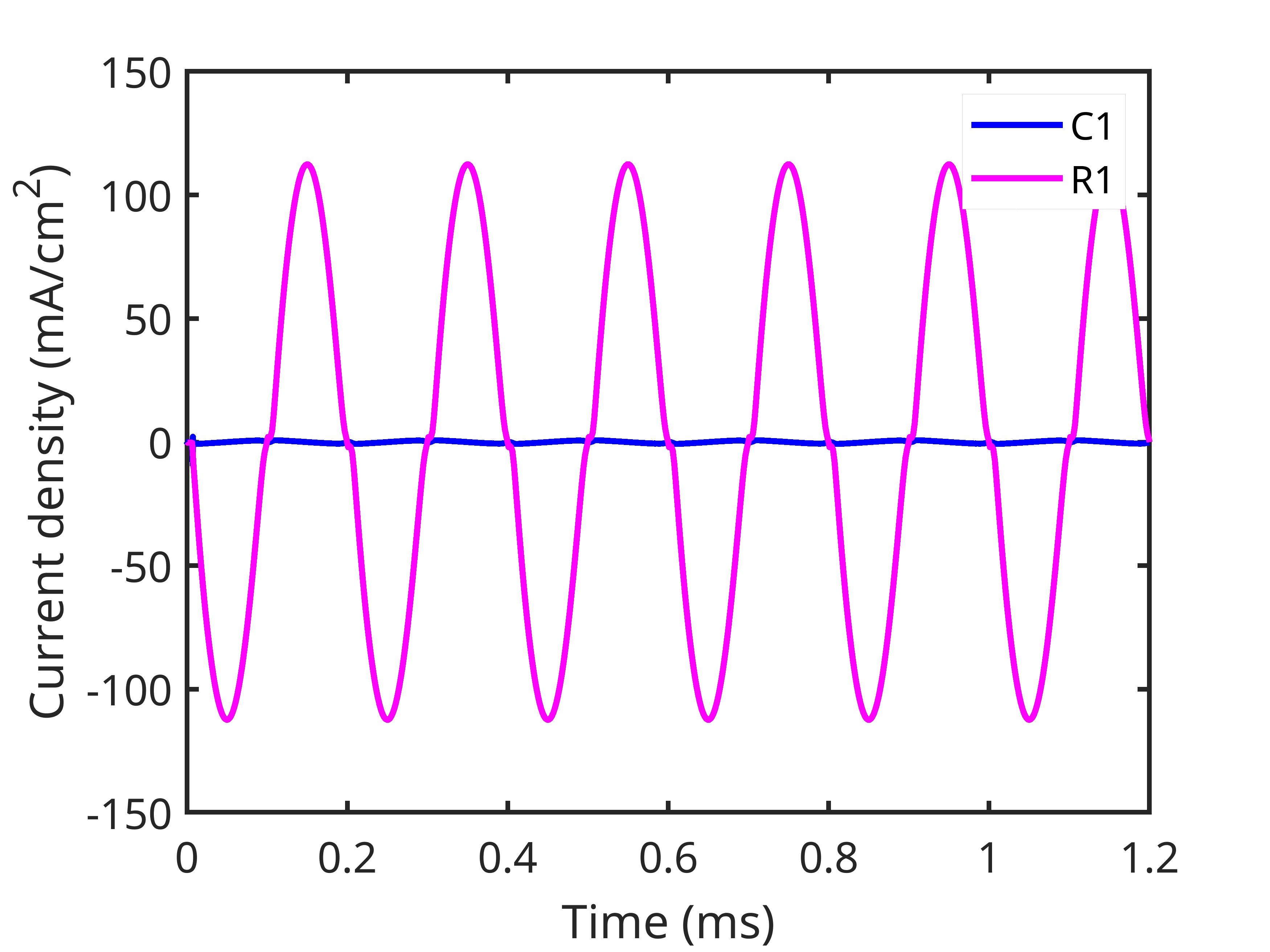} (a)
    \includegraphics[scale=.49]{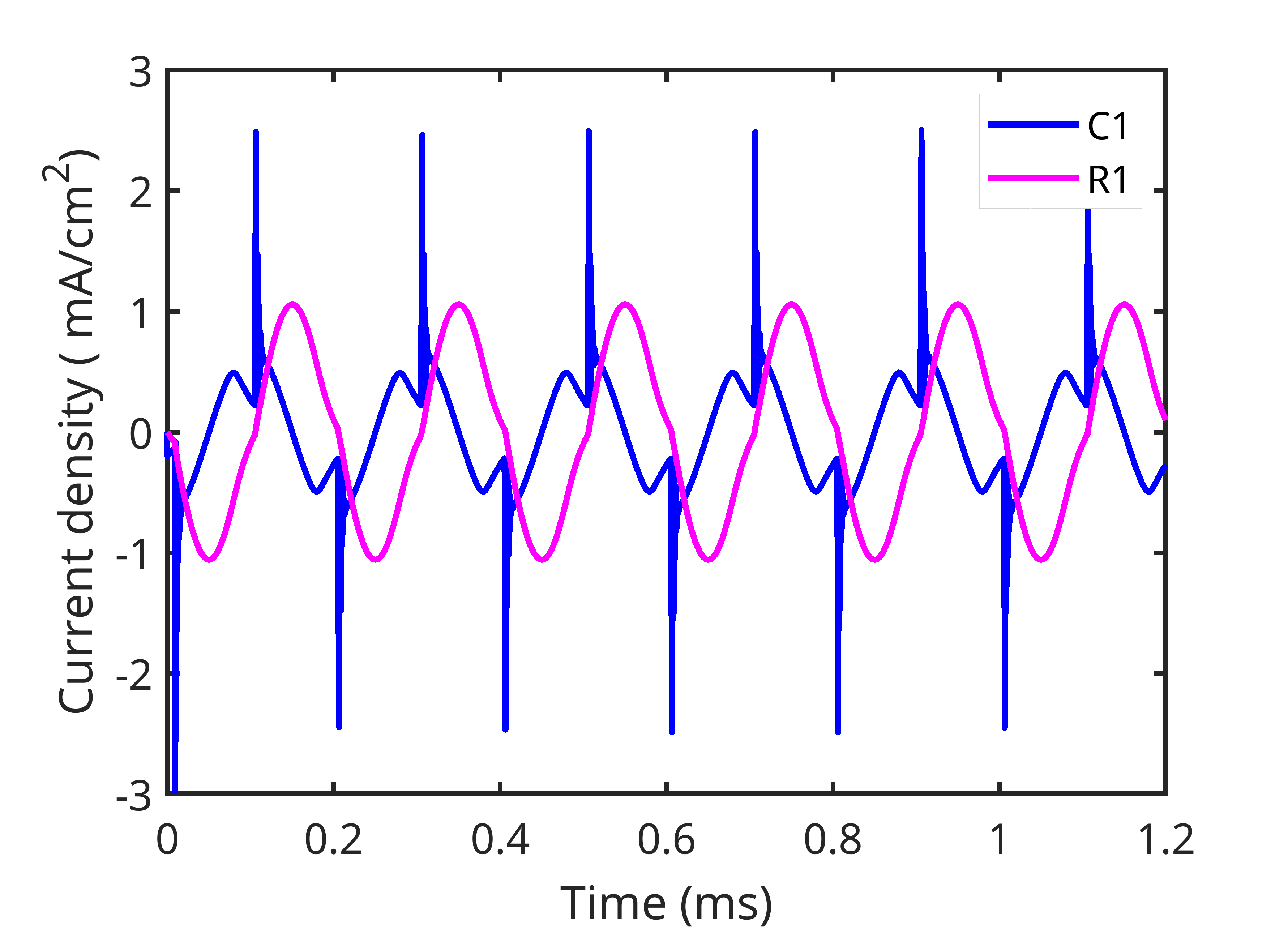} (b)
    \includegraphics[scale=.49]{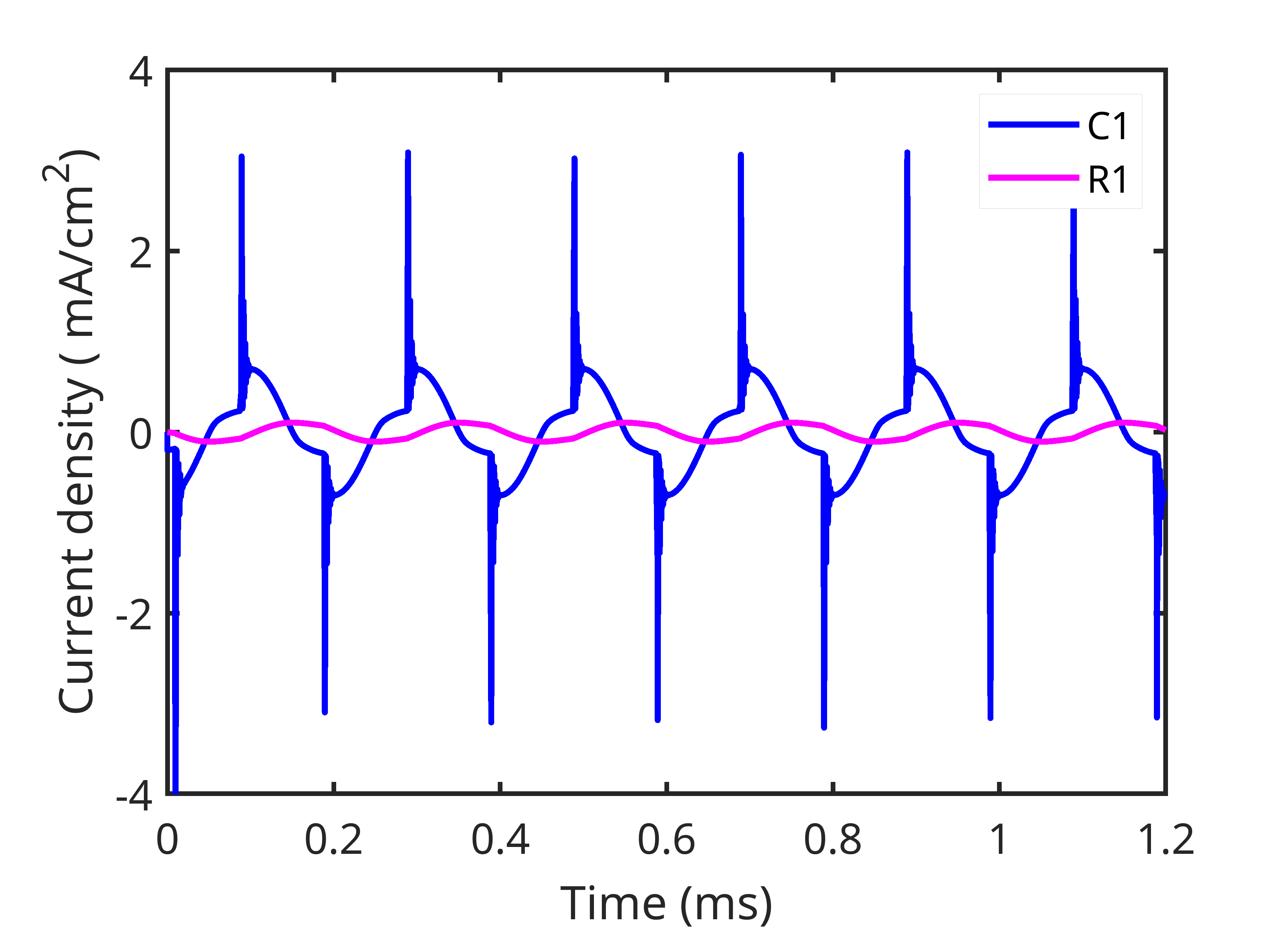} (c)
    \includegraphics[scale=.49]{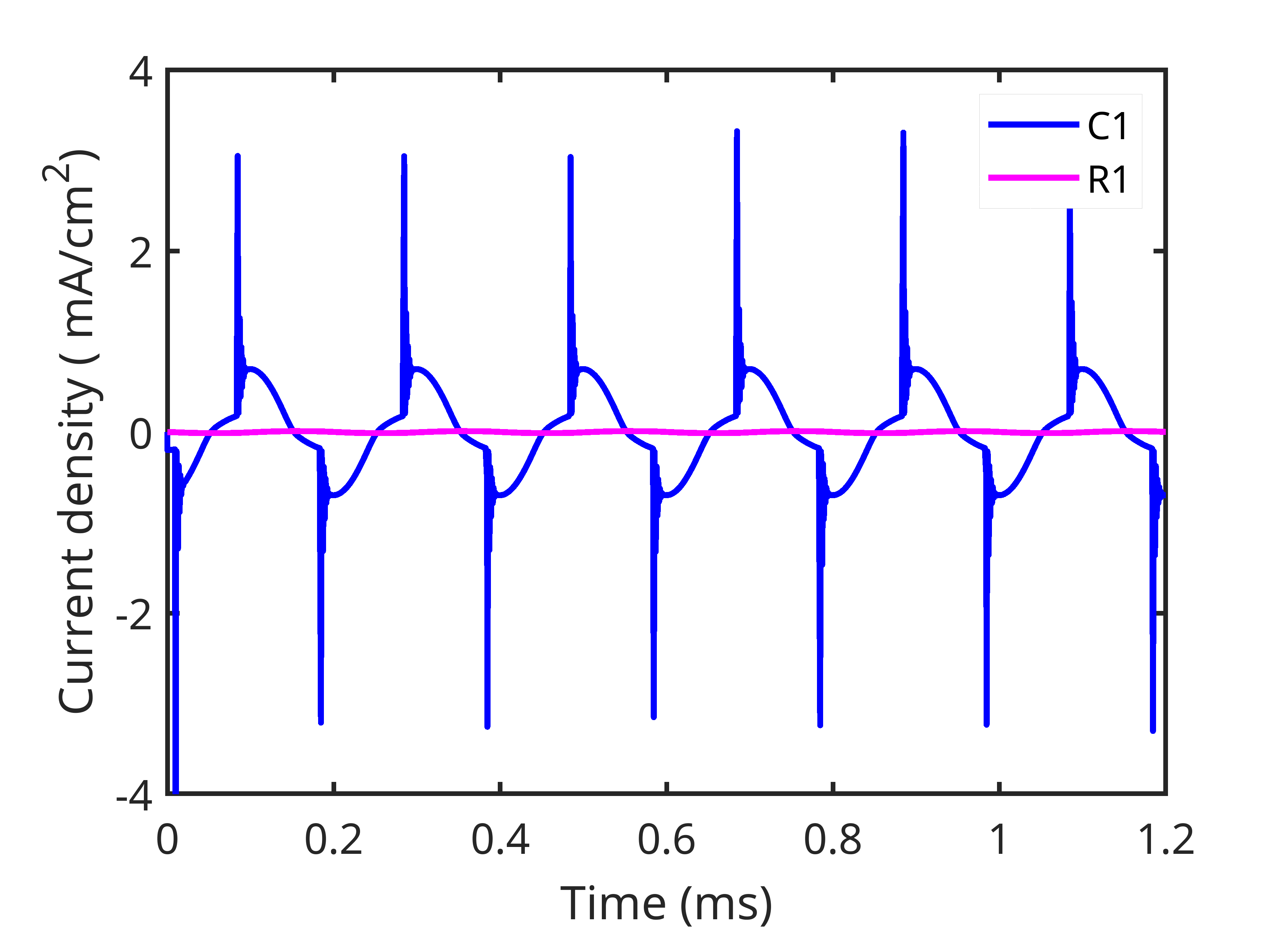}(d)
     \caption{The time evolution of the current through a capacitor and resistor in the lossy dielectric for different values of resistance: (a) 5 $\Omega$, (b) 5 $k\Omega$, (c) 50 $k\Omega$, and (d) 500 $k \Omega$.}
    \label{fig:Resistor_current_density}	
\end{figure}
When the resistance is very low (5 $\Omega$), no current passes through the capacitors, and all the electron current passes through the resistor as depicted in Figure \ref{fig:Resistor_current_density} (a).  The lossy dielectric circuit in this case behaves as a naked electrode DBD. As we increase the resistance from 5 $\Omega$ to 5 k$\Omega$ and 50 k$\Omega$, the electron current flowing through the resistor decreases while allowing the displacement current to pass through the capacitor (panels (b) and (c)). The circuit in this case behaves as a lossy dielectric. With a further increase in the resistance (to 500 k$\Omega$), the electron current through the resistance ultimately reaches zero (panel (d)), and all the displacement current passes through the capacitor. The lossy dielectric circuit in this case behaves as a DBD with a perfect dielectric.

\begin{figure}[h]
    \centering
    \includegraphics[scale=.49]{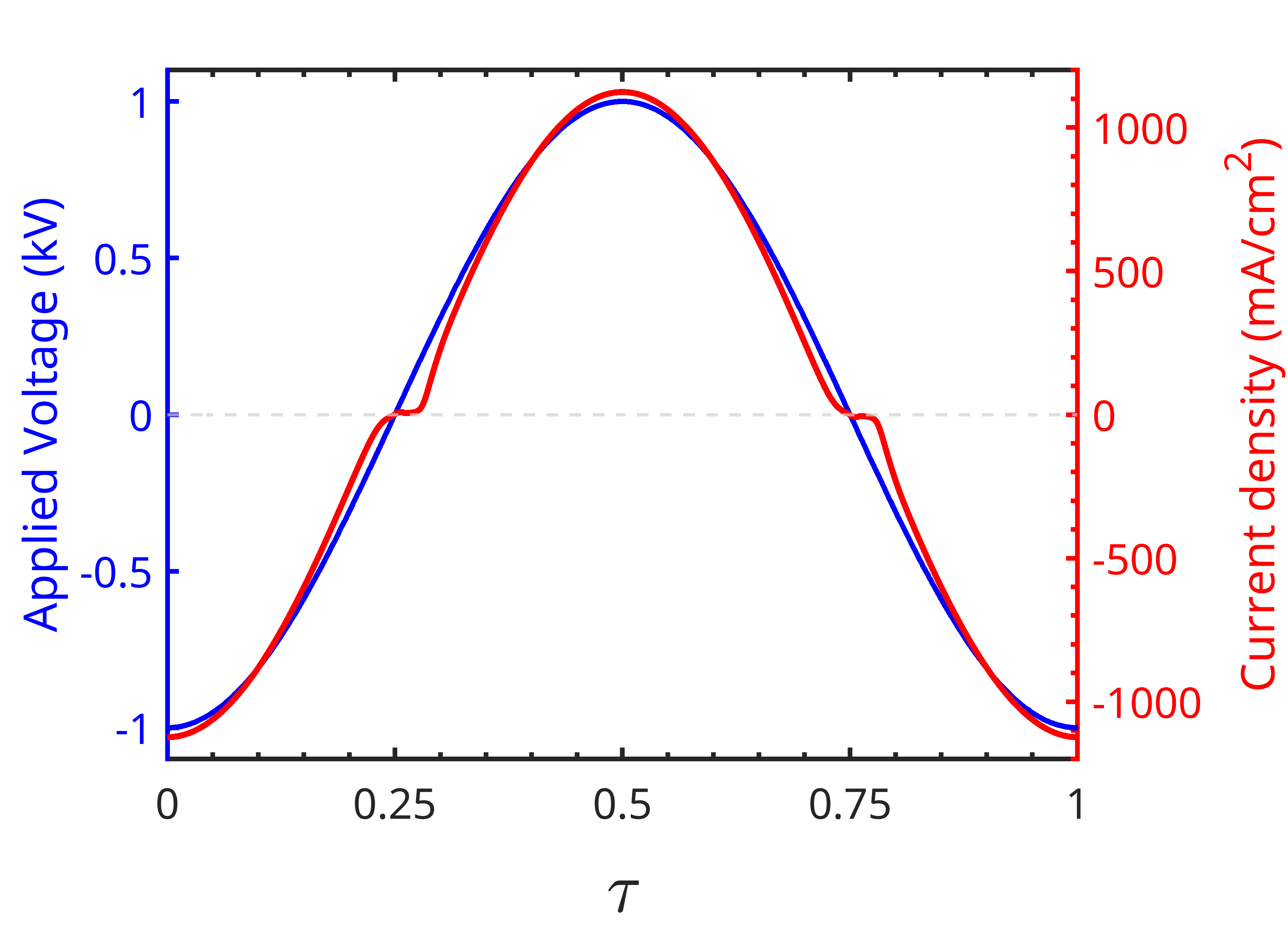}(a)
    \includegraphics[scale=.49]{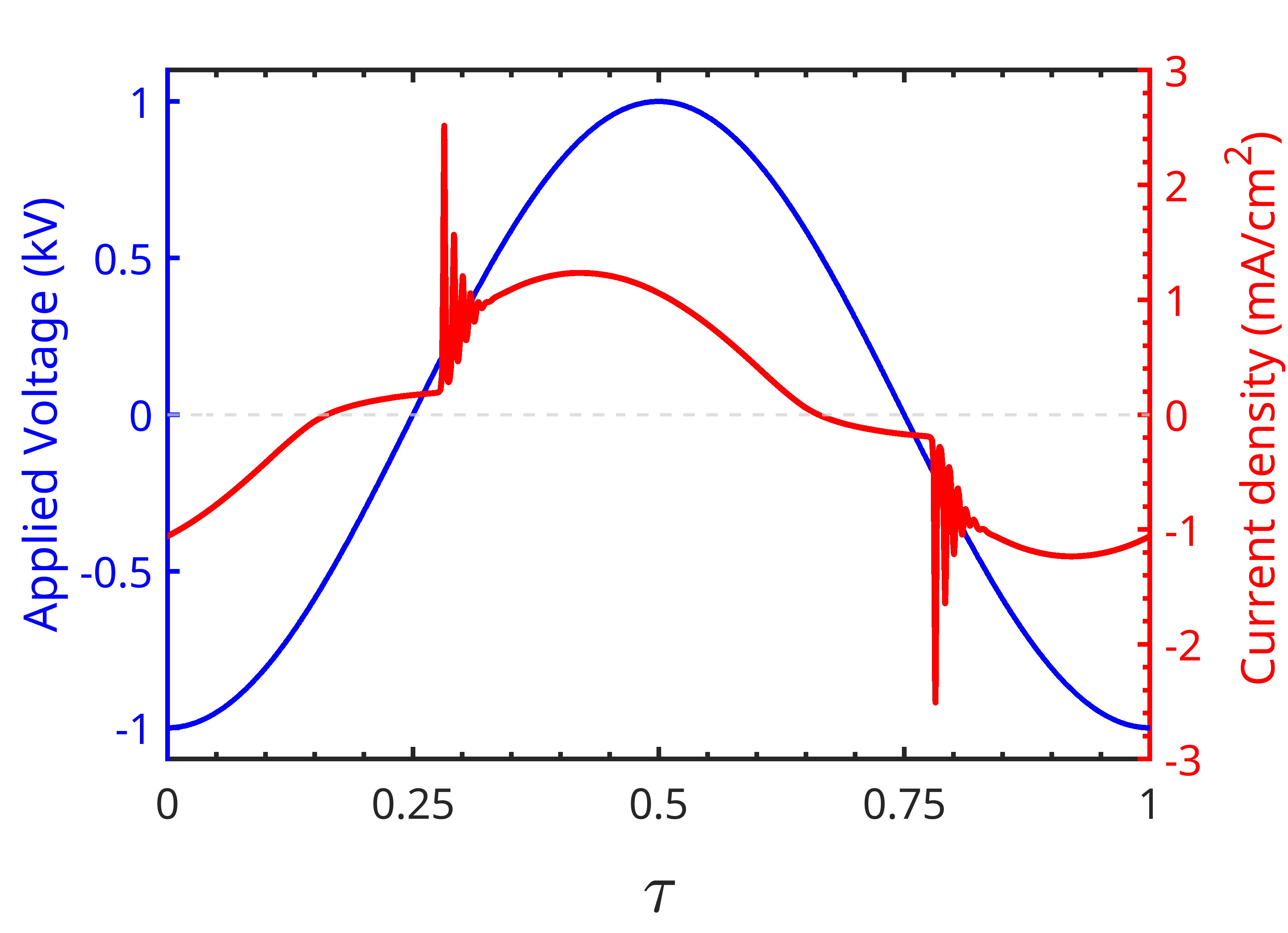}(b)\\
    \includegraphics[scale=.49]{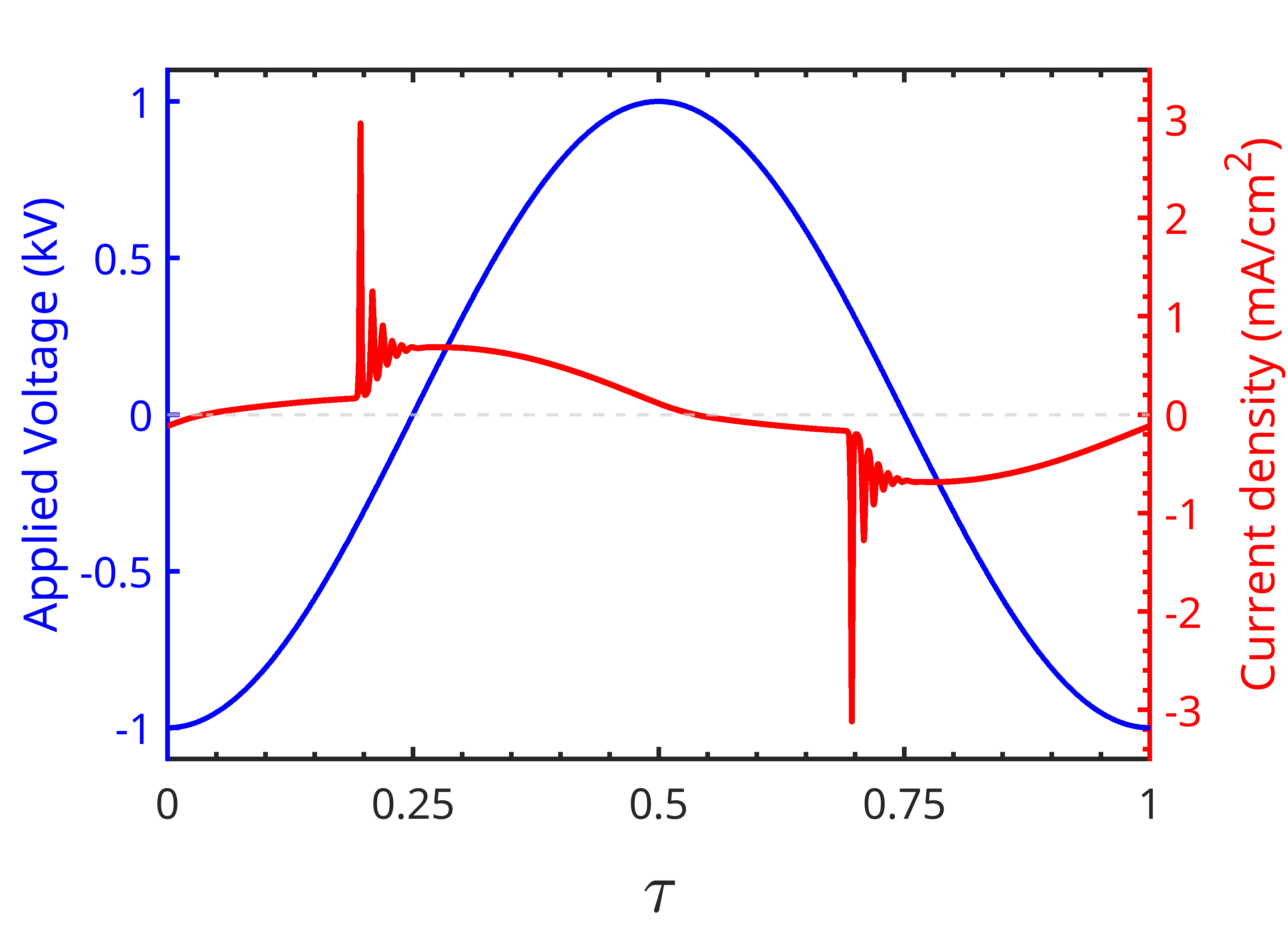}(c)
    \includegraphics[scale=.49]{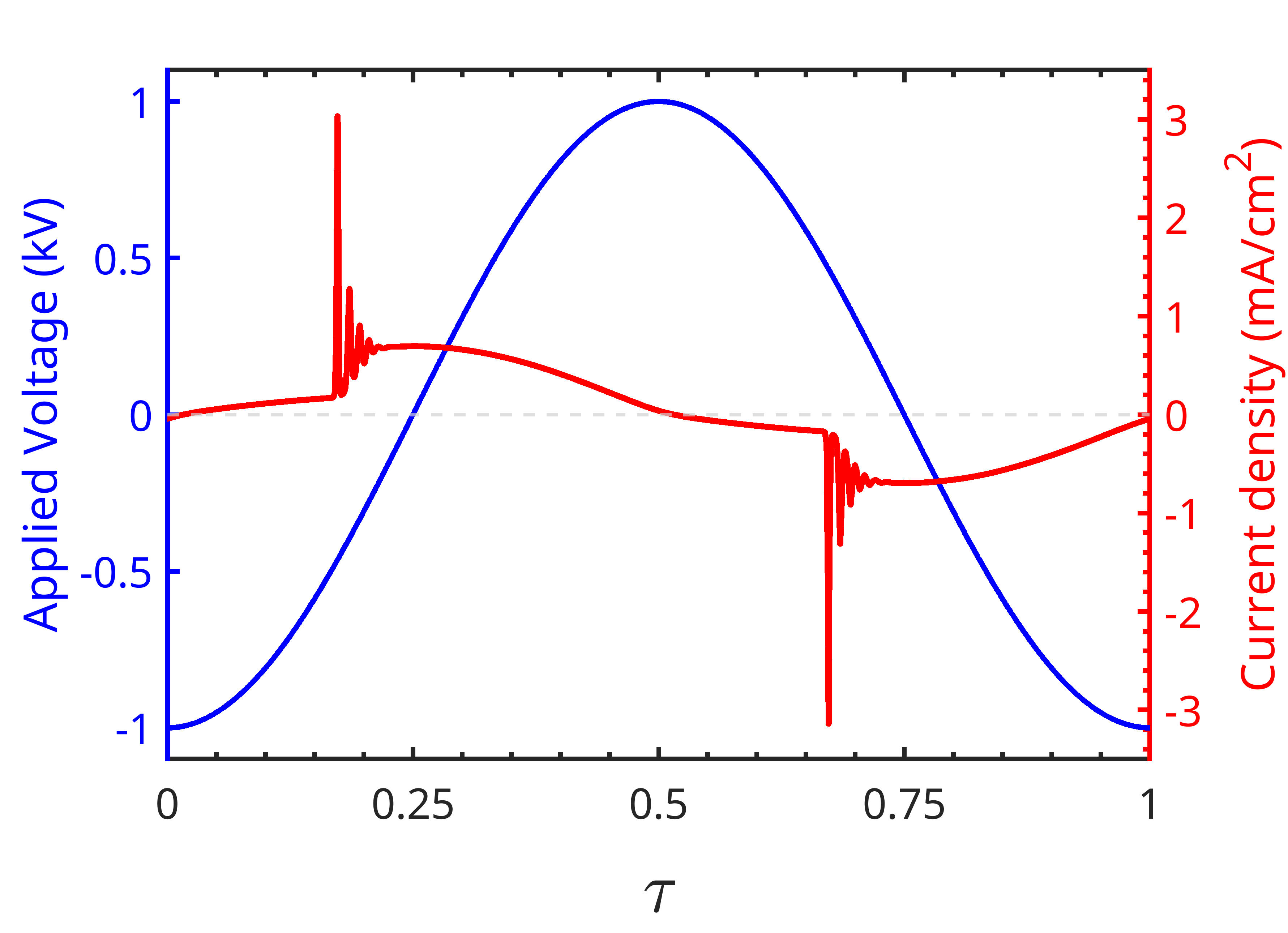}(d)
    \caption{Time evolution of the applied voltage and terminal current density (measured near the plasma domain) for different values of resistance in the RC parallel circuit for lossy dielectrics: (a) 5 $\Omega$, (b) 5 k$\Omega$, (c) 50 k$\Omega$, and (d) 500 k$\Omega$.}
    \label{fig:Resistor_current_volt}	
\end{figure}

The temporal variation of the current density for all four resistance values described above is shown in Figure \ref{fig:Resistor_current_volt}. For the lower value of resistance (R = 5 $\Omega$, panel (a)), the current density and applied voltage are in phase, and there are no current pulses in this case. This behavior is similar to a typical quasi-DC discharge. A similar glow mode of discharge at a low \textit{pd} value of 1.27 Torr cm is obtained by Denpoh et al.\cite{denpoh2020effects} for a CCP discharge. 
For higher resistance (R = 5 k$\Omega$, 50 k$\Omega$, and 500 k$\Omega$) in panels (b)-(d), the current leads the applied voltage by $90 ^\circ$, because of CCP coupling and multiple pulses are observed in all three cases. The magnitude of the current density is slightly increased with the increase in resistance. In short, we can conclude that a dielectric material with a low conductivity (higher resistance) is required to produce discharge in DBD more easily.

\begin{figure}[h!]
    \centering
    \includegraphics[scale=.5]{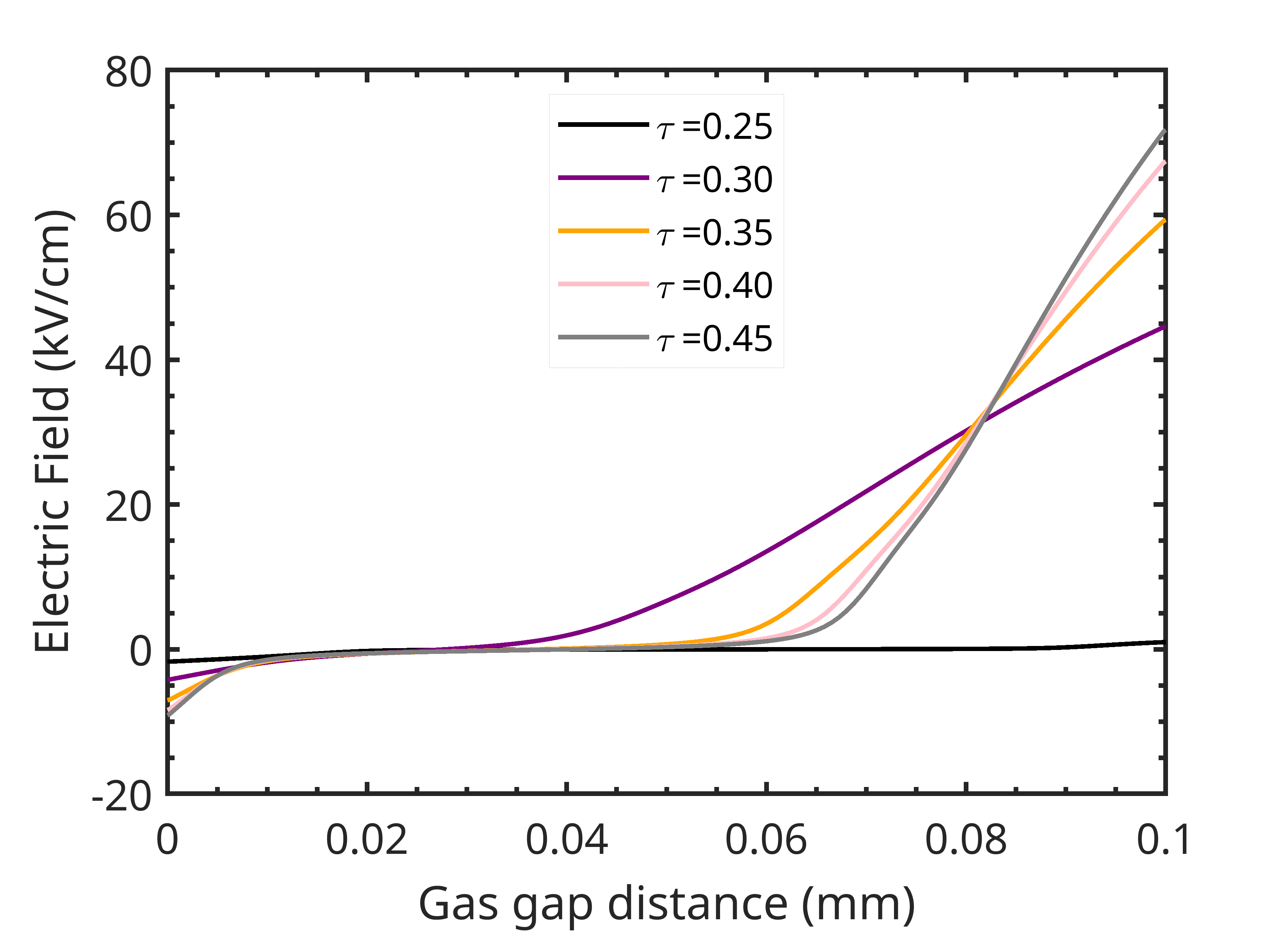}(a)
    \includegraphics[scale=.5]{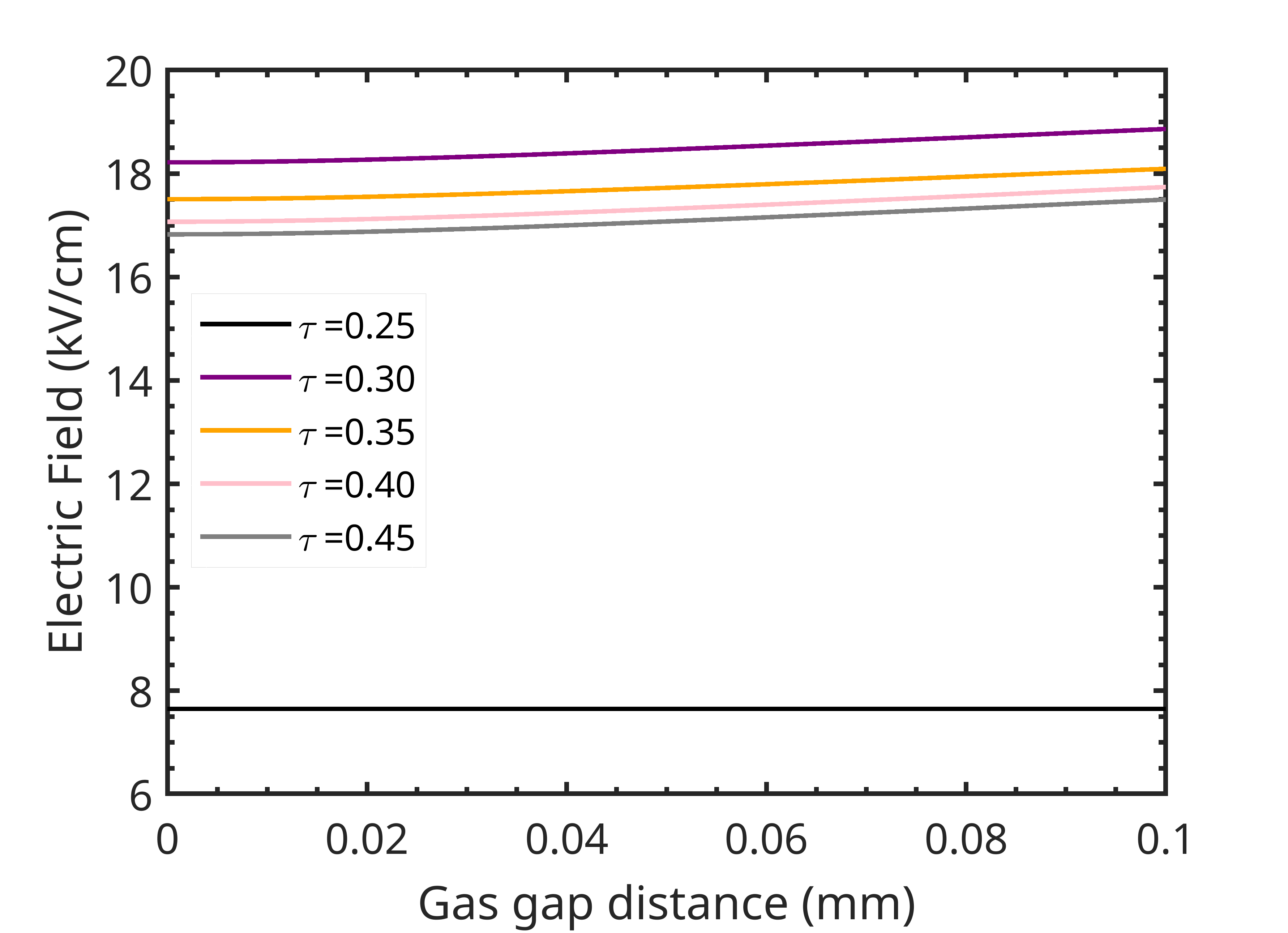}(b)\\
    \includegraphics[scale=.5]{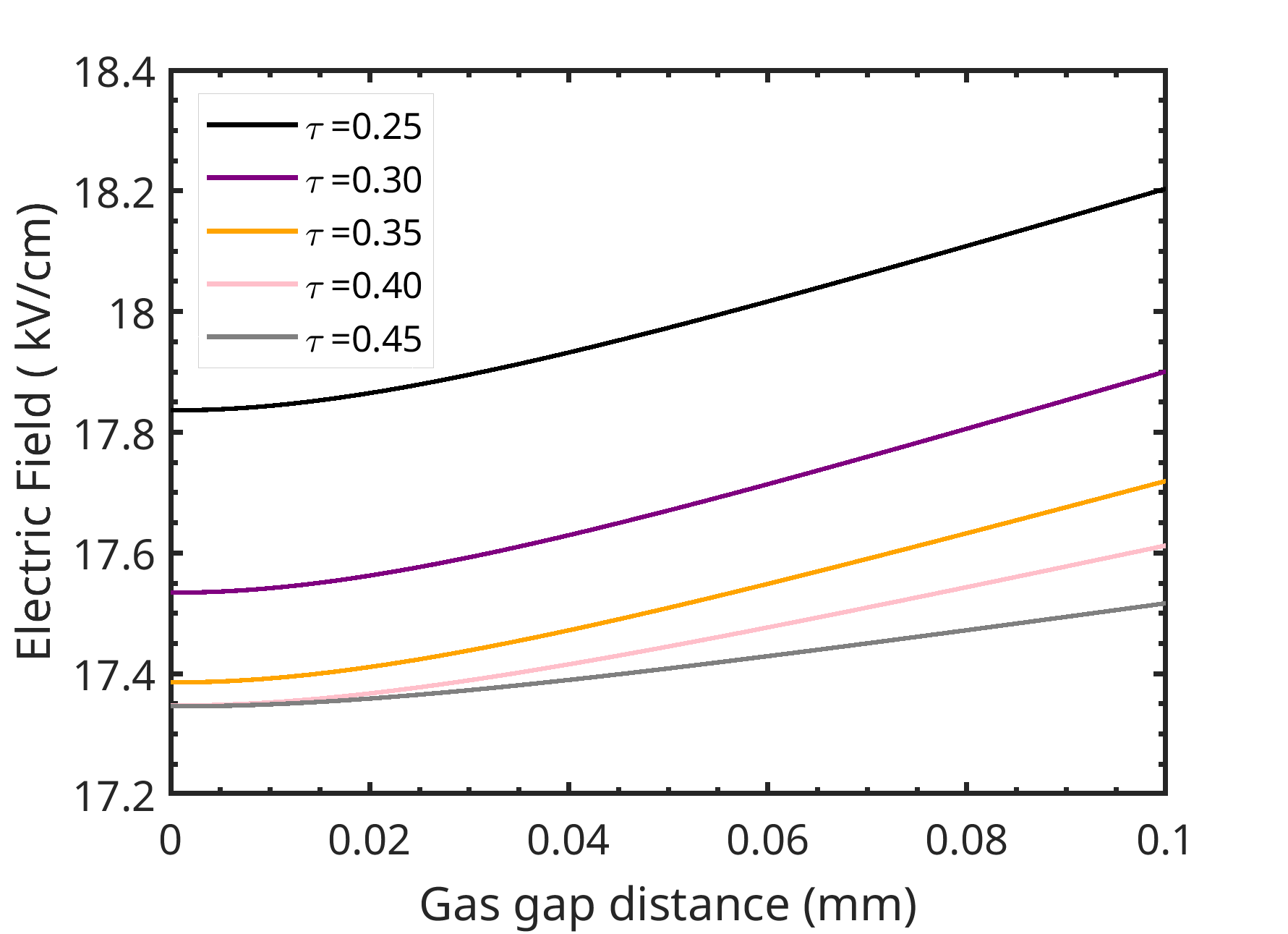}(c)
    \includegraphics[scale=.5]{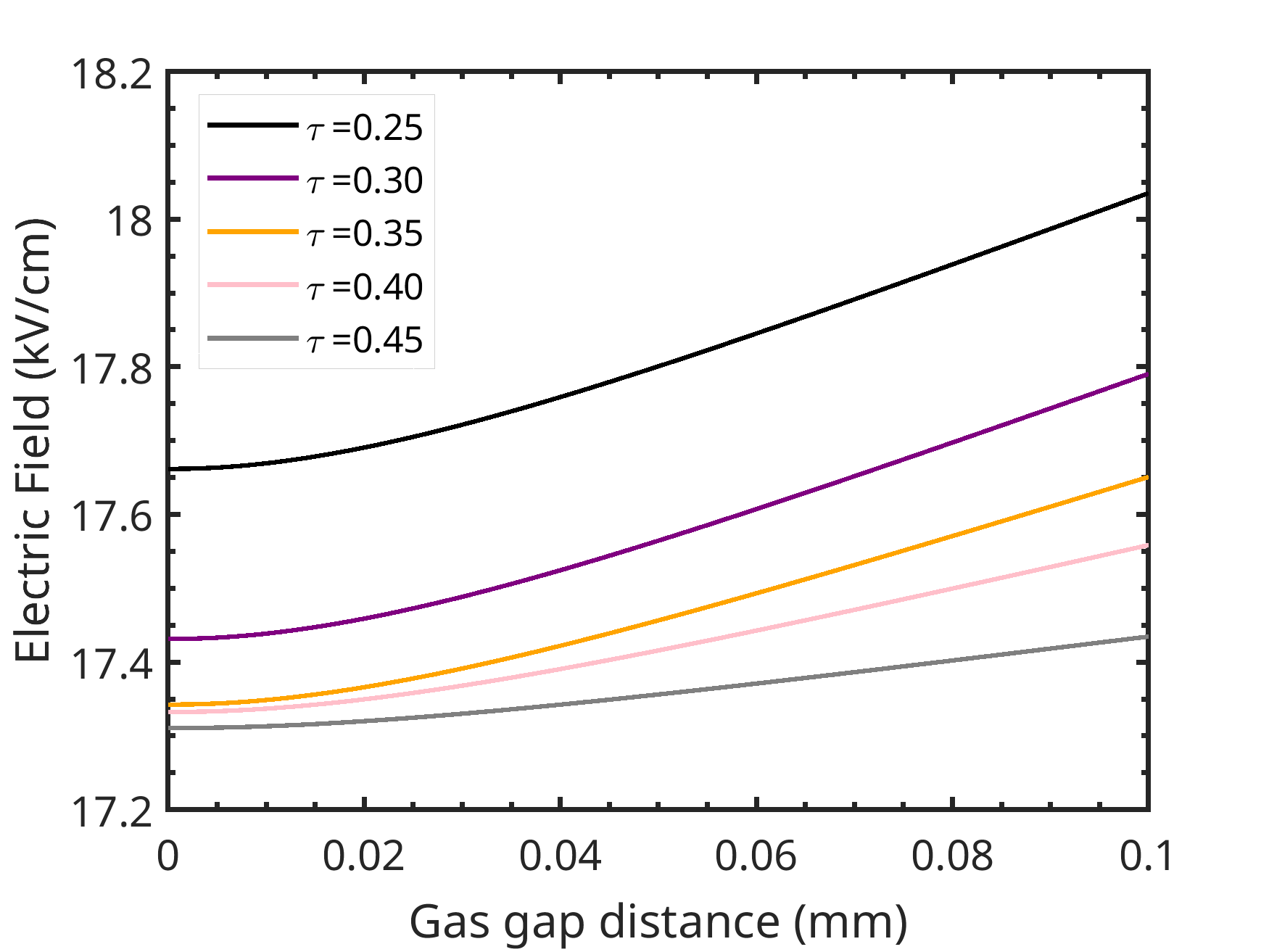}(d)
    \caption{The spatiotemporal evolution of the electric field at five different phases of the half cycle of the AC period of the third cycle for different values of resistance in lossy dielectrics: (a) 5 $\Omega$, (b) 5 k$\Omega$, (c) 50 k$\Omega$, and (d) 500 k$\Omega$.}
    \label{fig:Resistor_Electric_field}	
\end{figure}

\begin{figure}[h!]
    \centering
    \includegraphics[scale=.5]{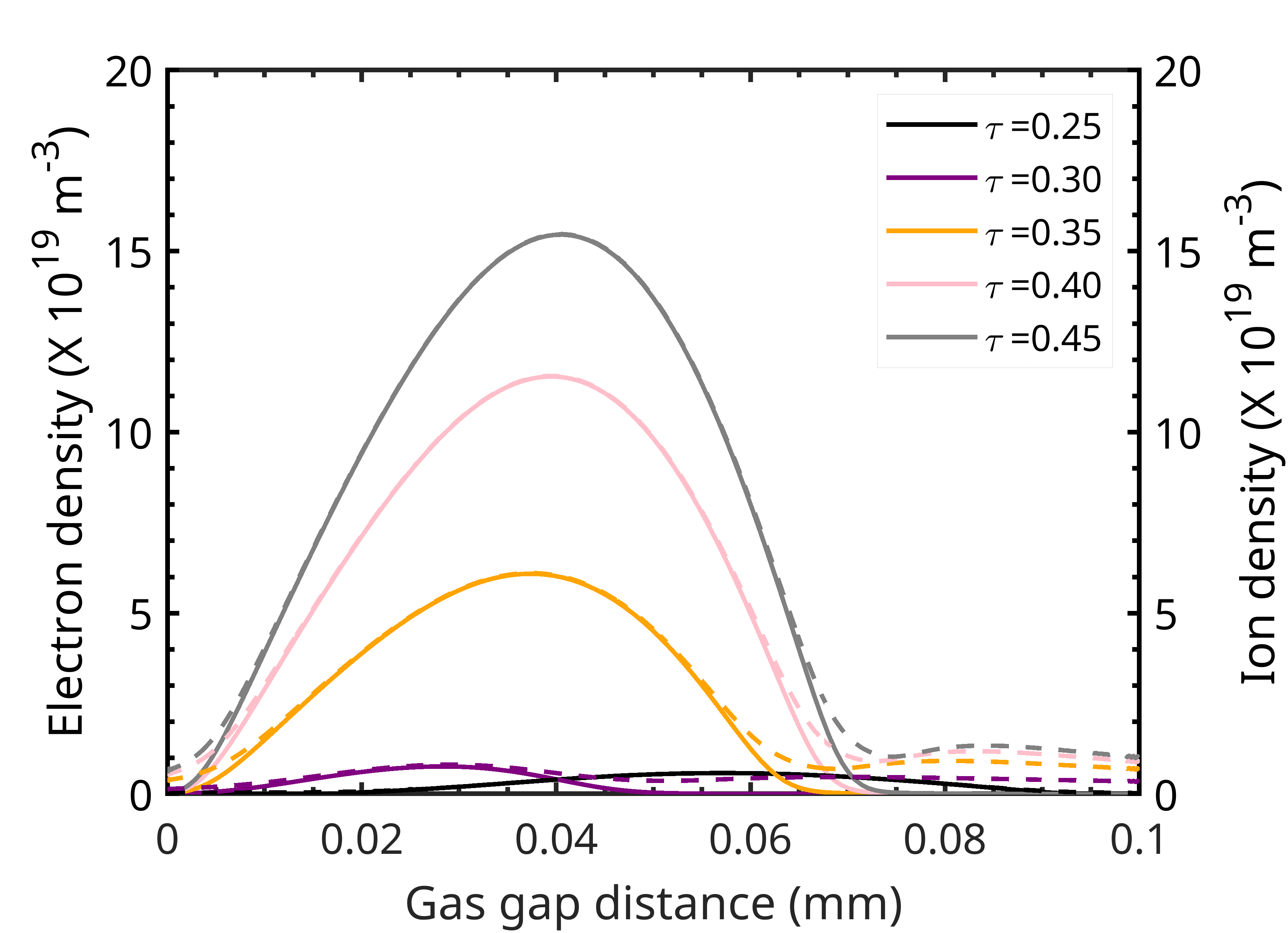}(a)
    \includegraphics[scale=.5]{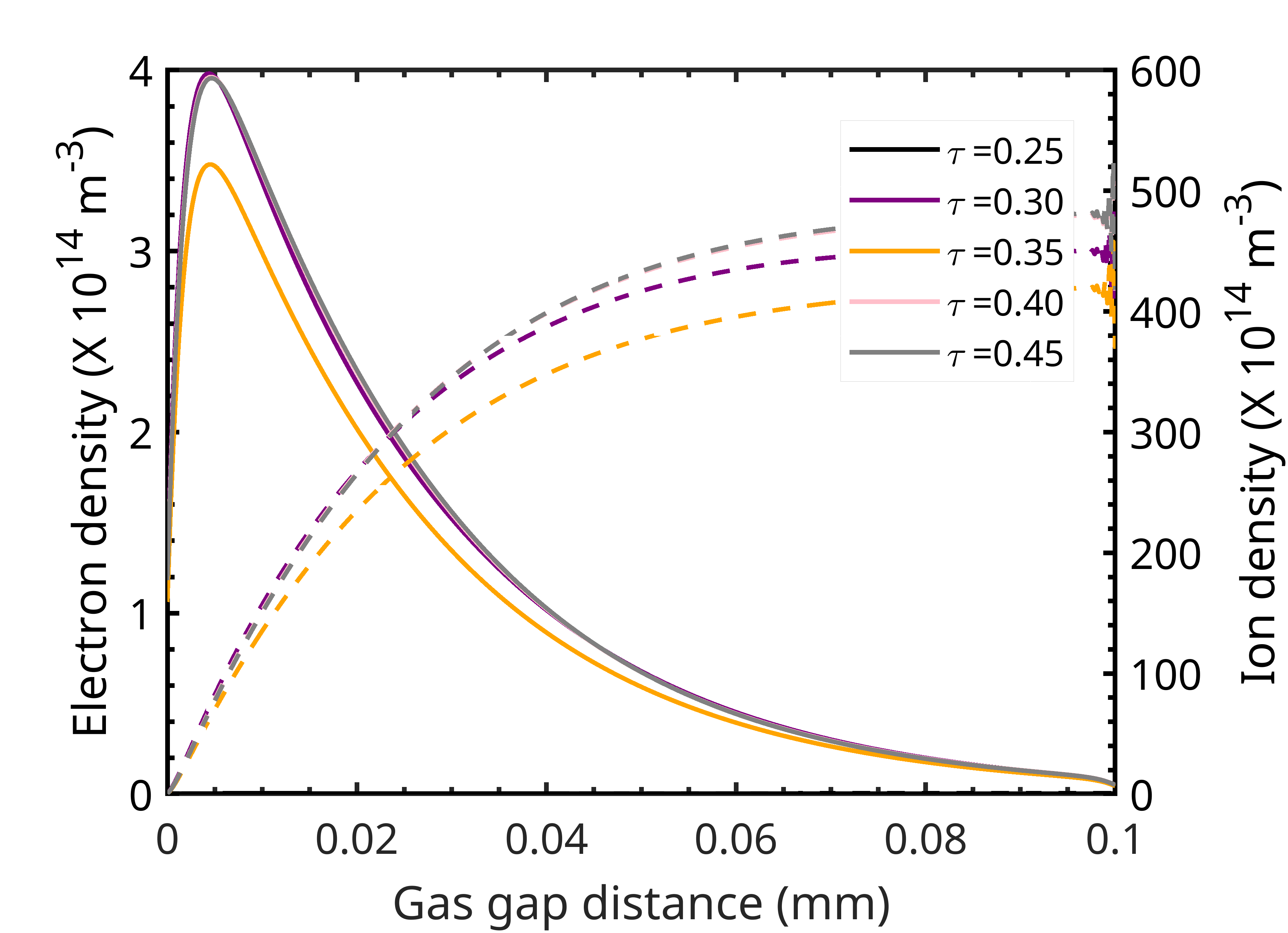}(b)\\
    \includegraphics[scale=.5]{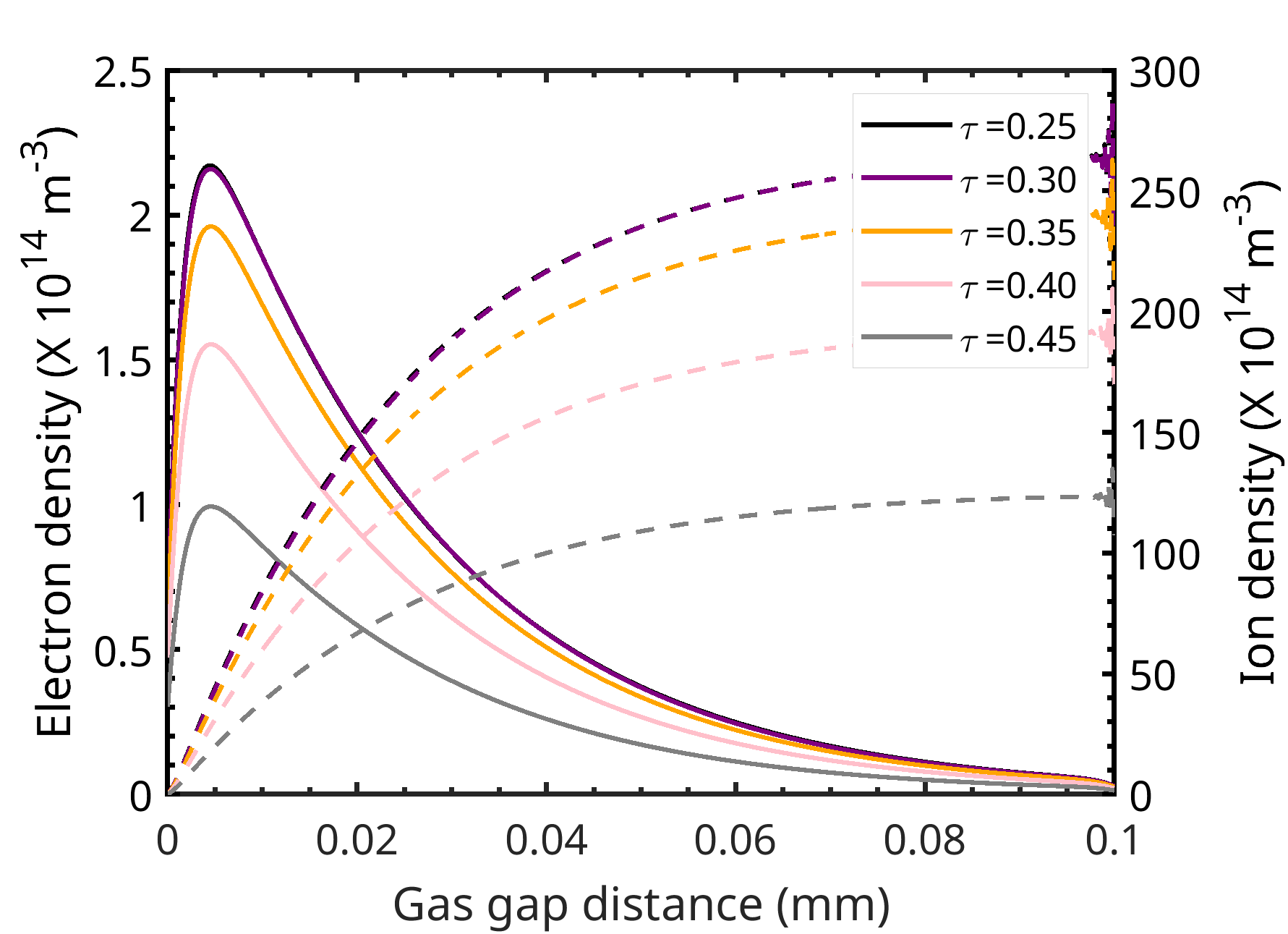}(c)
    \includegraphics[scale=.5]{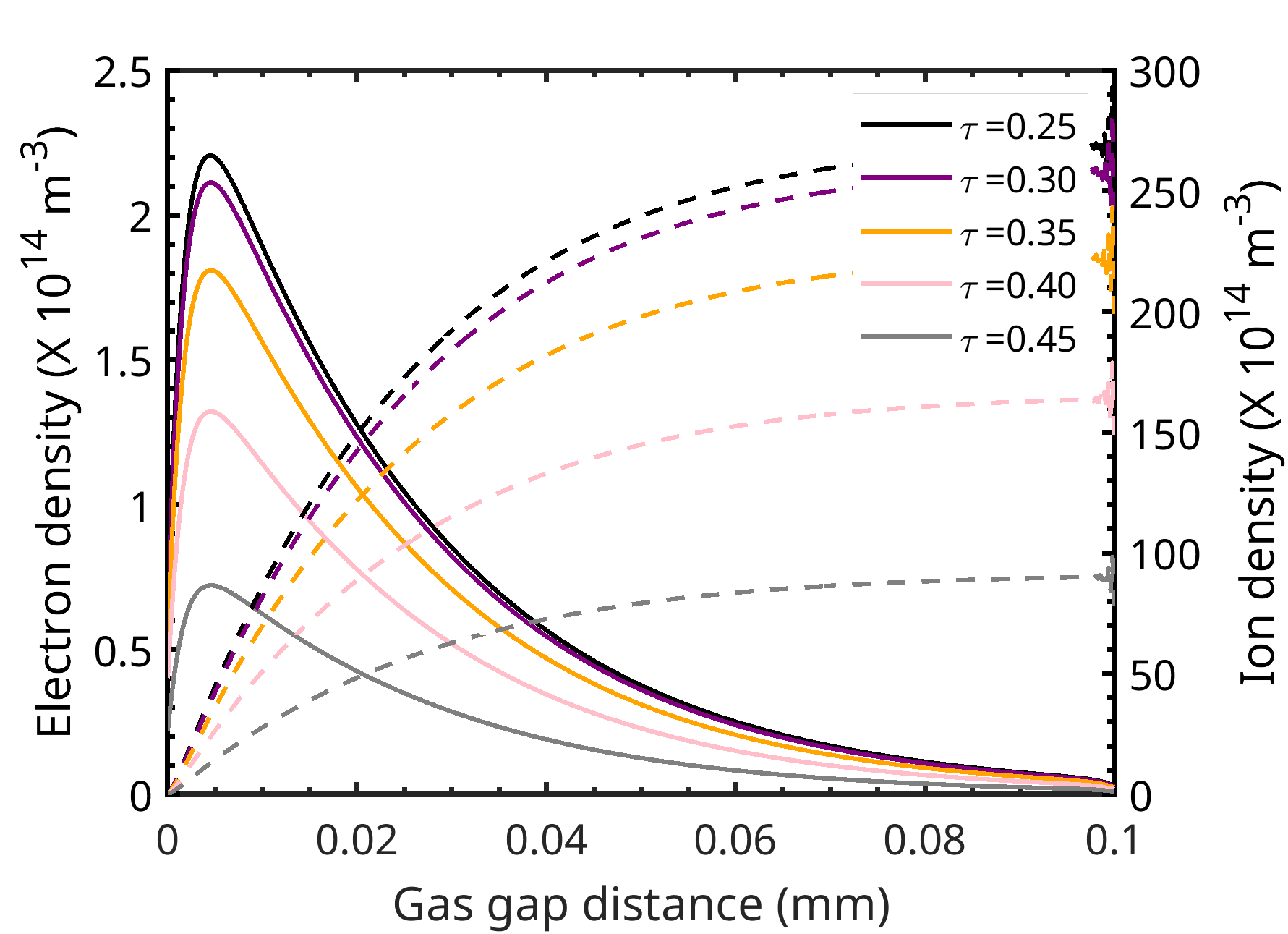}(d)
    \caption{The spatial distribution of the electron density (solid curve) and ion density (dotted curve) at five different phases of the half cycle of the AC period of the third cycle for different values of resistance in lossy dielectrics: (a) 5 $\Omega$, (b) 5 k$\Omega$, (c) 50 k$\Omega$, and (d) 500 k$\Omega$.}
    \label{fig:Resistor_electron_ion}	
\end{figure}

To investigate the mode of discharge in lossy dielectrics in a resistive discharge, we plot the spatiotemporal evolution of the electric field (Figure \ref{fig:Resistor_Electric_field}) and spatial evolution of the electron and ion densities during the first pulse of the positive half cycle (Figure \ref{fig:Resistor_electron_ion}). For the lowest resistance (5 $\Omega$) case, the electric field is not uniform along the gas gap and the electron (solid curve) and ion density (dotted curve) during the pulse are nearly equal $(n_{e} = n_{i})$ in a gas gap, except within the sheath region. Moreover, the electron density is almost zero within the sheath region and the ion density is significantly higher than the electron density. This indicates that the discharge in this case is similar to a quasi-DC glow mode.
For the higher value of resistance (5 k$\Omega$, 50 k$\Omega$, and 500 k$\Omega$ cases), the electric field is uniform in the gas gap. In addition, the electron density is less than the ion density in all three cases. This indicates that the discharge in all three cases of higher resistance is in the Townsend mode.

\subsubsection{Current Pulses for High Permittivity Ferroelectrics}
\begin{figure}[h!]
    \centering
    \includegraphics[scale=.49]{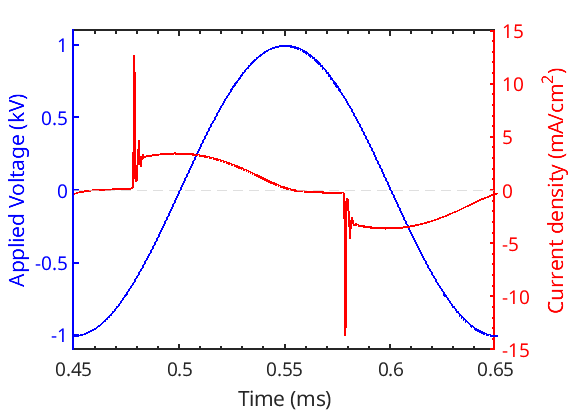}(a)
    \includegraphics[scale=.49]{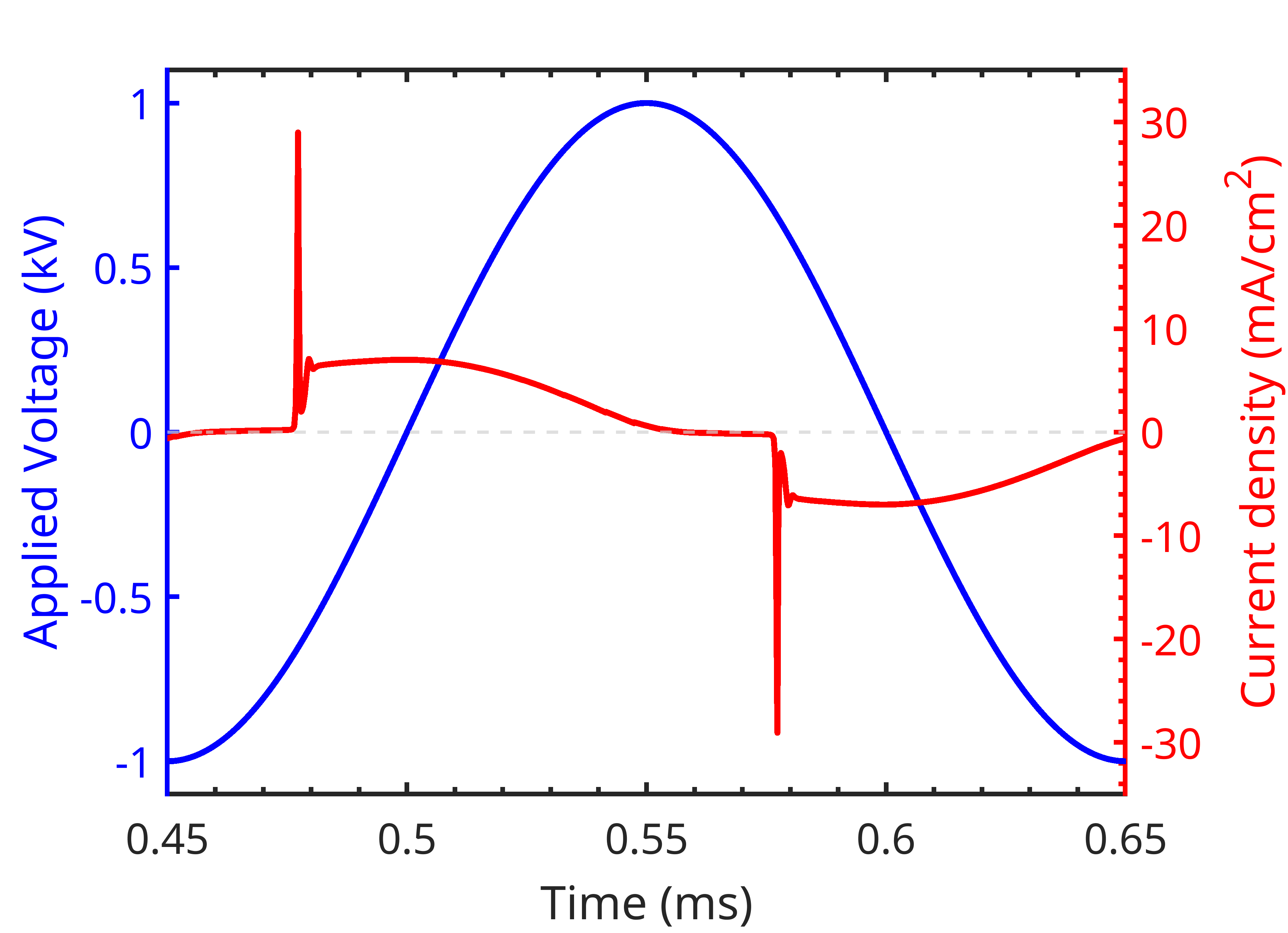}(b)\\
    \includegraphics[scale=.49]{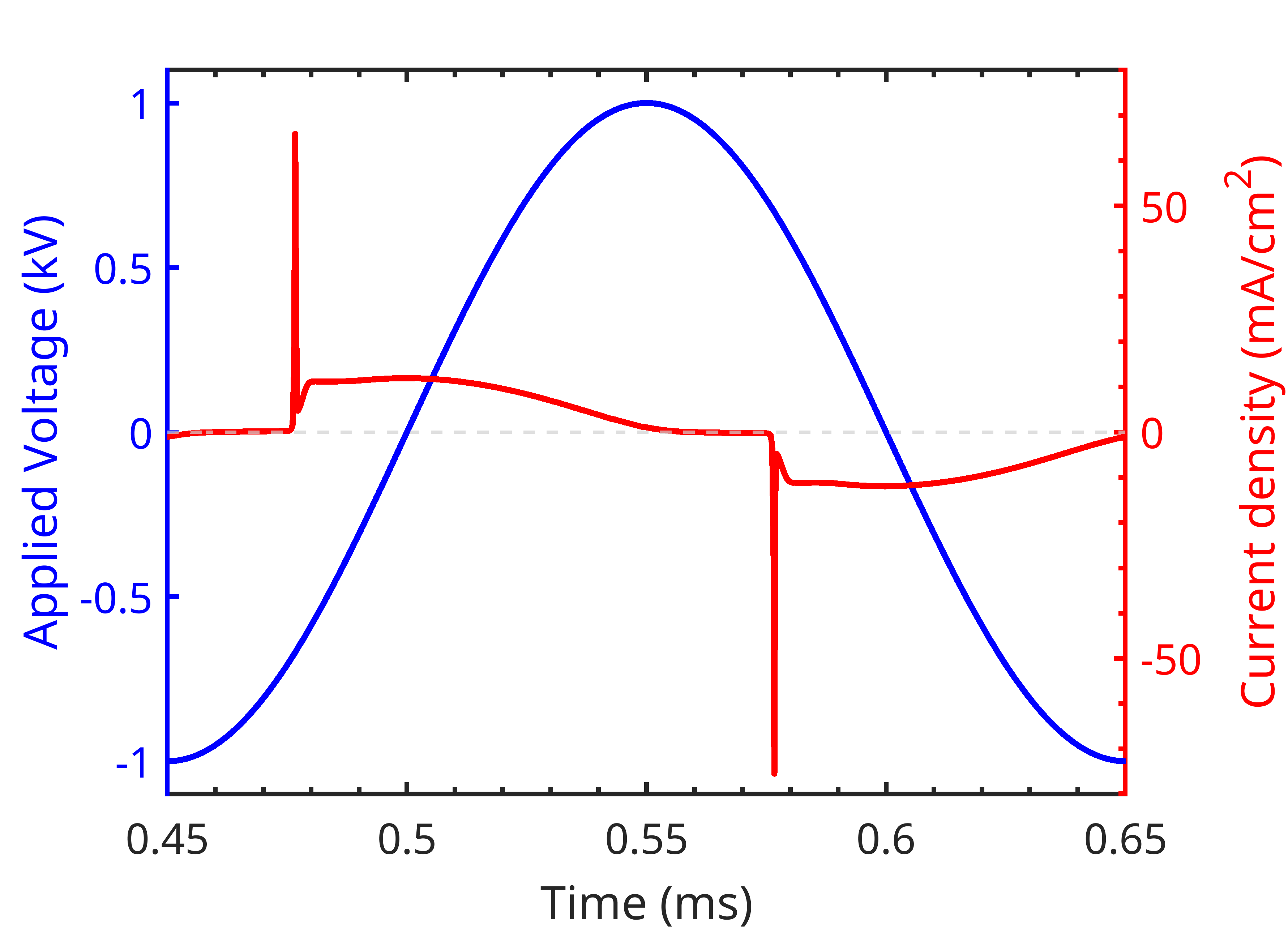}(c)
    \includegraphics[scale=.49]{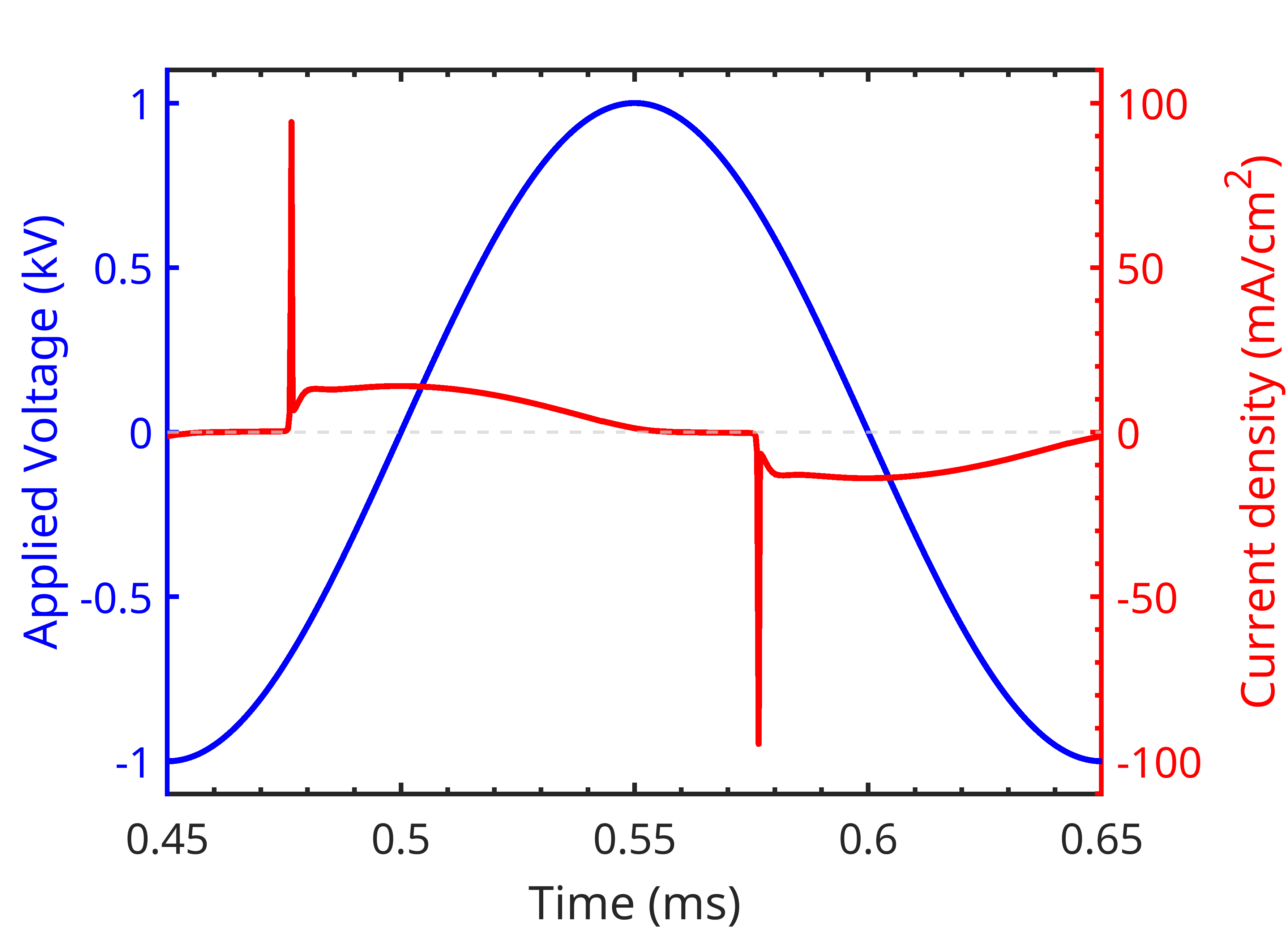}(d)\\
    \includegraphics[scale=.49]{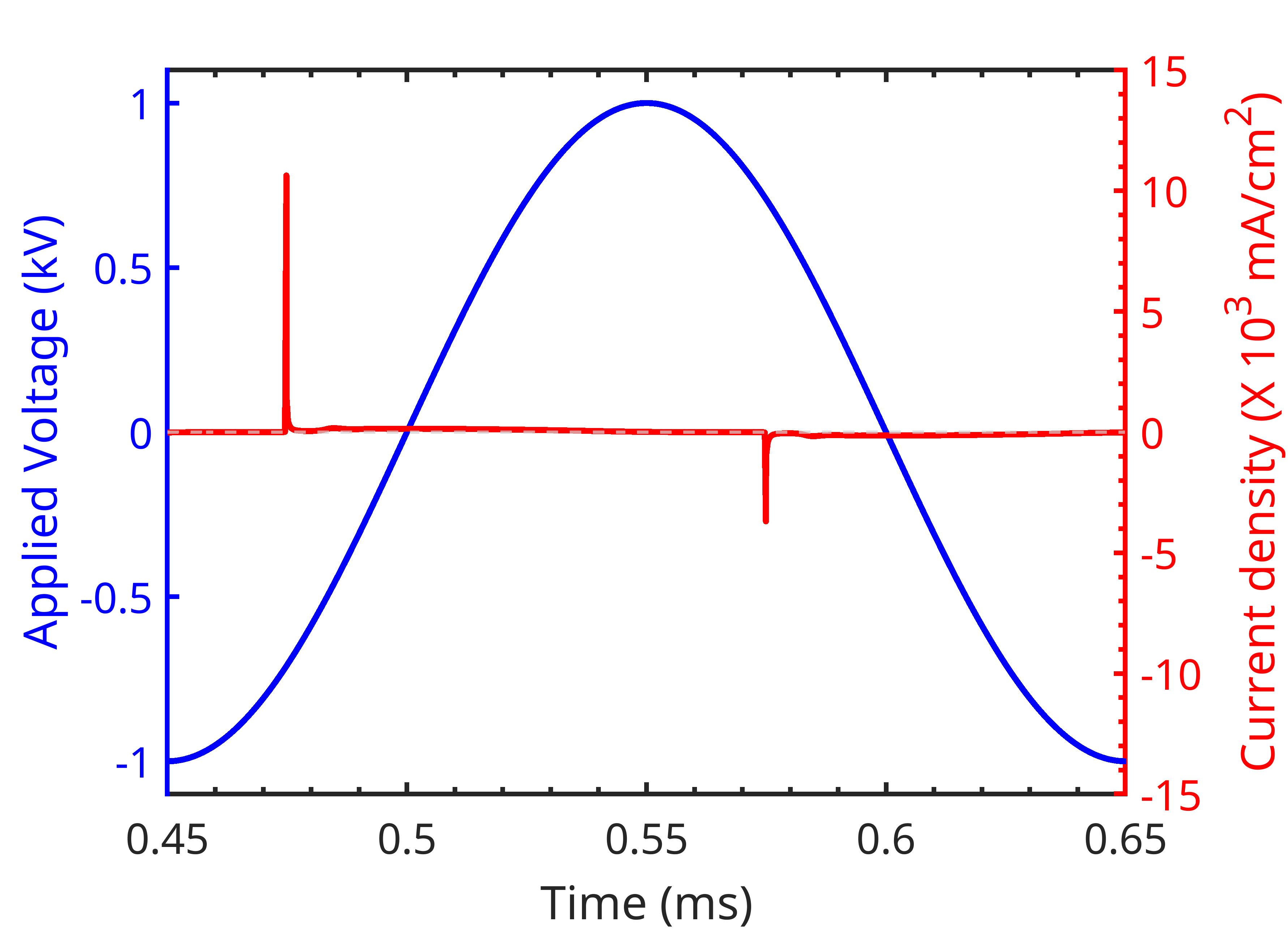}(e)
    \includegraphics[scale=.49]{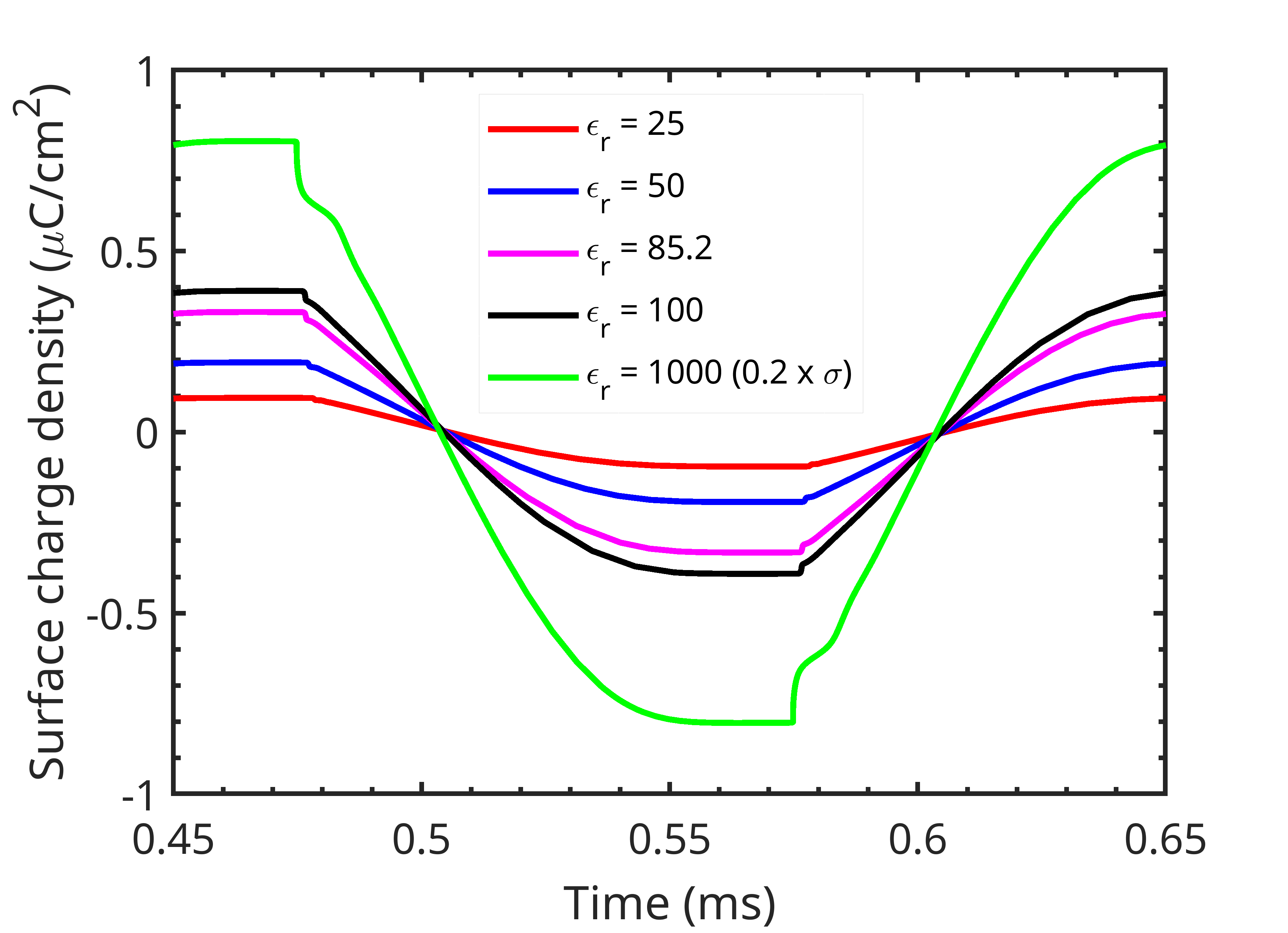}(f)
    \caption{Time evolution of the current density multi pulses in the DBD with large values of permittivity in the ferroelectrics, (a) $\epsilon_{r} = 25$, (b) $\epsilon_{r} = 50$, (c) $\epsilon_{r} = 85.2$ (water), (d) $\epsilon_{r} = 100$, and (e) $\epsilon_{r} = 1000$. (f) The temporal evolution of surface charge density on the powered electrode (left) for all three choices of permittivity in the ferroelectrics. Other simulation parameters are kept the same as in the base case. Note that the surface charge density is scaled by a factor of 0.2 for $\epsilon_{r} = 1000$ to highlight its greater magnitude within the Y-axis range.}
    \label{fig:Ferroelectrics}	
\end{figure}
In this section, we study the dynamics of the atmospheric pressure barrier argon plasma using ferroelectric materials as a dielectric layer ferroelectric barrier discharge (FBD)\cite{navascues2019large}. Outstanding properties of ferroelectrics are high permittivity and a large value of the secondary electron emission coefficient. For ferroelectric materials with relative permittivities from 25 to 1000, we observed a significant enhancement in the current density, even for the same value of $\gamma$ as shown in Figure \ref{fig:Ferroelectrics} ((a)-(e)). The higher discharge current density for a large value of permittivity is due to their stronger polarization charge \cite{yao2019transition} that facilitates a much higher charge accumulation onto the electrode surface as seen on panel (f) causing an enhancement in the discharge process and hence current density. Note that the ferroelectric with relative permittivity of  85.2 in Figure \ref{fig:Ferroelectrics} (c) is lithium niobate $(LiNbO_{3})$ and this is close to the permittivity value of water (80.10). Hence the result obtained in this case could be similar to using water as liquid dielectrics.

\newpage
\section{Conclusions} \label{sec:conclusions}
In this paper, we have investigated the peculiarities of atmospheric pressure DBD in Argon at low driving frequencies using a one-dimensional fluid plasma model solved with COMSOL. In particular, we have explored the dependence of current pulses on several key parameters, including \textit{pd} values, driving frequency, and properties of dielectrics (lossy dielectric in resistive barrier discharges, and large value of dielectric permittivity in ferroelectrics ). In addition, we have observed the transition of the discharge mode from Townsend to dynamic CCP by changing \textit{pd} values.

We performed our simulations for a wide range of \textit{pd} values ranging from 7.6 to 760 Torr cm (all on the right branch of the Paschen curve) and the gap distance from 0.1 mm to 10 mm.
Our results show that several current pulses per half-cycle are observed over a wide range of \textit{pd} values. The discharge is in the Townsend mode for the lowest \textit{pd} value of 7.6 Torr cm (close to the Paschen minimum). For the higher \textit{pd} values of  228 and 760 Torr cm, the dynamic regime of the CCP has been observed with significant oscillations in electron temperature over an AC period. Our study illustrates that multiple current pulses per AC period can occur in the Townsend discharge and the dynamic CCP mode because of the decoupling of electron and ion motion in the absence of quasineutrality in the gap (in Townsend mode) or the sheath (in CCP). We explain that the effect is more easily observed in Helium than in Argon because of the different ion mobilities in these gases. We argue that the negative differential resistance \citep{akishev2001pulsed} may not be necessary for self-pulsing, and expect that this effect can also be observed on the left branch of the Paschen curve where kinetic models for electrons and ions have to be used.

We also investigated the impact of driving frequency on the discharge dynamics and the transition from the Townsend mode to the glow mode. For the frequencies of 5 kHz and 10 kHz, multiple current pulses are observed and the number of current pulses is proportional to $\omega^{-\frac{1}{2}}$. For these two cases, the discharge is in the Townsend mode. However, at a higher frequency of 25 MHz, a single pulse-like feature is observed, and the discharge is in the high-frequency CCP mode, with the electron and ion densities in plasma remaining nearly constant.  

In addition, we performed simulations of resistive discharges with lossy dielectrics using an RC circuit equivalent of a lossy dielectric. By varying the value of resistance, we can replicate the lossy nature of dielectrics with (i) the lowest resistance of 5 $\Omega$ representing naked electrodes where all the current passes through the resistor, and (ii) the highest resistance of 500 k$\Omega$ representing a pure dielectric where all current passes through the capacitor. For the lowest resistance of $5 \Omega$, the current density and voltage are in phase, and the densities of ions $n_{i}$ and electrons $n_{e}$ are nearly equal within the gas gap, except for the sheath region where $n_{e}$ is zero. These characteristics suggest that the discharge behavior closely resembles that of a quasi-DC discharge in the glow mode. As the resistance is increased, the RC circuit starts behaving like a lossy dielectric in a resistive discharge, and ultimately, at 500 k$\Omega$, the circuit exhibits typical DBD behavior. The discharge in the case of higher resistance is in the Townsend mode as indicated by a uniform electric field along the gas gap and much lower electron density during the pulse compared to the ion density. Surface charges are essential for the formation of current pulses. In our paper, we demonstrate that the number of pulses gradually decreases with increasing electric conductivity of lossy dielectrics. We also demonstrate the transition from quasi-static to dynamic and high-frequency regimes of discharge operation and the effects of the \textit{pd} values on plasma dynamics. Finally, we performed simulations of DBD for large values of permittivity corresponding to ferroelectric materials.  Higher current density was observed and the number of pulses decreased with the increase in the permittivity.

Based on our analysis, self-pulsing of DBD appears similar to the subnormal oscillations in DC discharges and Trichel pulses in corona discharge. They all happen in the regime, where quasineutrality is absent and the electron and ion transport is decoupled because the electric field is weakly perturbed by space charges \cite{zhang2023numerical}. However, additional studies are desirable because the simplest fluid model with a Maxwellian EEDF used in our studies has limited applicability and accuracy. We expect that using fluid models with non-Maxwellian EEDF will not change our results quantitatively. The analysis of non-local electron kinetic effects on self-pulsing phenomena could be the subject of future studies.

\begin{acknowledgments}
This work was supported by the NSF EPSCOR projects OIA-1655280 and OIA-2148653. 
S.T. thanks Jun-Chieh (Jerry) Wang and Bhagirath Ghimire for the helpful discussions.
S.T. also thanks the COMSOL support team for their technical support.
\end{acknowledgments}
\section*{AUTHOR DECLARATIONS}
\subsection*{Conflict of Interest}
The authors have no conflicts to disclose.
\subsection*{Author Contributions}
\textbf{Shanti K. Thagunna}: Conceptualization (supporting); Formal analysis (lead); Writing – original draft (lead). \textbf{Vladimir I. Kolobov}: Conceptualization (lead); Formal analysis (supporting); Writing – original draft (supporting). \textbf{Gary P. Zank}: Funding (lead); Formal analysis (supporting); Writing – original draft (supporting).
\section*{Data Availability}
The data that support the findings of this study are available from the corresponding author upon reasonable request.

\bibliographystyle{apsrev4-1}
\bibliography{Shanti_aip_pop}
\end{document}